\documentclass[12pt]{article}
\usepackage{a4p}
\usepackage{epsfig}
\usepackage{array}
\usepackage{cite}
\usepackage{pennames}
\usepackage{rotating}
\newcommand{\etal}      {{\it et~al.}}
\newcommand{\PhysLett}  {Phys.~Lett.}
\newcommand{\PhysRep}   {Phys.~Rep.}
\newcommand{\PhysRev}   {Phys.~Rev.}
\newcommand{\PhysRevl}   {Phys.~Rev.~Lett.}
\newcommand{\NPhys}  {Nucl.~Phys.}
\newcommand{\NIM} {Nucl.~Instr. and Meth.}
\newcommand{\CPC} {Comput.~Phys.\ Commun.}

\def\Z0{\mbox{Z}^0}

\begin{document}
%
%
\begin{titlepage}
\begin{center}
{\Large  EUROPEAN ORGANIZATION FOR NUCLEAR RESEARCH} 
\end{center}
\bigskip \bigskip
\begin{flushright}
 {\large  CERN-EP/2003-021} \\ 
          30 April 2003      \\
\end{flushright}
\bigskip \bigskip \bigskip \bigskip
\begin{center}
    {\Huge\bf  Search for pair-produced }
\end{center}
\begin{center}
    {\Huge\bf  leptoquarks in e$^+$e$^-$ interactions}
\end{center}
\begin{center}
    {\Huge\bf  at  {\boldmath $\sqrt{s}~\simeq~$}189~--~209~GeV}
\end{center}
\bigskip \bigskip \bigskip
\begin{center}
        {\Large The OPAL Collaboration}
\end{center}
\bigskip \bigskip \bigskip
\begin{center}
  {\large \bf  Abstract}
\end{center}
%
%
A search for pair-produced leptoquarks 
is performed using  $\mbox{e}^+\mbox{e}^-$ 
collision events collected by the OPAL detector at LEP at  
centre-of-mass energies between 189 and 209~GeV.
The data sample corresponds to a total integrated luminosity of 
 596~pb$^{-1}$.
The leptoquarks are assumed to be produced via couplings 
to the photon and the Z$^0$. For a given search channel only leptoquark 
decays involving a single lepton generation are considered.
No evidence for leptoquark pair production is observed.
Lower limits on masses for scalar and vector leptoquarks are calculated.
The results improve most of the LEP limits derived from previous searches for 
the pair production process 
by 10--25 GeV, depending on the leptoquark quantum numbers.  
\bigskip \bigskip \bigskip
\bigskip \bigskip \bigskip
\begin{center}
    {\large \bf (Submitted to European Physical Journal C) }  
\end{center}
\end{titlepage}
%
%
\begin{center}{\Large        The OPAL Collaboration
}\end{center}\bigskip
\begin{center}{
G.\thinspace Abbiendi$^{  2}$,
C.\thinspace Ainsley$^{  5}$,
P.F.\thinspace {\AA}kesson$^{  3}$,
G.\thinspace Alexander$^{ 22}$,
J.\thinspace Allison$^{ 16}$,
P.\thinspace Amaral$^{  9}$, 
G.\thinspace Anagnostou$^{  1}$,
K.J.\thinspace Anderson$^{  9}$,
S.\thinspace Arcelli$^{  2}$,
S.\thinspace Asai$^{ 23}$
% ...
}\end{center}\bigskip
\bigskip
$^{  1}$School of Physics and Astronomy, University of Birmingham,
Birmingham B15 2TT, UK
\newline
$^{  2}$Dipartimento di Fisica dell' Universit\`a di Bologna and INFN,
I-40126 Bologna, Italy
\newline
$^{  3}$Physikalisches Institut, Universit\"at Bonn,
D-53115 Bonn, Germany
\newline
% ...
$^{  5}$Cavendish Laboratory, Cambridge CB3 0HE, UK
\newline
% ...
$^{  9}$Enrico Fermi Institute and Department of Physics,
University of Chicago, Chicago IL 60637, USA
\newline
% ...
$^{ 16}$Department of Physics, Schuster Laboratory, The University,
Manchester M13 9PL, UK
\newline
% ...
$^{ 22}$Department of Physics and Astronomy, Tel Aviv University,
Tel Aviv 69978, Israel
\newline
$^{ 23}$International Centre for Elementary Particle Physics and
Department of Physics, University of Tokyo, Tokyo 113-0033, and
Kobe University, Kobe 657-8501, Japan
\newline
% ...
$^{  e}$ and Department of Experimental Physics, Lajos Kossuth University,
 Debrecen, Hungary
\newline
% ...
$^{  s}$ now at Universitaire Instelling Antwerpen, Physics Department, 
B-2610 Antwerpen, Belgium
\newline
% ...
%
%
\newpage
%
\section{Introduction}
\label{sec:introduction}
%
\indent
In the Standard Model (SM) quarks and leptons appear as formally 
independent components.
However, they show an apparent symmetry  
in terms of the
family and multiplet structure of the electroweak interactions.
Some theories beyond the SM~\cite{GUTS_COMP} therefore
predict the existence of new bosonic fields, called leptoquarks (LQ),
mediating interactions between quarks and leptons.
The interactions of leptoquarks with the known particles
are usually described by an effective Lagrangian that satisfies 
the requirement of baryon and lepton number conservation and respects the
SU(3)$_{\mathrm C}~\otimes$~SU(2)$_{\mathrm L}~\otimes$~U(1)$_{\mathrm Y}$ 
symmetry of the SM~\cite{LEPTOQ,lowenergy}.
This results in nine scalar ($S$) and nine vector ($V$) leptoquarks 
which are colour triplets or antitriplets and are grouped into 
weak isospin triplets ($S_1$ and $V_1$), doublets 
($S_{1/2}$, $\tilde{S}_{1/2}$, $V_{1/2}$ and $\tilde{V}_{1/2}$)
and singlets ($S_{0}$, $\tilde{S}_{0}$, 
$V_{0}$ and $\tilde{V}_{0}$)\footnote{In this paper the 
notation used in~\cite{lowenergy} is adopted and
a scalar multiplet of weak isospin $I$ is denoted $S_{I}$ and 
a vector multiplet $V_{I}$.
This is slightly different from the notation used 
in~\cite{LEPTOQ} where, on the contrary,
the multiplets are denoted by their multiplicity, 
i.e. $S_{2I+1}$ or $V_{2I+1}$, 
and different symbols are used for leptoquarks with fermion 
number, $F$, equal to 2 ($S$, $V$) or 0 ($R$, $U$). }.
Their properties are shown in Tables~\ref{tab:scalarleptoquarks} 
and~\ref{tab:vectorleptoquarks}. 
A charge eigenstate within a multiplet will be referred to as a ``state''
and denoted by $S_{I}(Q_{{\rm em}})$  or $V_{I}(Q_{{\rm em}})$, 
where $Q_{{\rm em}}$ is the electric
charge in units of $e$. \\

\begin{table}[t]
{\footnotesize
\begin{center}
\begin{tabular}{|ccccrcrcccccc|}
   \hline
   \hline
\rule[-3mm]{0mm}{9mm}~~LQ~~& & $F$ & & $I_3$ & & $Q_{{\rm em}}$ & & decay 
                                       & & coupling  & & $\beta$  \\
   \hline
\rule[-3mm]{0mm}{9mm}  & &  & & & & & &$e^{-}_{L} u_{L}$ & &
                        $\lambda_{LS_{0}}$& &  \\
$S_0$   & & 2 & &  0   & &$-1/3$  & &$e^{-}_{R} u_{R}$ & &
                        $\lambda_{RS_{0}}$ & &
                     $\frac{\lambda_{LS_{0}}^{2}+\lambda_{RS_{0}}^{2}}
                           {2\lambda_{LS_{0}}^{2}+\lambda_{RS_{0}}^{2}}$ \\
\rule[-3mm]{0mm}{0mm}   & &  & &     & &      & &$\nu_{e} d_{L}$   & &
                        $-\lambda_{LS_{0}}$ & & \\
   \hline
\rule[-3mm]{0mm}{9mm} $\tilde{S}_{0}$& & 2 & & 0 
                 & &$-4/3$& &$e^{-}_{R} d_{R}$& &$\lambda_{R\tilde{S}_{0}}$
                                                                   & & 1 \\
   \hline
\rule[-3mm]{0mm}{9mm}  & &   &  &  1 & &2/3   
                    & & $\nu_{e} u_{L}$& &$\sqrt{2}\lambda_{LS_{1}}$& & 0 \\
$S_1$  &  & 0 & & 0 & &$-1/3$& &
                     $\left\{ \begin{array}{r} \nu_{e} d_{L}\\
                                               e^{-}_{L} u_{L}
                              \end{array} \right.$                 &
                            $ \begin{array}{r}  \\  \end{array}$ &
                             $ \begin{array}{r} -\lambda_{LS_{1}} \\ 
                                                -\lambda_{LS_{1}}
                              \end{array} $
                            & & 1/2  \\
\rule[-3mm]{0mm}{0mm} & &    &  &$-1$ & &$-4/3$ 
                        & & $e^{-}_{L} d_{L}$& &$-\sqrt{2}\lambda_{LS_{1}}$
                              & & 1\\
   \hline
\rule[-3mm]{0mm}{9mm}   & &     & &$1/2$ & &$-2/3$ & &
                     $\left\{ \begin{array}{r} \nu_{e} \overline{u}_{L} \\
                                               e^{-}_{R} \overline{d}_{R}
                              \end{array} \right.$         &
                     $\begin{array}{r}   \\  \end{array}$ &
                     $\begin{array}{r}  \lambda_{LS_{1/2}} \\ 
                                        -\lambda_{RS_{1/2}} \end{array}$ & &
                      $\frac{\lambda_{RS_{1/2}}^{2}}
                            {\lambda_{LS_{1/2}}^{2}+\lambda_{RS_{1/2}}^{2}}$ \\
$S_{1/2}$& & 0 & &     & &             & &       & &          & &         \\
\rule[-3mm]{0mm}{0mm}  & &    & &$-1/2$& &$-5/3$& &
                     $\left\{ \begin{array}{r} e^{-}_{L} \overline{u}_{L} \\
                                               e^{-}_{R} \overline{u}_{R}
                      \end{array} \right.$  &
                     $\begin{array}{r}   \\  \end{array}$ &
                     $\begin{array}{r}  \lambda_{LS_{1/2}} \\ 
                                        \lambda_{RS_{1/2}}
                      \end{array}$ & & 1  \\
   \hline
\rule[-3mm]{0mm}{9mm}     & &   & &$1/2$ &   &$1/3$ & &
                     $ \nu_{e} \overline{d}_{L}$ &  &
                                    $\lambda_{L\tilde{S}_{1/2}}$ & & 0\\
$\tilde{S}_{1/2}$& & 0 & &     & &   & &   & &   & &  \\
\rule[-3mm]{0mm}{0mm}    & &    & &$-1/2$& &$-2/3$& &
                   $ e^{-}_{L} \overline{d}_{L}$ &  & 
                                     $\lambda_{L\tilde{S}_{1/2}}$ & & 1\\
   \hline
   \hline
\end{tabular}
\end{center}
\caption{ {\sl Quantum numbers and couplings for scalar leptoquarks.
         $F=3B+L$ is the fermion number which is a function of the 
         baryon and lepton numbers $B$ and $L$,  
         $Q_{{\rm em}}$ is the electric charge in units of $e$,
         $I_3$ is the third component of the weak isospin
         and $\beta$ is the branching ratio of the decay to a charged
         lepton and a quark of any flavour. 
         Under the assumption of non-zero couplings only within
         a single generation of leptons, $u$ and $d$ denote {\it up}- and 
         {\it down}-type quarks respectively, and
         the same Table applies to second and
         third generation leptoquarks with the obvious 
         substitutions $e\rightarrow
         \mu,\tau$. In the last column,
          $ \lambda^{2}$ is a shorthand for  $\sum_{j}(\lambda^{ij})^{2}$,
          where $i$ denotes the lepton generation and j the quark flavour. 
          }}
\label{tab:scalarleptoquarks}
}
\end{table} 
\begin{table}[t]
{\footnotesize
\begin{center}
\begin{tabular}{|ccccrcrcccccc|}
   \hline
   \hline
\rule[-3mm]{0mm}{9mm}~~LQ~~& & $F$ & & $I_3$ & & $Q_{{\rm em}}$ & & decay 
                                  & & coupling  & &
                                                                    $\beta$  \\
   \hline
\rule[-3mm]{0mm}{9mm}  & &  & &     & & & &$e^{-}_{L}\overline{d}_{R}$ & &
                         $\lambda_{LV_{0}}$& &  \\
$V_0$ & & 0 & &0 & &$-2/3$  & &$e^{-}_{R}\overline{d}_{L}$ & &
                         $\lambda_{RV_{0}}$& &
                     $\frac{\lambda_{LV_{0}}^{2}+\lambda_{RV_{0}}^{2}}
                           {2\lambda_{LV_{0}}^{2}+\lambda_{RV_{0}}^{2}}$ \\
\rule[-3mm]{0mm}{0mm}  & &  & &  & &  & &$\nu_{e}\overline{u}_{R}$ & &
                     $\lambda_{LV_{0}}$ & & \\
   \hline
\rule[-3mm]{0mm}{9mm} $\tilde{V}_{0}$& &0 & & 0 & &$-5/3$& &
                      $e^{-}_{R}\overline{u}_{L}$
                        & &$\lambda_{R\tilde{V}_{0}}$
                        & & 1 \\
   \hline
\rule[-3mm]{0mm}{9mm}  & &  &  &  1 & &1/3& & $\nu_{e}\overline{d}_{R}$& &
                       $\sqrt{2}\lambda_{LV_{1}}$
                                                                       & &0 \\
$V_1$  &  & 0 & & 0 & &$-2/3$& &
                     $\left\{ \begin{array}{r} \nu_{e}\overline{u}_{R}\\
                                               e^{-}_{L}\overline{d}_{R}
                              \end{array} \right.$                 &
                            $ \begin{array}{r}  \\  \end{array}$ &
                             $ \begin{array}{r} \lambda_{LV_{1}} \\ 
                                                -\lambda_{LV_{1}}
                              \end{array} $
                            & & 1/2  \\
\rule[-3mm]{0mm}{0mm} & &  &  &$-1$& &$-5/3$ & & $e^{-}_{L}\overline{u}_{R}$
                                 & &$\sqrt{2}\lambda_{LV_{1}}$
                                                                      & & 1\\
   \hline
\rule[-3mm]{0mm}{9mm}    & &    & &$1/2$ & &$-1/3$ & &
                     $\left\{ \begin{array}{r} \nu_{e} d_{R} \\
                                               e^{-}_{R} u_{L}
                              \end{array} \right.$         &
                     $\begin{array}{r}   \\  \end{array}$ &
                     $\begin{array}{r}  \lambda_{LV_{1/2}} \\ 
                                        \lambda_{RV_{1/2}} \end{array}$ & &
                      $\frac{\lambda_{RV_{1/2}}^{2}}
                            {\lambda_{LV_{1/2}}^{2}+\lambda_{RV_{1/2}}^{2}}$ \\
$V_{1/2}$& & 2 & &    & &             & &       & &          & &         \\
\rule[-3mm]{0mm}{0mm}     & &   & &$-1/2$& &$-4/3$& &
                     $\left\{ \begin{array}{r} e^{-}_{L} d_{R} \\
                                               e^{-}_{R} d_{L}
                      \end{array} \right.$  &
                     $\begin{array}{r}   \\  \end{array}$ &
                     $\begin{array}{r}  \lambda_{LV_{1/2}} \\ 
                                        \lambda_{RV_{1/2}}
                      \end{array}$ & & 1  \\
   \hline
\rule[-3mm]{0mm}{9mm}     & &   & &$1/2$ & &$2/3$ & &
                     $ \nu_{e} u_{R}$ &  &
                                    $\lambda_{L\tilde{V}_{1/2}}$ & & 0\\
$\tilde{V}_{1/2}$& & 2 & &    & &   & &   & &   & &  \\
\rule[-3mm]{0mm}{0mm}    & &    & &$-1/2$& &$-1/3$& &
                   $ e^{-}_{L} u_{R}$ &  & 
                                     $\lambda_{L\tilde{V}_{1/2}}$ & & 1\\
   \hline
   \hline
\end{tabular}
\end{center}
\caption{{\sl Same as Table~\ref{tab:scalarleptoquarks}, but for 
              vector leptoquarks.
          }}
\label{tab:vectorleptoquarks}
}
\end{table} 
Under these assumptions, only the masses and the couplings to right-handed
and left-handed leptons, denoted by $\lambda_{\mathrm{R}}$ and
$\lambda_{\mathrm{L}}$, remain free parameters, since
the couplings to the electroweak gauge bosons are completely determined by 
the electric charge and the third component of the weak isospin, while the
interactions with gluons are given by the colour charge.
Each coupling can carry generation indices for 
the two fermions~\cite{lowenergy}, so that $\lambda^{ij}$ couples a leptoquark
to an $i^{th}$ generation lepton and a $j^{th}$ generation quark.
In this note only leptoquark decays involving a single family of leptons
are searched for, while no distinction is made
between quarks from different generations. 
This corresponds to the simplifying assumption that   
$\lambda^{ij}\cdot \lambda^{mn}=0$ if $i \neq m$. 
The states with couplings both to
right-handed charged leptons and left-handed neutrinos have an unknown  
branching ratio 
into a charged lepton and a quark, $\beta$, depending
on the relative values of the couplings, 
while for all the other states $\beta$ has a known fixed value.  
Some leptoquarks with couplings to left-handed leptons have the same
properties  as scalar quarks in supersymmetric models
with R-parity violation~\cite{RPV}. This is the case for $S_{0}(-1/3)$, 
$\tilde{S}_{1/2}(1/3)$ and 
$\tilde{S}_{1/2}(-2/3)$. The results obtained in this analysis can 
therefore also be
interpreted in terms of these models. \\

Several experimental results constrain theories that predict the 
existence of leptoquarks.
Searches for events with leptoquark single production, 
where a first generation leptoquark 
could be formed as a resonance
between an electron\footnote{Charge conjugation is implied 
throughout this paper for all
particles, e.g. \hspace*{-2.5mm} positrons are referred 
to as electrons, etc.}  
and a quark, were performed by
the ZEUS and H1 experiments at the ep collider HERA~\cite{ZEUS_H1} 
and by the DELPHI and OPAL experiments at LEP~\cite{lep_single_lq}.
Leptoquark masses, M$_{{\rm LQ}}$, of $\mathcal{O}(100$~GeV)  are excluded
for $\lambda$ values greater than  ${ \cal O}(10^{-2})$.
All LEP and Tevatron experiments have searched for events with leptoquark pair
production~\cite{LEP1,mypaper,CDF_D0}, setting limits
on M$_{\mathrm{LQ}}$ as a function of  
the branching ratio 
for decay into a charged lepton and a quark. The values of these limits
range from 99 GeV to 275 GeV depending on the decay channel and the spin of the
leptoquarks.  
\newline
\newline

In this paper a search is presented for pair-produced scalar and vector 
leptoquarks of all three generations performed with the OPAL detector.
Compared  to single production by electron-quark interactions, 
pair production has the advantage that all states can 
be produced, including leptoquarks that decay only into a neutrino 
and a quark. 
Searches for this channel at LEP are able to explore 
the region of large decay
branching ratio into quark-neutrino final states, where the 
Fermilab experiments
have reduced sensitivity.   
The study is based on data recorded during the LEP runs from 1998 to 2000 
at centre-of-mass energies, $\sqrt{s}$, between 189 and 209 GeV.
The different values of $\sqrt{s}$ and the corresponding integrated
luminosities are listed in Table~\ref{tab:lumin}.
\newline

In principle, leptoquarks of all three generations can be pair-produced 
in e$^+$e$^-$ collisions at LEP, 
by {\it s}-channel 
$\gamma$ or Z$^0$ exchange and, in the case of first generation 
leptoquarks, by the exchange of a quark in the 
{\it t-} or {\it u}-channel~\cite{LEPTOQEE}.
The current upper limits on the couplings $\lambda$ to fermions
are $\mathcal{O}(10^{-2})$ in the mass range 
kinematically accessible;
this makes the {\it t}- or {\it u}-channel 
contribution to the first generation production 
cross-section (including interference between this channel and 
the $s$-channel) 
less than 1$\%$ of the pure $s$-channel contribution.
Therefore, in the present analysis, only $s$-channel leptoquark 
production is considered. 
Consequently, for a given state
the cross-section depends on the mass, the electric charge and the third 
component of the weak isospin, but is independent 
of the $\lambda$ couplings.
On the other hand, for couplings smaller than ${\cal O}(10^{-6})$
the lifetime of leptoquarks would be sufficiently long to have interactions
with the material of detector and to produce a
secondary decay vertex, clearly separated from the interaction region
of the electron beams. 
This topology is not considered here as
the tracks of charged particles are required to come from the 
primary interaction vertex so that,  
to summarize, the present analysis covers the region 
$10^{-6} < \lambda < 10^{-2}$. \\

The decay of a heavy leptoquark into a charged lepton and a quark
leads to final states characterized by an isolated energetic 
charged lepton and a hadronic jet, 
while for decays into a quark and a neutrino, the final state would have large
missing energy and a jet.
Given the assumptions about the $\lambda$ couplings, 
the following topologies are considered for events 
resulting from the
decay of a leptoquark-antileptoquark pair:\newline
{\bf Class A}: Two hadronic jets and two neutrinos; it consists 
               of the final states 
               $\nu_{l} \overline{\nu_{l}} u^{j} \overline{u}$$^{k}$ and
               $\nu_{l} \overline{\nu_{l}} d^{j} \overline{d}$$^{k}$,
               where $l= {\rm e},\mu,\tau$ and
               $u^{j}$, $d^{j}$ are {\it up}- and {\it down}-type quarks
               of the $j^{th}$ generation.
\newline
{\bf Class B}: Two hadronic jets, one neutrino and one charged lepton of the 
same generation ($\nu_{l} l^{\pm} u^{j} d^{k}$).
\newline
{\bf Class C}: Two hadronic jets and one pair of oppositely charged leptons
of the same generation
 (${l^{+} l^{-}} u^{j} \overline{u}$$^{k}$,
  ${l^{+} l^{-}} d^{j} \overline{d}$$^{k}$).\\ 

\begin{table}[h]
{\footnotesize
\begin{center}
\begin{tabular}{|c|c|r|}
      \hline \hline
  YEAR  &  ~~~$\langle~\sqrt{s}~\rangle$~~(GeV)~~~   &
     \multicolumn{1}{c|}{~~~~{\rule[-4mm]{0mm}{10mm}   
               ${\displaystyle \int} {\cal  L}~dt$~~~(pb$^{-1}$)~~~~~} 
                                                           }\\
\hline 
 \rule[-3mm]{0mm}{9mm}  1998       &   188.6   &   169.1~~~~~~~~~~~     \\
\hline 
 \rule[-3mm]{0mm}{9mm}   1999      &   191.6   &    28.9~~~~~~~~~~~     \\
 \rule[-3mm]{0mm}{6mm}             &   195.5   &    72.3~~~~~~~~~~~     \\
 \rule[-3mm]{0mm}{6mm}             &   199.5   &    74.7~~~~~~~~~~~     \\
 \rule[-3mm]{0mm}{6mm}             &   201.7   &    39.2~~~~~~~~~~~      \\
\hline 
 \rule[-3mm]{0mm}{9mm}   2000      &   203.8   &     8.5~~~~~~~~~~~      \\
 \rule[-3mm]{0mm}{6mm}             &   205.1   &    69.6~~~~~~~~~~~      \\
 \rule[-3mm]{0mm}{6mm}             &   206.3   &    63.1~~~~~~~~~~~      \\
 \rule[-3mm]{0mm}{6mm}             &   206.6   &    63.8~~~~~~~~~~~      \\
 \rule[-3mm]{0mm}{6mm}             &   208.0   &     6.7~~~~~~~~~~~      \\
\hline 
 \multicolumn{2}{|c|}{ \rule[-3mm]{0mm}{9mm} TOTAL}   &  595.9~~~~~~~~~~~ \\
\hline \hline
\end{tabular}
\end{center}
 \caption{\sl Average centre-of-mass energies and corresponding integrated 
           luminosities for the data samples used in the
           analysis.  
           The search for vector leptoquarks
           includes only the data with  
           $\sqrt{s} > 195$~GeV.}   
\label{tab:lumin}   
}
\end{table} 
%
%
\section{The OPAL Detector}
%
\indent
The OPAL detector is described in detail in~\cite{opaltechnicalpaper}. 
It was a multi-purpose apparatus having nearly complete solid angle 
coverage\footnote{The right-handed coordinate system is defined so that 
the positive direction 
of the $z$ axis  is along the e$^-$ beam; $r$ is the
coordinate normal to the beam axis, $\phi$ is the azimuthal angle with respect
to the positive direction of the $x$-axis (pointing towards the 
centre of LEP) and
$\theta$ is the polar angle with respect to +$z$.}.
The central detector consisted of a system of tracking chambers 
inside a 0.435~T solenoidal magnetic field
as well as  of two layers of silicon microstrip detectors 
\cite{simvtx} surrounding the beam-pipe.
The  tracking chambers included  
a high-precision drift chamber, a large-volume 
jet chamber and a set of {\it z}-chambers measuring the track 
coordinates along the beam direction. The resolution on 
the transverse momentum of a track was
given by 
$\sigma_{p_{t}}/p_{t} \simeq 
\sqrt{(0.02)^{2}+(0.0015\cdot p_{t})^{2}/({\rm GeV})^{2}}$ 
and the average angular
resolution was about 0.3~mrad in $\phi$ and 1~mrad in $\theta$.
A lead-glass electromagnetic calorimeter 
was located outside the magnet coil and covered the full azimuthal range
for polar angles in the range of $\mid\cos(\theta)\mid < 0.984$.
It was divided into two regions: the barrel ($\mid\cos(\theta)\mid<0.82$)
and the endcaps ($|\cos(\theta)| >~0.81$). The energy resolution for
high momentum electrons was around 3\%.
The magnet return yoke, divided into barrel and endcap sections along with
pole tips, was instrumented for hadron calorimetry in the region
$\mid\cos(\theta)\mid < 0.99$.
Four layers of muon chambers covered the outside of the hadron calorimeter.
Close to the beam axis the forward calorimeter and gamma catcher, 
together with the silicon-tungsten 
luminometer~\cite{bib-siw} and the forward scintillating tile 
counter~\cite{MIP_plug}, completed the geometrical 
acceptance down to 33 mrad from the beam direction.
%
\section{Monte Carlo simulations}
%
At lowest order the $s$-channel contribution to the differential cross 
section for the production of a pair of scalar leptoquarks of mass 
M$_{\mathrm{LQ}}$ in $\rm{e^{+}e^{-}}$ collisions at a 
centre-of-mass energy $\sqrt{s}$ is given by~\cite{LEPTOQEE}
{\small
\begin{equation}
%
%
 \frac{d\sigma_{S}}{d\cos{\theta}}~=~
\frac{3\pi\alpha^2}{8s}
\left(1~-~4\mathrm{M}_{\mathrm{LQ}}^{2}/s\right)^{\frac{3}{2}}
\sin^{2}\theta
                     \sum_{a=L,R} |k_a(s)|^2~,
\label{eq:totscalar}
\end{equation}}
while for vector leptoquarks one has
 {\small
 \begin{equation}
%
%
 \frac{d\sigma_{V}}{d\cos{\theta}}~=~
  \frac{3\pi\alpha^2}{8s}
     \left(1~-~4\mathrm{M}_{\mathrm{LQ}}^{2}/s\right)^{\frac{3}{2}}
     \left\{ \frac{4 + \left[ 1~-~3 
              \left( 1 - 4\mathrm{M}_{\mathrm{LQ}}^{2}/s \right) 
                      \right]\sin^{2}\theta} 
                  { 1 -  \left(1~-~4\mathrm{M}_{\mathrm{LQ}}^{2}/s\right)}
     \right\}          
  \sum_{a=L,R} |k_a(s)|^2~,
 \label{eq:totvector}
 \end{equation}}
\newline
where $\alpha$ is the electromagnetic coupling and
{\small
\begin{equation}
 k_a(s)~=~-Q_{{\rm em}}~+~
      Q_{a}^{{\rm Z}}({\mathrm e}) \frac{s}{s~-~{\rm M}_{{\rm Z}}^{2}~
       +~i{\rm M}_{{\rm Z}}\Gamma_{{\rm Z}}}
       Q^{{\rm Z}}({\mathrm{LQ}}).
\end{equation}}
\newline
Here $Q_{{\rm em}}$ is the electric charge of the leptoquark,
${\mathrm M}_{{\rm Z}}$ and $\Gamma_{{\rm Z}}$ are the mass and 
the width of the ${\rm Z}^{0}$ boson,
and  the couplings are given by 
{\small
\begin{eqnarray}
     Q^{{\rm Z}}(\mathrm{LQ})     &=& \frac{I_3~-~Q_{{\rm em}}\sin^2\theta_W}
                    {\cos{\theta_W}\sin{\theta_W}}~,   \nonumber \\
     Q^{{\rm Z}}_{L}(\mathrm{e})  &=& \frac{-\frac{1}{2}~+~\sin^2\theta_W}
                                     {\cos{\theta_W}\sin{\theta_W}}~, 
                                      \label{eq:qzcouplings} \\
     Q^{{\rm Z}}_{R}(\mathrm{e})  &=& \tan{\theta_W}~,  \nonumber
\end{eqnarray}}
\newline
with $I_3$  being the third component of the leptoquark weak 
isospin and 
$\theta_W$ the Weinberg angle. \newline

The Monte Carlo generator LQ2~\cite{LQ2} is used to simulate
leptoquark pair events.
Initial state QED radiation is included.
In LQ2 scalar leptoquarks  
decay isotropically in their rest frame,
while the angular distribution of decay products of vector leptoquarks depends
on the helicity state.  
The hadronization of the
final state quark 
pair is performed by JETSET~\cite{PYTHIA}.
For scalar leptoquarks, samples of at least 1000 signal events are 
generated for each value 
of the leptoquark mass  
from M$_{\mathrm{LQ}} = 50$~GeV to the kinematic limit 
in steps of 10 GeV or less 
for all the different decay topologies at the centre-of-mass energies with the
highest integrated luminosities (189, 196, 200 and 206 GeV). 
The search for vector 
leptoquarks includes only data
with $\sqrt{s}>195$ GeV and the signal was simulated
for M$_{\mathrm{LQ}} \geq 70$~GeV. 
Since leptoquarks carry colour, they may hadronize before decaying
if their couplings to fermions are small.
This effect  is evaluated from Monte
Carlo samples of pair-produced  
scalar quarks decaying via  R-parity violating couplings.  
These events have features similar to events of class {\bf C} and allow
the impact of this effect on the detection efficiencies and on the leptoquark 
mass reconstruction to be estimated 
and taken into account as a  systematic uncertainty.
\newline

All relevant SM background processes are studied using  
various samples of simulated Monte Carlo events for
each centre-of-mass energy in the data. 
Two-fermion events 
(Z$^{0\ast}$/$\gamma^{\ast}$~$\rightarrow$~{\rm f\={f}}($\gamma$), 
with {\rm f}~=~q,$\tau$ and denoted by 2f), 
are simulated with KK2f~\cite{kk2f}. 
The Monte Carlo programs HERWIG~\cite{HERWIG}, PHOJET~\cite{PHOJET},
and BDK~\cite{BDK} 
are used to generate two-photon ($\gamma \gamma$) events with
hadronic and leptonic final states. 
Other processes with four fermions in the final state, 4$\rm{f}$, including 
W$^{\pm}$ and Z$^{0}$ pair production, are simulated with grc4f~\cite{GRACE}
and KORALW~\cite{KORALW}.  JETSET~\cite{PYTHIA} is used as the principal
model for the hadronization.
Besides the main samples alternative generators or hadronization
models such as KORALZ~\cite{KORALZ}, Vermaseren \cite{VERMASEREN} and
HERWIG~\cite{HERWIG}, 
are used to check the expectation from the SM
background.
Generally, at each centre-of-mass energy, the number of 
simulated events for the background processes   
corresponds to at least fifteen times the integrated 
luminosity of the data, except for the 
$\gamma \gamma$ process where, at some
centre-of-mass energies, Monte Carlo events  corresponding to only about
 three times the data integrated luminosity  are available. \\  

The full  response of the OPAL detector~\cite{GOPAL}  
is simulated for all the Monte Carlo events.
%
\section{Analysis}
%
All the leptoquark event topologies (classes {\bf A} to {\bf C} as
defined in Section~\ref{sec:introduction}) 
are characterized by large charged track multiplicities and large number 
of energy 
deposits (clusters) in the calorimeters due to the hadronization  
of the quark pair. Moreover, in events of
classes {\bf B} and {\bf C}, energetic and well-isolated 
charged leptons are present.
\newline

The tracks of charged particles reconstructed in the tracking system and 
the clusters in the electromagnetic and hadronic
calorimeters  
are required to satisfy the same quality criteria 
as in~\cite{mypaper}. 
To avoid double counting, calculations of  
quantities such as visible energy and transverse 
momentum  are performed from charged particle tracks and from clusters 
in the electromagnetic and
hadron calorimeters following 
the method explained in~\cite{MT}.
Electron and muon identification is performed using  
standard OPAL algorithms~\cite{WW_161}.
The electron identification is based on the match between 
the momentum of a track 
and the energy of a
cluster in the electromagnetic calorimeter associated to the track; moreover
the value of the ionization energy loss, ${\rm d}E/{\rm d}x$, 
measured for the track in the OPAL jet
chamber must be in agreement with what
expected for an electron. 
The muon identification requires at least two hits 
corresponding to the direction of the track in the muon
chambers
and minimum energy deposition for clusters in the hadron calorimeter
associated to the track.
The energy of identified electrons is given by the energy of the 
electromagnetic calorimeter cluster, while for muons the momentum of the track 
is used to calculate the energy.
Tau lepton identification is performed using an artificial neural 
network algorithm described in detail in~\cite{taunn}.
The hadronic jets are reconstructed using the Durham 
algorithm~\cite{DURHAM}. The resolution on the direction of a jet is 
about 25 mrad, while the resolution on the jet energy is
10--20\%, depending on the energy itself, 
the jet shape and the detector region.
%

%
\subsection{Event preselection}
\label{subsection_preselection}
%
\indent
Several preselection requirements are applied to all classes of events.
To reduce the number of events due to interactions of the LEP beams
with residual gas in the beam-pipe or with its material, 
at least $20\%$ of the reconstructed tracks are required to satisfy the 
track quality criteria. 
There must be at least four accepted tracks and at least four accepted
electromagnetic clusters not associated to any track.
Finally, the total visible energy, $E_{\mathrm{vis}}$,
is required to be
greater than 0.25$\sqrt{s}$ and smaller than 1.25$\sqrt{s}$ and its fraction
deposited in the region $|\cos(\theta)| >~0.9$ must be less than 50$\%$. \\

After the preselection 51218 events are observed in the data and 
49690 are expected from SM background, mostly from two fermion events.
The efficiencies  for signal events  
range from 86\% to 99\% for both scalar and vector leptoquarks 
at all centre-of-mass energies, 
depending on the 
leptoquark mass and the decay channel.

%
\subsection{The {\boldmath $\nu \nu {\rm q q}$} channel (class A)}
 \label{subsection_nunuchannel}
%
\indent
Signal events of class {\bf A} are characterized by a 
pair of hadronic jets and large missing energy due to the neutrinos escaping 
detection.  
The following cuts were applied to the data:
\begin{itemize}
\item[{\bf (A1)}] The total visible energy, $E_{\rm{vis}}$, 
                 has to be in the range
                 $0.25 < E_{\rm{vis}} / \sqrt{s} < 0.75$.
\item[{\bf (A2)}] Neutrinos or 
                  particles escaping along the beam pipe are not detected
                  resulting in a total reconstructed 
                  momentum vector of the event, $\vec{p}_{{\rm tot}}$,
                  different from the expected value of $\vec{0}$. 
                  The missing momentum of the event is then
                  defined as 
                  $\vec{p}_{{\rm miss}} \equiv - \vec{p}_{{\rm tot}}$.    
                  The component of the missing momentum in the
                  direction transverse to the beam axis,  
                  $p_{\mathrm{t}}^{\mathrm{miss}}$, is required to be
                  larger than  0.2$\sqrt{s}$.
\item[{\bf (A3)}] Events are required to contain
                  no isolated electron or muon with an energy, 
                  $E_{{\rm e}}$ or $E_{\mu}$, 
                  larger than 0.15$\sqrt{s}$, where the isolation criterion  
                  requires that the angle between the lepton and the
                  nearest charged track is larger              
                  than 10$^{\circ}$.
                  The events must also contain 
                  no tau lepton with an associated output from the neural net
                  used for the identification, $\cal{O}_{\tau}$, 
                  larger than 0.75.
\item[{\bf (A4)}] The events are forced into two jets. 
                  The angle between the directions of the jets, 
                  $\theta_{\mathrm{jj}}$,
                  is required to be such that
                 $\cos(\theta_{\mathrm{jj}}) > -0.1$.
\item[{\bf (A5)}] The invariant mass of the two jets, M$_{\mathrm{jj}}$,
                 has to be smaller than 70~GeV.
\end{itemize}

Table~\ref{tab:nevA} shows the numbers of events after each cut, 
together with the numbers of background events predicted
from SM Monte Carlo samples, and the efficiencies for signal
events corresponding to M$_{\mathrm{LQ}}$~=~90~GeV at $\sqrt{s}= 206$~GeV. 
Cut  {\bf (A2)} greatly reduces 
the  $\gamma\gamma$ and  2f  
backgrounds.
Cuts {\bf (A3)}{\bf --}{\bf (A5)} reject almost completely
$\gamma\gamma$ and 2f  
events, and are very efficient against 4f background.
In the whole data sample, 28 events survive the selection,
while $22.8^{+2.7}_{-1.3}$ (stat.)\footnote{
The statistical error on the expected background
is calculated by considering the 68.27\% confidence band around the
number of events surviving the selection,
following~\cite{ref:feldcous}. }
events
are expected from Standard Model processes, with the largest contribution, 
about $40\%$, due to
events with a single ${\rm W}$-boson  (${\rm We} \nu$).
At $\sqrt{s} = 206$ GeV the selection efficiency for signal events 
for leptoquarks
of mass M$_{{\rm LQ}}$ = 90 GeV is $(31.3 \pm 0.7({\rm stat.}))\%$.\\ 
   
\begin{figure}[p]
  \begin{center}
    \begin{minipage}[t]{0.45\textwidth}
            \begin{center}
                \vspace*{-2cm}
                  \epsfig{file=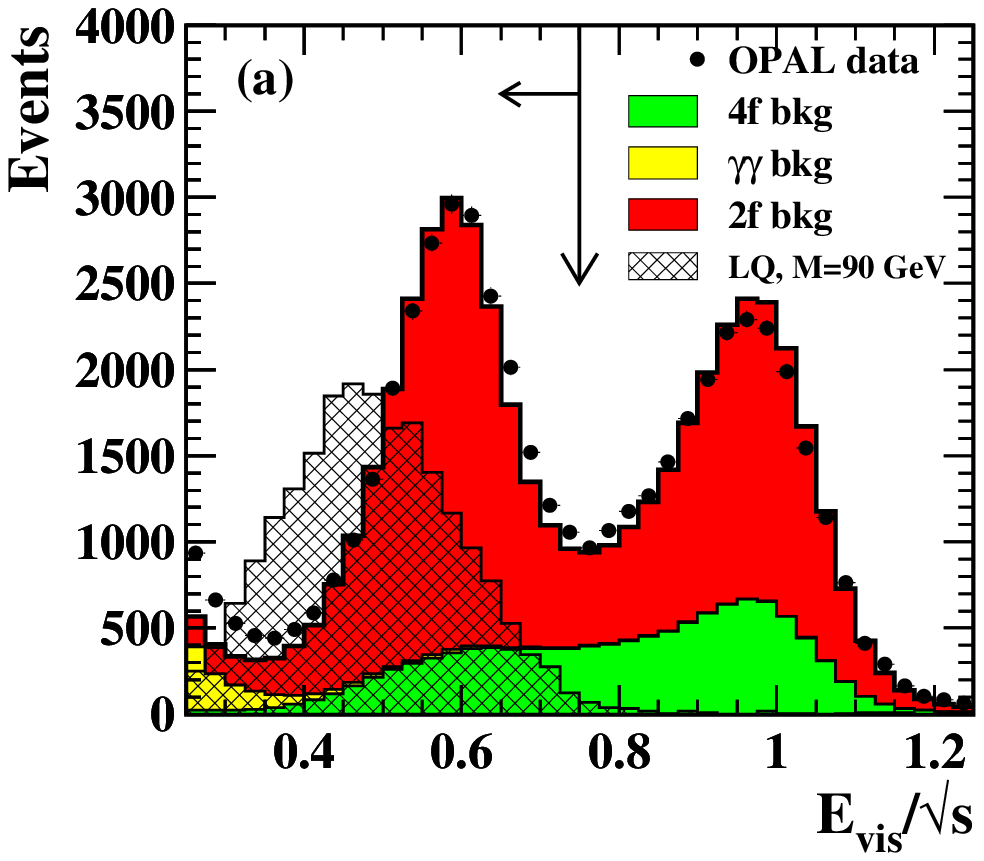,width=8cm,
                                bbllx=5pt,bblly=270pt,bburx=295pt,bbury=525pt}
             \end{center}
             \vspace*{-0.7cm}
    \end{minipage}
    \hfill
    \begin{minipage}[t]{0.45\textwidth}
          \begin{center}
             \vspace*{-2cm}
                 \epsfig{file=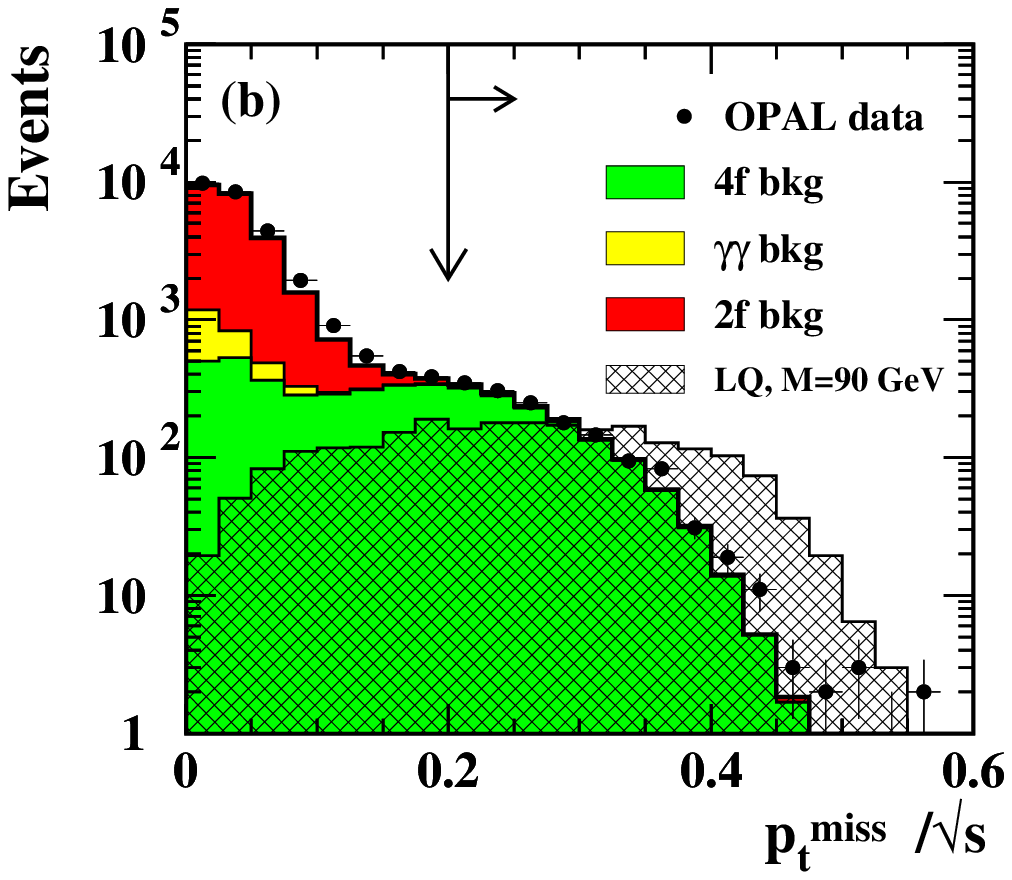,width=8cm,
                         bbllx=5pt,bblly=270pt,bburx=295pt,bbury=525pt}
          \end{center}
          \vspace*{-0.7cm}
     \end{minipage}
  \end{center}
  \begin{center}
    \begin{minipage}[t]{0.45\textwidth}
         \begin{center}
           \vspace*{-0.5cm}
             \epsfig{file=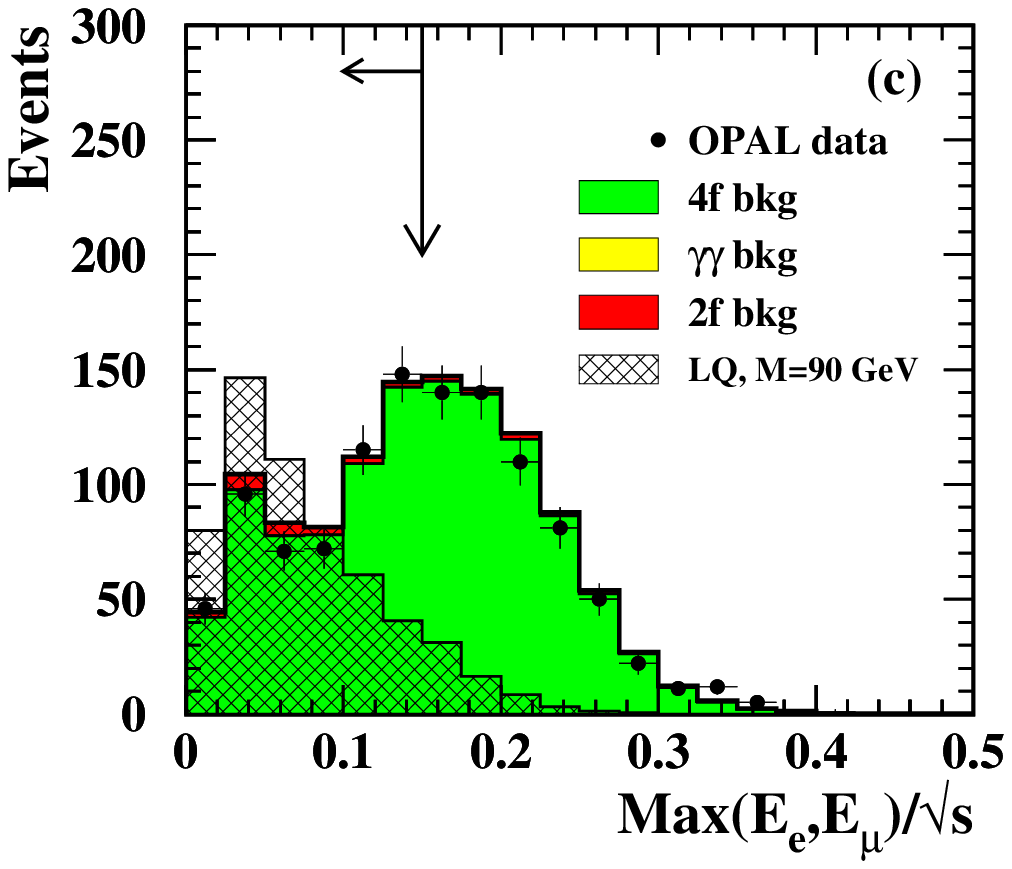,width=8cm,
                      bbllx=5pt,bblly=270pt,bburx=295pt,bbury=525pt}
          \end{center}
          \vspace*{-0.7cm}
    \end{minipage}
    \hfill
    \begin{minipage}[t]{0.45\textwidth}
        \begin{center}
          \vspace*{-0.5cm}
            \epsfig{file=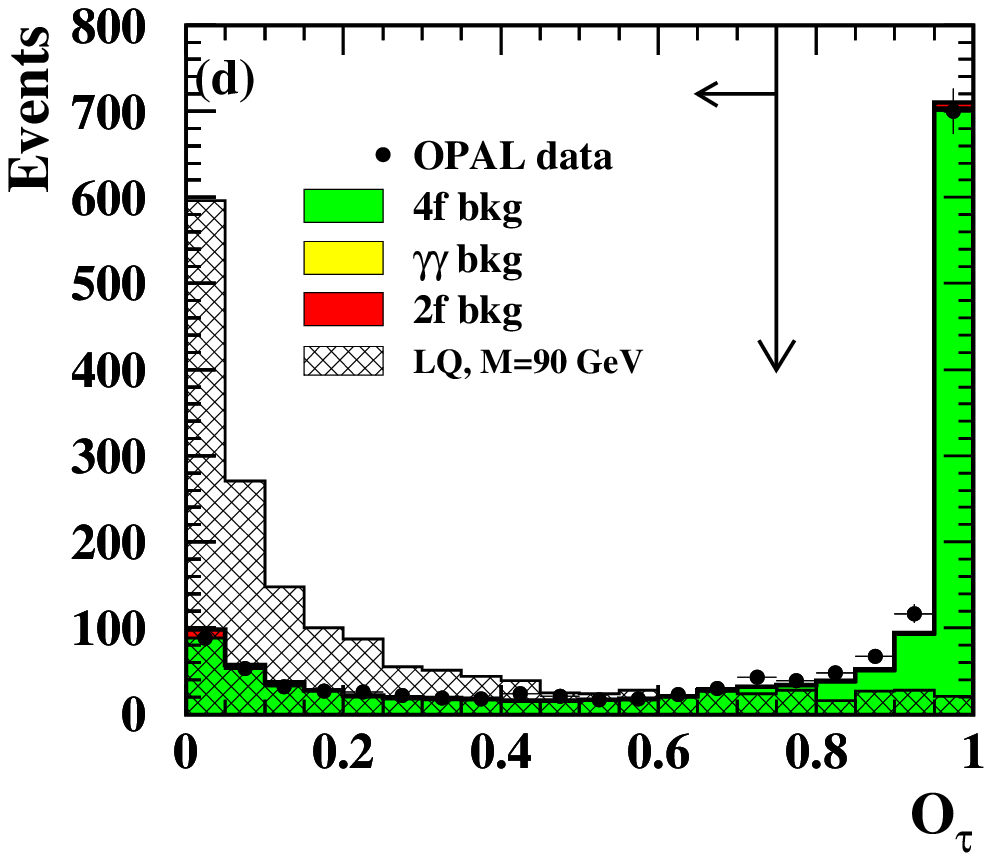,width=8cm,
                          bbllx=5pt,bblly=270pt,bburx=295pt,bbury=525pt}
         \end{center}
          \vspace*{-0.7cm}
    \end{minipage}
   \end{center}
  \begin{center}
     \begin{minipage}[t]{0.45\textwidth}
         \begin{center}
            \vspace*{-0.5cm}
             \epsfig{file=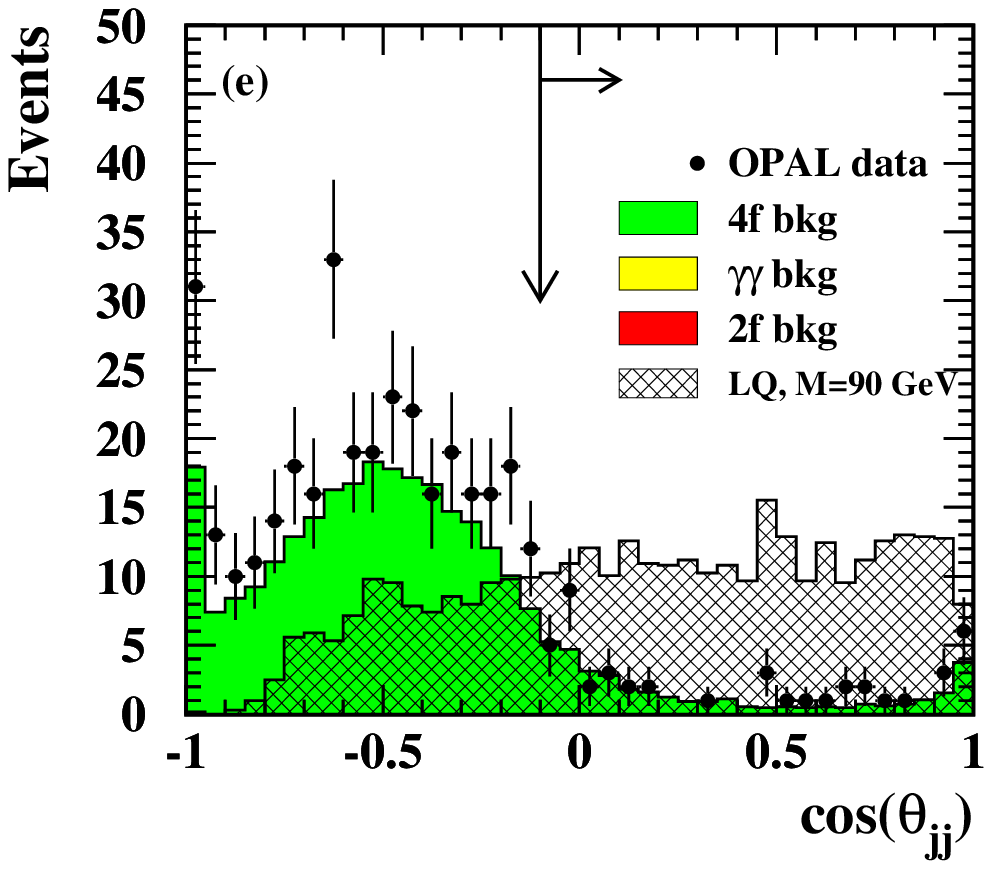,width=8cm,
                     bbllx=5pt,bblly=270pt,bburx=295pt,bbury=525pt}
         \end{center}
         \vspace*{-0.9cm}
     \end{minipage}
     \hfill
     \begin{minipage}[t]{0.45\textwidth}
       \begin{center}
          \vspace*{-0.5cm}
            \epsfig{file=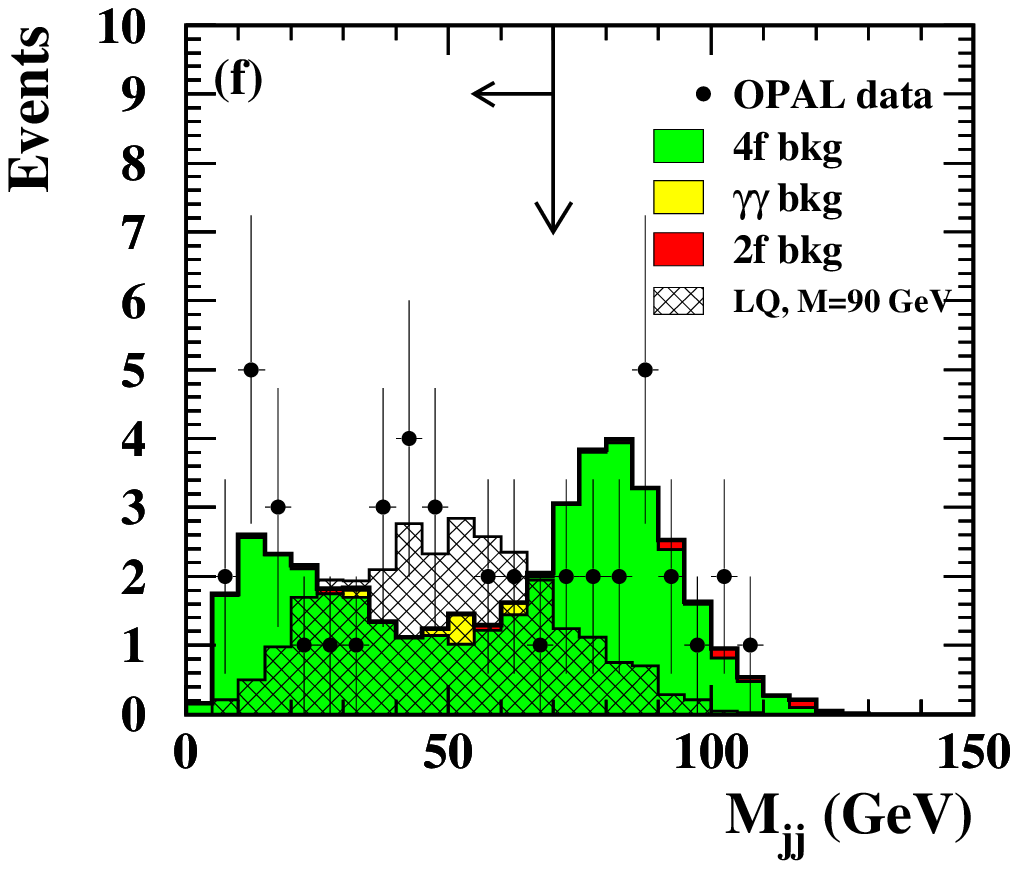,width=8cm,
                    bbllx=5pt,bblly=270pt,bburx=295pt,bbury=525pt}
       \end{center}
       \vspace*{-0.9cm}
      \end{minipage}
   \end{center}
\vspace*{-0.3cm}
\caption[1]{\sl Class {\bf A}, the $\nu \nu {\rm q q}$ channel: 
                distributions of the selection variables  
                for the data 
                (points with bars denoting the statistical error), 
                the estimated Standard Model background 
                (filled histogram)
                and a simulated signal (hatched histogram), with 
                arbitrary normalization, 
                 corresponding to scalar leptoquarks of mass
                 M$_{\mathrm{LQ}}=90$~GeV at $\sqrt{s} = 206$ GeV.
                 All the distributions are shown for the events surviving 
                 all the cuts applied before the cut on the 
                 plotted variable, following 
                 the description of the selection in the text.  
                 The arrows indicate the positions of the cuts and 
                  the accepted regions.
             \newline
                 {\bf (a)}~The scaled visible energy.  
                 {\bf (b)}~The scaled transverse missing momentum.  
                 {\bf (c)}~The scaled energy of the most energetic lepton
                           (electron or muon), if a lepton is found.
                 {\bf (d)}~The output from the neural net, 
                           $\cal{O}_{\tau}$, for the
                           tau with the highest value in the event, 
                           after cut {\bf (A-2)}.
                 {\bf (e)}~The cosine of the angle between the two  
                           reconstructed jets, 
                           $\theta_{{\rm jj}}$.
                 {\bf (f)}~The invariant mass of the two  reconstructed jets, 
                           ${\rm M_{jj}}$.
                   }  
\label{fig_vv}
\end{figure}

Figure~\ref{fig_vv} shows the distributions of the variables
used in the selection for events of class {\bf A}. 
The discrepancies between the observed data and the expected
SM events
in the distribution of the scaled visible energy $E_{\rm vis} / \sqrt{s}$, 
Figure~\ref{fig_vv}(a), 
are related to the bad modelling of Monte Carlo $\gamma\gamma$ events 
and of the emission of
photons in the initial state (initial state radiation).  

\vspace*{0.5cm}
\begin{table}[htbp]
\begin{center}
{\footnotesize
\begin{tabular}{|c||c||c||c|c|c||c|}
\hline
\hline
 \multicolumn{7}{|c|}{\rule[-3mm]{0pt}{9mm}  
                {\normalsize {\boldmath  $ \nu \nu {\rm q q} $}}}  \\
\hline
\rule[-4mm]{0pt}{10mm} {\bf Cut} &  {\bf Data}    &  
                                    {\bf Background }  & 
                                     4f  &
                                     $\gamma \gamma$ &
                                     2f & 
                                     {\boldmath $\varepsilon~(\%)$ } \\
\hline
%
\rule[-4mm]{0pt}{10mm}   (A1) & 28313 & 26596.0  & 4299.0 & 1175.0  & 21122.0
                              &   94.1 \\
\rule[-4mm]{0pt}{7mm}    (A2) & ~1474 & ~1404.0  & 1353.0 & ~~5.5  & ~~~45.4 
                              &   61.9 \\
\rule[-4mm]{0pt}{7mm}    (A3) & ~~371 & ~~340.1  & ~313.7 & ~~2.3  & ~~~24.1 
                              &    52.9 \\
\rule[-4mm]{0pt}{7mm}    (A4) & ~~~45 & ~~~43.0  & ~~41.3 & ~~0.8  & ~~~~0.9 
                              &     37.3  \\
\rule[-4mm]{0pt}{7mm}    (A5) & ~~~28 & ~~~22.8  & ~~21.6 & ~~0.8  & ~~~~0.4 
                              &     31.3  \\
\hline \hline
\end{tabular}
 }
\caption[]{\sl The remaining numbers of events after each cut of 
                selection {\bf A} ($\nu\nu {\rm qq}$ channel) for 
                various background processes 
                compared with the observed number of events in the whole 
                data sample. 
                The last column contains the signal efficiency
                for events with M$_{\mathrm{LQ}}$~=~90~GeV at 
                $\sqrt{s} =  206$~GeV.
                Within the statistical errors the efficiencies for scalar
                and vector leptoquarks are the same, so the mean value
                is quoted. 
           }
\label{tab:nevA}
\end{center}
\end{table}
%
%
\subsection{The {\boldmath$l^{\pm} \nu {\rm q q}$} channel (class B)}
%
\indent
The selection of signal events of class {\bf B} is different  
for final states with an electron or muon (class {\bf B1}) and those 
with a tau lepton (class {\bf B2}).
%
\subsubsection{Electron and muon channels (class B1)}
%
\begin{itemize}
\item[{\bf ~~~(B1-1)}]  The visible energy must lie in the range 
                        $0.5 < E_{{\rm vis}} / \sqrt{s} < 1.0$.
\item[{\bf ~~~(B1-2)}]  The direction of the missing momentum must satisfy
                        $|\cos(\theta_{{\rm miss}})|<0.9$.
\item[{\bf ~~~(B1-3)}]  The event is required to contain at least one 
                        identified charged lepton (an electron for the first 
                        generation, a muon for the second). 
\item[{\bf ~~~(B1-4)}]  The most energetic charged lepton in the event
                        is considered to be the one produced in the decays of 
                        the leptoquark pair. 
                        The energy and momentum of the escaping neutrino are 
                        calculated from the
                        missing momentum of the event.
                        The energy of the most energetic lepton
                        (the charged lepton or the neutrino)
                        has to be larger than $0.15\sqrt{s}$, while the energy
                        of the second one
                        has to be larger than $0.10\sqrt{s}$.     
\item[{\bf ~~~(B1-5)}] The charged lepton and the neutrino are required to
                       be isolated from other tracks in the event
                       by requiring that the angle
                       between each of them and the nearest charged track, 
                       $\theta_{{\rm e,ct}}$ or  $\theta_{{\rm \mu,ct}}$ for
                       the charged leptons of the 
                       first and second generation respectively,
                       $\theta_{{\rm \nu,ct}}$ for the neutrino,
                       must be at least 10$^{\circ}$. 
\item[{\bf ~~~(B1-6)}] The event is forced into two jets after removing
                       the track corresponding to the charged
                       lepton. The angle between the jets
                       is required to satisfy
                       $\cos(\theta_{\mathrm{jj}}) > -0.1$.
\item[{\bf ~~~(B1-7)}] To reject ${\rm W}$-pair events, 
                       a five constraint  
                       kinematic fit is applied, where energy and momentum
                      conservation is required and  the two-jet 
                      system and the two-lepton system are constrained to have 
                      the same mass, M$_{{\rm jj, fit}}$. 
                      As the momentum of the neutrino 
                      is not measured, the effective number of constraints 
                      in the fit is two.
                      Events with a fit probability 
                      larger than 0.1 and, at the same time, a fitted mass  
                      M$_{{\rm jj, fit}}$ larger than 75~GeV
                      are rejected.
\item[{\bf ~~~(B1-8)}] Finally, to reconstruct the leptoquark mass, 
                       a second kinematic fit
                       is applied with the same constraints as in cut 
                       {\bf (B1-7)}, 
                       but this time  pairing the leptons with the jets.
                       Of the two possible combinations, the one with 
                       the higher fit probability  is considered. 
                       The events are selected if the fitted mass 
                       M$_{{\rm LQ}}$ is larger
                       than 50 GeV and the fit probability P$_{{\rm fit}}$ is
                       larger than 10$^{-3}$.
\end{itemize}

In Table \ref{tab:nevB1} the numbers of events after each cut are shown,
together with the numbers of predicted background events and the 
efficiencies for signal events corresponding to 
${\mathrm{M_{LQ}}}$~=~90~GeV at $\sqrt{s}= 206$~GeV.
The contribution of $\gamma \gamma$ events is negligible after cut 
{\bf (B1-4)}.
Cuts {\bf (B1-4)} and {\bf (B1-5)} are particularly efficient in reducing  
2f events.
The numbers of events observed in the data, 13 for the first generation
and 26 for the second, are in agreement with the expectation from 
Standard Model processes, $13.7^{+2.4}_{-1.0}$ (stat.) 
and  $24.5^{+2.5}_{-1.3}$ (stat.), respectively. 
About $80\%$ of the expected background is due to 
${\rm W}^{+}{\rm W}^{-}$ events.  
At $\sqrt{s} = 206$ GeV the selection 
efficiencies for signal events for leptoquarks with
M$_{{\rm LQ}}=90$ GeV are $(28.0 \pm 1.0({\rm stat.}))\%$ and 
$(35.8 \pm 1.1({\rm stat.}))\%$ for the first and second generation 
respectively. \\

\begin{figure}[p]
 \begin{center}
      \begin{minipage}[t]{0.45\textwidth}
          \begin{center}
             \vspace*{-2cm}
               \epsfig{file=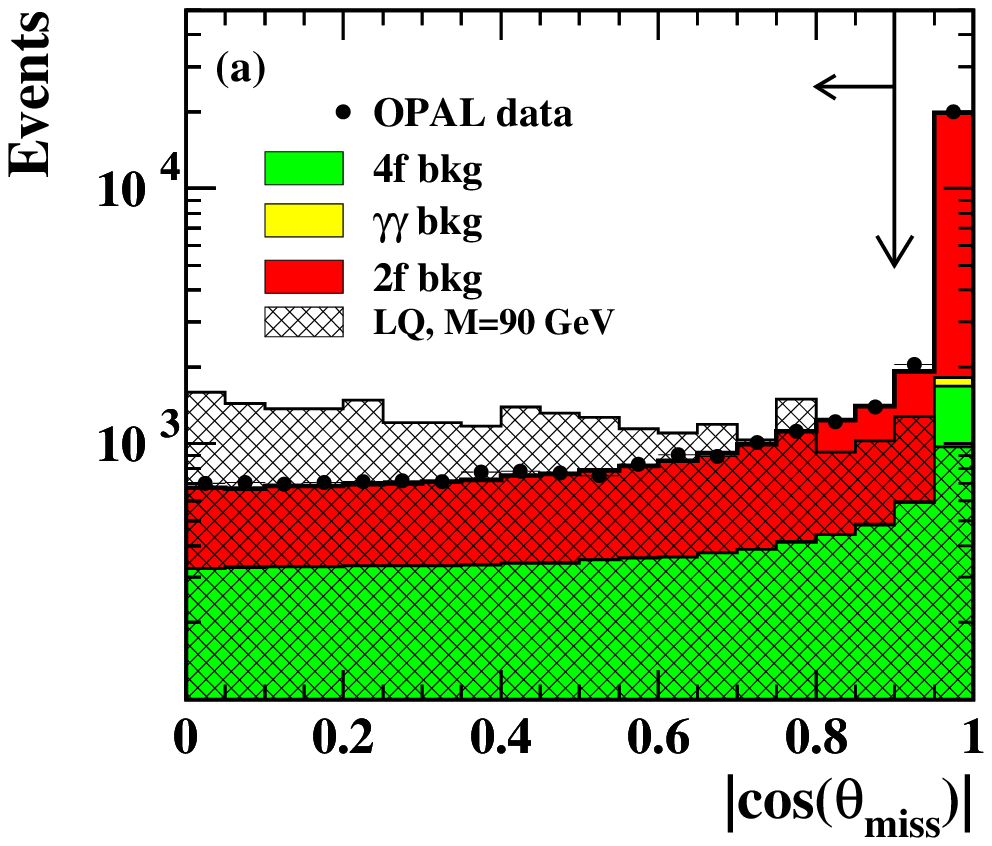,width=8cm,
                         bbllx=5pt,bblly=270pt,bburx=295pt,bbury=525pt}
             \end{center}
             \vspace*{-0.7cm}
         \end{minipage}
\hfill
       \begin{minipage}[t]{0.45\textwidth}
          \begin{center}
             \vspace*{-2cm}
                \epsfig{file=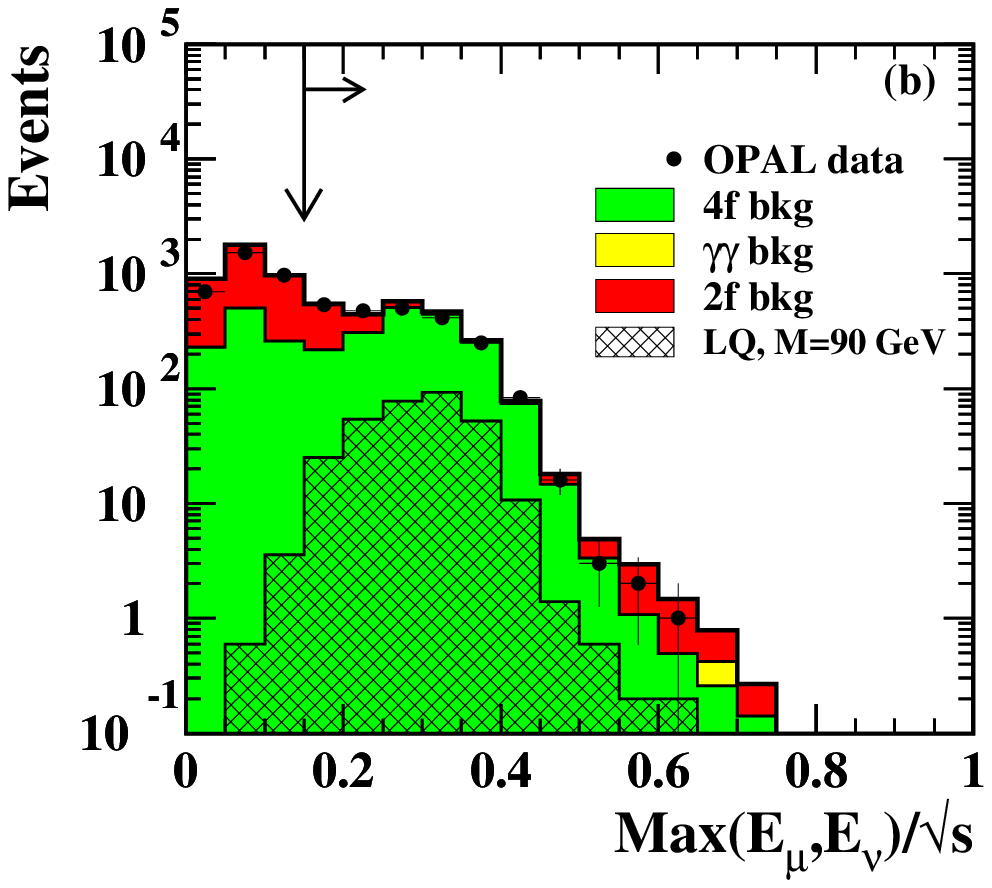,width=8cm,
                           bbllx=5pt,bblly=270pt,bburx=295pt,bbury=525pt}
           \end{center}
           \vspace*{-0.7cm}
        \end{minipage}
   \end{center}
   \begin{center}
       \begin{minipage}[t]{0.45\textwidth}
          \begin{center}
            \vspace*{-0.5cm}
               \epsfig{file=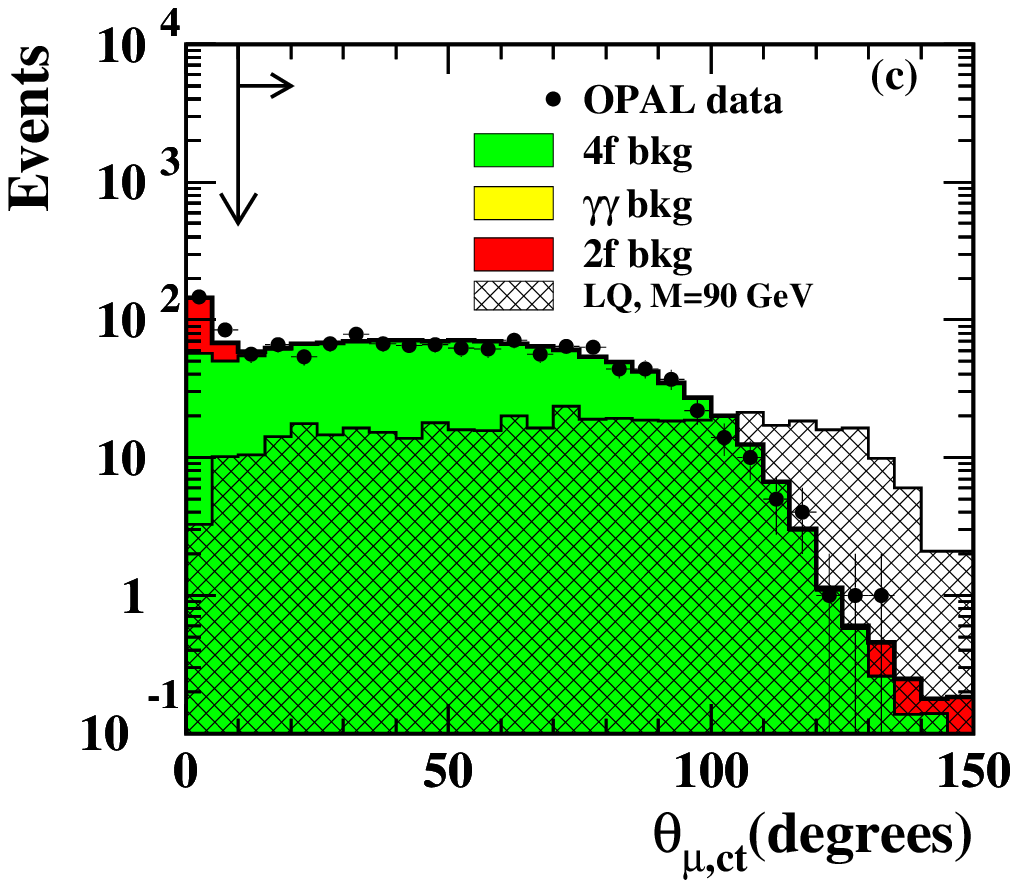,width=8cm,
                       bbllx=5pt,bblly=270pt,bburx=295pt,bbury=525pt}
             \end{center}
             \vspace*{-0.7cm}
         \end{minipage}
\hfill
       \begin{minipage}[t]{0.45\textwidth}
          \begin{center}
            \vspace*{-0.5cm}
               \epsfig{file=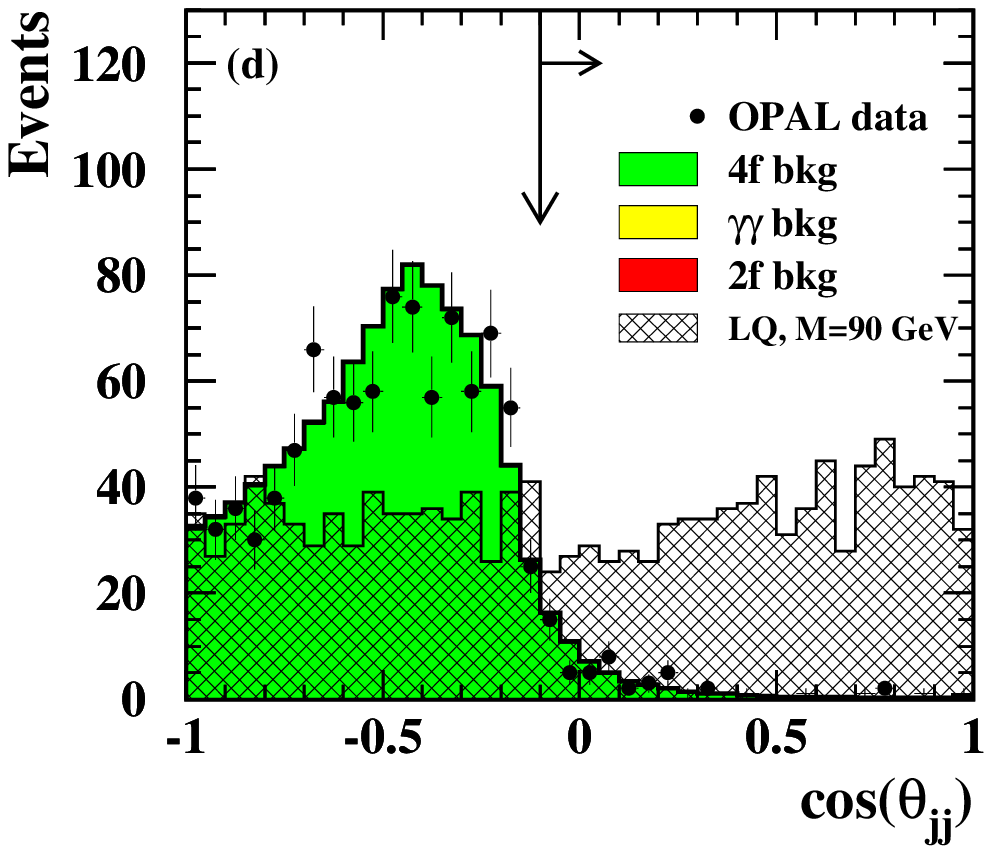,width=8cm,
                           bbllx=5pt,bblly=270pt,bburx=295pt,bbury=525pt}
          \end{center}
          \vspace*{-0.7cm}
        \end{minipage}
    \end{center}
     \begin{center}
         \begin{minipage}[t]{0.45\textwidth}
            \begin{center}
               \vspace*{-0.5cm}
                 \epsfig{file=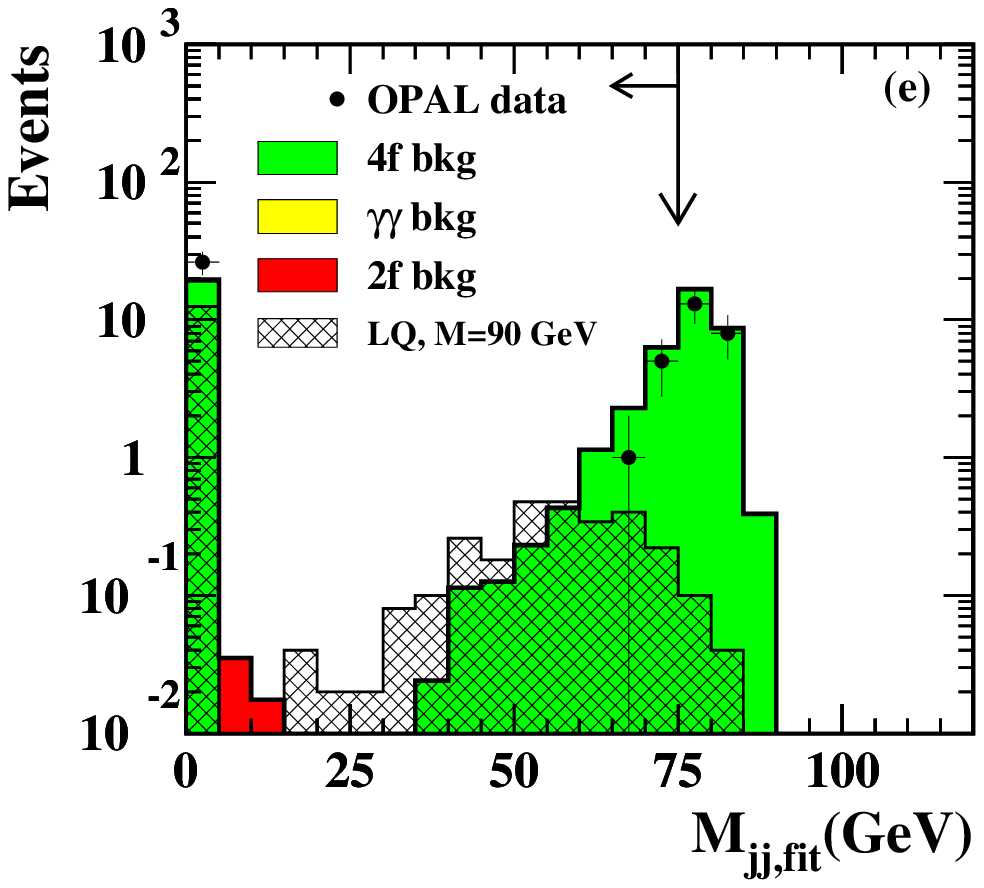,width=8cm,
                         bbllx=5pt,bblly=270pt,bburx=295pt,bbury=525pt}
             \end{center}
             \vspace*{-0.9cm}
         \end{minipage}
\hfill
         \begin{minipage}[t]{0.45\textwidth}
            \begin{center}
                \vspace*{-0.5cm}
                 \epsfig{file=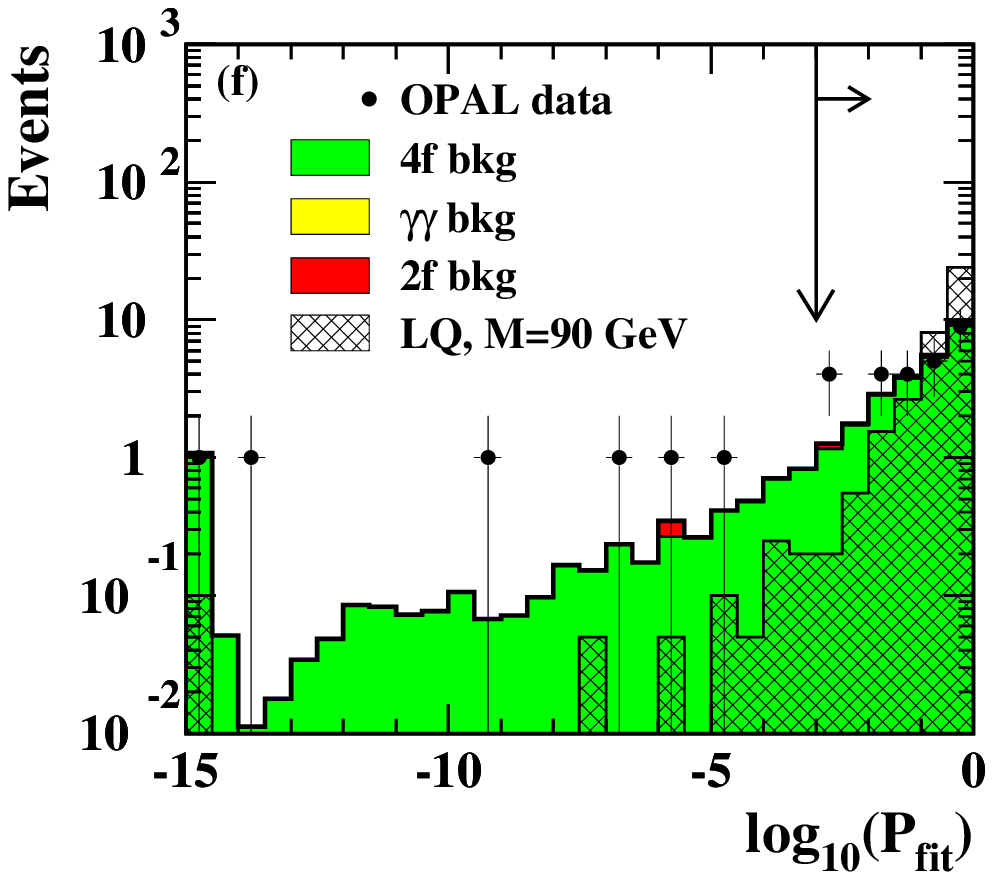,width=8cm,
                           bbllx=5pt,bblly=270pt,bburx=295pt,bbury=525pt}
            \end{center}
             \vspace*{-0.9cm}
         \end{minipage}
     \end{center}
 \caption[1]{\sl Class {\bf B1}, the $l^{\pm} \nu {\rm q q}$ channel:
                 same as Figure~\ref{fig_vv}, but for class {\bf B1}.
                 All the distributions refer to the selection for 
                 second generation leptoquarks.
              \newline
                 {\bf (a)}~The absolute value of the cosine of 
                           ~$\theta_{{\rm miss}}$, 
                          the angle between 
                          the direction of the missing momentum and the 
                          $z$-axis. 
                 {\bf (b)}~The scaled energy of the most energetic lepton 
                           (${\mu}$ or $\nu$).
                 {\bf (c)}~The angle between the direction of the most 
                            energetic muon in the event
                            and the  nearest charged track.
                 {\bf (d)}~The cosine of the angle between 
                          the directions of the two reconstructed jets, 
                          $\theta_{jj}$. 
                  {\bf (e)}~The invariant mass, ${\rm M_{jj,fit}}$, 
                            of the jet-jet system reconstructed
                            by the kinematic fit described in cut {\bf (B1-7)}.
                            The first bin also contains the events 
                            failing the fit or
                            with a probability smaller than 0.1.
                  {\bf (f)}~The logarithm of the fit probability, 
                            P$_{{\rm fit}}$,
                            used to reconstruct the leptoquark mass.  
                            The first bin also contains
                            the events failing the fit or with a
                            probability 
                             smaller than $10^{-15}$.  
             }
\label{fig_lv}
\end{figure}

The distributions of some of the variables used in the selection for
class {\bf B1} are presented in Figure~\ref{fig_lv} and show a
good agreement between the  data and the simulated background. 
\begin{table}[htbp]
\begin{center}
\hspace*{-0.2cm}
{\footnotesize
\begin{tabular}{|c||c||c||c|c|c||c|}
\hline
\hline
  \multicolumn{7}{|c|}{\rule[-3mm]{0pt}{9mm}  
              {\normalsize {\boldmath  $ {\rm e}^{\pm}\nu {\rm q q} $}}}  \\
\hline
\rule[-4mm]{0pt}{10mm} {\bf Cut} & {\bf Data}    &  {\bf Background}       
                      & 4f  &
                      $\gamma \gamma$  & 2f & 
                     {\boldmath $ \varepsilon~(\%) $ }\\
\hline
\rule[-4mm]{0pt}{10mm} (B1-1) & 37400 & 36904.0   & 8767.0 & 149.0  & 27988.0 
                              &  94.3  \\
\rule[-4mm]{0pt}{7mm} (B1-2)  & 15413 & 15192.0   & 6492.0 & ~11.9  & ~8688.0 
                              &  86.3  \\ 
\rule[-4mm]{0pt}{7mm} (B1-3)  & ~9832 & 10152.0   & 4270.0 & ~~6.2  & ~5876.0 
                              &  82.9  \\
\rule[-4mm]{0pt}{7mm} (B1-4)  & ~2533 & ~2529.0   & 1714.0 & ~~1.9  & ~~813.0 
                              &  79.5  \\
\rule[-4mm]{0pt}{7mm} (B1-5)  & ~1042 & ~1121.0   & 1109.0 & ~~1.1  & ~~~10.4 
                              &  70.5  \\
\rule[-4mm]{0pt}{7mm} (B1-6)  & ~~~32 & ~~~36.4   & ~~35.9 & ~~0.3  & ~~~~0.2 
                              &  30.0 \\
\rule[-4mm]{0pt}{7mm} (B1-7)  & ~~~17 & ~~~18.4   & ~~17.9 & ~~0.3  & ~~~~0.2 
                              &  28.6  \\
\rule[-4mm]{0pt}{7mm} (B1-8)  & ~~~13 & ~~~13.7   & ~~13.4 & ~~0.1  & ~~~~0.2 
                              &  28.0  \\
\hline \hline
   \multicolumn{7}{|c|}{\rule[-3mm]{0pt}{9mm}  
             {\normalsize {\boldmath  $\mu^{\pm}\nu {\rm q q}$}} }  \\
\hline
\rule[-4mm]{0pt}{10mm} {\bf Cut} & {\bf Data}    &  {\bf Background}     
                        & 4f  &
                         $\gamma \gamma$  & 2f & 
                         {\boldmath $ \varepsilon~(\%) $ }\\
\hline
\rule[-4mm]{0pt}{10mm} (B1-1) & 37400 & 36904.0 & 8767.0 &  149.0 & 27988.0  
                              &   92.6 \\
\rule[-4mm]{0pt}{7mm} (B1-2)  & 15413 & 15192.0 & 6492.0 &  ~11.9  & ~8688.0  
                              &   84.3 \\ 
\rule[-4mm]{0pt}{7mm} (B1-3)  & ~5478 & ~6064.0 & 2820.0 &  ~~1.0  & ~3243.0  
                              &   80.0 \\
\rule[-4mm]{0pt}{7mm} (B1-4)  & ~1311 & ~1336.0 & 1223.0 & ~$< 0.1$& ~~113.1  
                              &   76.4 \\
\rule[-4mm]{0pt}{7mm} (B1-5)  & ~~997 & ~1044.0 & ~1040.0& ~$< 0.1$& ~~~~3.5  
                              &   70.6 \\
\rule[-4mm]{0pt}{7mm} (B1-6)  & ~~~53 & ~~~56.2 & ~~~55.4&  ~~0.0  & ~~~~0.8  
                              &   37.4 \\
\rule[-4mm]{0pt}{7mm} (B1-7)  &  ~~~32 & ~~~30.4& ~~~29.6& ~~0.0  & ~~~~0.8  
                              &   36.7 \\
\rule[-4mm]{0pt}{7mm} (B1-8)  & ~~~26 & ~~~24.5 & ~~~23.9 & ~~0.0  & ~~~~0.6  
                              &  35.8 \\
\hline
\hline
\end{tabular}
 }
\caption[]{\sl Same as Table~\ref{tab:nevA}, but for  
                selection {\bf B1} (${\rm e}^{\pm}\nu {\rm qq}$ 
                and  $\mu^{\pm}\nu {\rm q q}$ channels). 
                If the number of expected events is smaller than 0.1,
                but still different from 0, the notation
                ``$< 0.1$'' is used. }
\label{tab:nevB1}
\end{center}
\end{table}
%
%
 \subsubsection{Tau channel (class B2)}
%
 \label{sec:B2}
\indent
 \begin{itemize}
\item[{\bf ~~~(B2-1)}] To account for the additional neutrinos from 
                       the tau decay,
                       the total visible energy required is smaller 
                      than in class {\bf B1}:  
                      $0.35 < E_{{\rm vis}} / \sqrt{s} < 0.85$.
\item[{\bf ~~~(B2-2)}] The direction of the missing momentum is
                       required to satisfy
                       $|\cos(\theta_{{\rm miss}})|<0.9$.
\item[{\bf ~~~(B2-3)}] The events are required to contain 
                       at least one identified 
                       tau lepton.
\item[{\bf ~~~(B2-4)}] The tau with the highest value of the output 
                       from the neural
                       network algorithm, $\mathcal{O}_{\tau}$, 
                       is chosen as the one
                       coming from the decay of the leptoquark pair. 
                       Events are accepted if  $\mathcal{O}_{\tau} > 0.75$.
\item[{\bf ~~~(B2-5)}] The energy and momentum of the tau candidate are
                       calculated using the tracks associated to the tau 
                       by the neural network algorithm and all the  
                       clusters in the calorimeters within  a cone 
                       of 10$^{\circ}$
                       around the track with the largest momentum. 
                       Due to the missing
                       energy and momentum carried away by the neutrinos
                       produced in the tau decay, the measured tau energy
                       cannot be used as an input in kinematic fits.
                       However it is rescaled using 
                       coefficents obtained by solving the following 
                       equation to require energy and momentum
                       conservation:
                   \bigskip

 \begin{center}

                   $
                      c_{1} p_{{\rm j_{1}}} +
                      c_{2} p_{{\rm j_{2}}} +
                      c_{3} p_{ \tau}        +
                       c_{4} {p}_{{\rm miss}}  =
                      (\vec{0},\sqrt{s})
                   $                      
 \end{center}         
      where $p_{{\rm j_{i}}} \equiv (\vec{p}_{{\rm j_{i}}}, E_{{\rm j_{i}}})$,
          $i=1, 2$, 
          are the measured momentum and energy 
          of the jets obtained by forcing 
          the event into two jets after having removed all the tracks 
          and clusters
                       belonging to the
                       tau, 
          $p_{\tau} \equiv (\vec{p}_{\tau}, E_{\tau}  )$ are the
                       same quantities for the tau, and 
          $p_{{\rm miss}}\equiv(\vec{p}_{{\rm miss}},|\vec{p}_{{\rm miss}}| )$ 
                       is calculated from the missing momentum of the event.
                       The $c$ coefficients are required to be positive.
                       The energy of the tau is then taken to be 
                       $E_{\tau,{\rm fit}} = c_{3}E_{\tau}$ and events with 
                       $ E_{\tau,{\rm fit}} < m_{\tau}$, where $m_{\tau}$
                       denotes the nominal mass of the tau lepton,
                        are rejected.
                       The momentum of the tau is recalculated using
                       $|\vec{p}_{\tau,{\rm fit}}| = 
                       \sqrt{(E_{\tau,{\rm fit}})^{2}-m_{\tau}^{2}}$
                       and the original measured momentum direction.
                       
                       The unchanged jet momenta and the rescaled tau momentum
                       are used as inputs to kinematic fits as
                       described in cuts {\bf (B1-7)} and {\bf (B1-8)} and
                       the events are accepted 
                       or rejected by the same criteria.
                       Since the energy of the tau is rescaled using 
                       energy and momentum
                       conservation, and since the momentum of the neutrino
                       is unmeasured, the fit has only one effective 
                       constraint. 
\end{itemize}
After cut {\bf (B2-5)}, cuts similar to {\bf (B1-4)}-{\bf (B1-6)} are
applied by using the energies and momenta of the leptons and the jets
as obtained from the kinematic fit:
\begin{itemize}

\item[{\bf ~~~(B2-6)}] The energies of the leptons, $E_{\tau, {\rm fit}}$ 
                       and  $E_{\nu, {\rm fit}}$ 
                       for the tau and the neutrino respectively, 
                       have to satisfy 
                       $0.1 < E_{\tau,{\rm fit}} / \sqrt{s}  < 0.3$ and
                       $0.2 < E_{\nu,{\rm fit}} / \sqrt{s} < 0.4 $.
\item[{\bf ~~~(B2-7)}] The angle between the tau momentum and the nearest 
                       track not
                       belonging to the tau candidate is required to be 
                       at least
                       20$^{\circ}$. The corresponding  angle for the 
                       neutrino has to be at least 10$^{\circ}$.
\item[{\bf ~~~(B2-8)}] The angle between the jets is required to satisfy
                       $\cos(\theta_{\mathrm{jj}}) > -0.1$.
\end{itemize}

In Table \ref{tab:nevB2} the numbers of events after each cut are shown,
together with the numbers of predicted background events and the 
efficiencies for signal events corresponding to 
${\mathrm{M_{LQ}}}$~=~90~GeV. 
Cuts {\bf (B2-2)} and {\bf (B2-4)} are particularly efficient in rejecting  
2f events. 4f events are especially reduced by cuts {\bf (B2-5)} and
{\bf (B2-8)}. 
At the end of the selection 35 events are observed in the data, in agreement
with the $36.0^{+2.7}_{-1.6}$ (stat.) events expected from Standard Model
processes, about $90\%$ from ${\rm W}^{\pm}$ boson pair production events. 
At  $\sqrt{s} = 206$ GeV the selection efficiency for signal events
for  leptoquarks of mass M$_{{\rm LQ}}$ = 90 GeV 
is  $(18.0 \pm 0.9({\rm stat.}))\%$.\\

\begin{figure}[p]
    \begin{center}
         \begin{minipage}[t]{0.45\textwidth}
            \begin{center}
               \vspace*{-2cm}
                 \epsfig{file=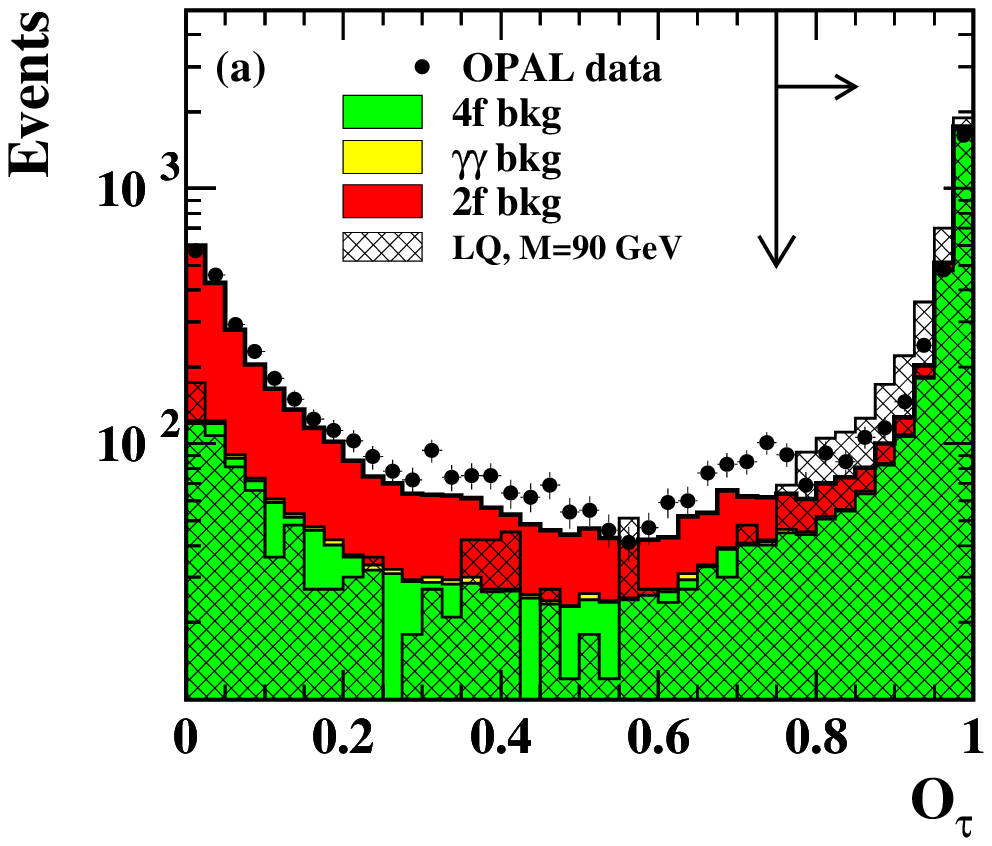,width=8cm,
                           bbllx=5pt,bblly=270pt,bburx=295pt,bbury=525pt}
             \end{center}
             \vspace*{-0.7cm}
         \end{minipage}
\hfill
       \begin{minipage}[t]{0.45\textwidth}
          \begin{center}
              \vspace*{-2cm}
                \epsfig{file=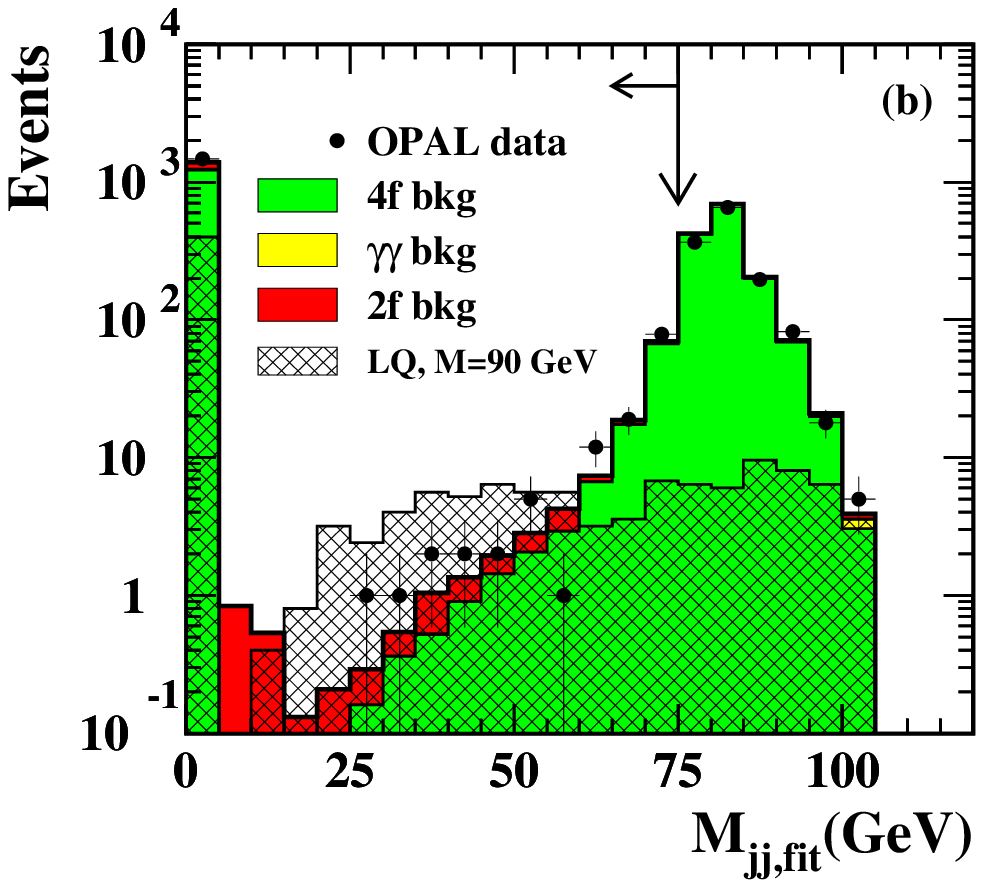,width=8cm,
                           bbllx=5pt,bblly=270pt,bburx=295pt,bbury=525pt}
           \end{center}
          \vspace*{-0.7cm}
        \end{minipage}
   \end{center}
   \begin{center}
        \begin{minipage}[t]{0.45\textwidth}
            \begin{center}
               \vspace*{-0.5cm}
                 \epsfig{file=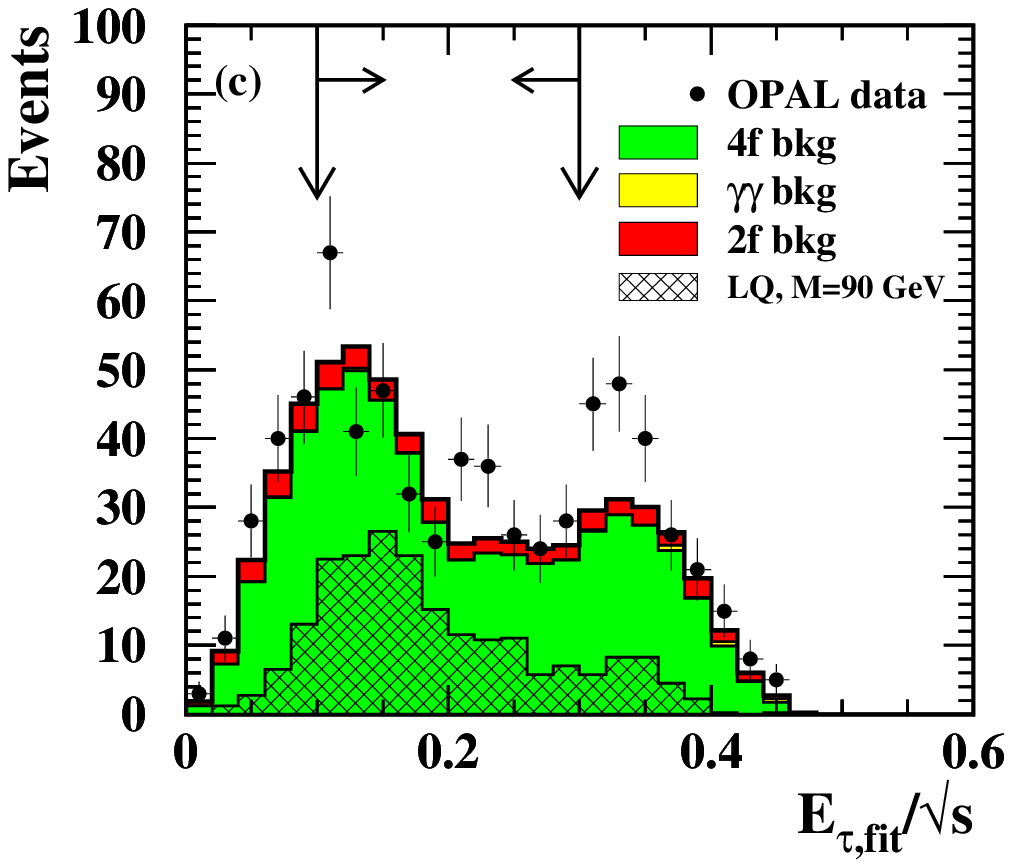,width=8cm,
                           bbllx=5pt,bblly=270pt,bburx=295pt,bbury=525pt}
             \end{center}
             \vspace*{-0.7cm}
        \end{minipage}
\hfill
       \begin{minipage}[t]{0.45\textwidth}
          \begin{center}
              \vspace*{-0.5cm}
                 \epsfig{file=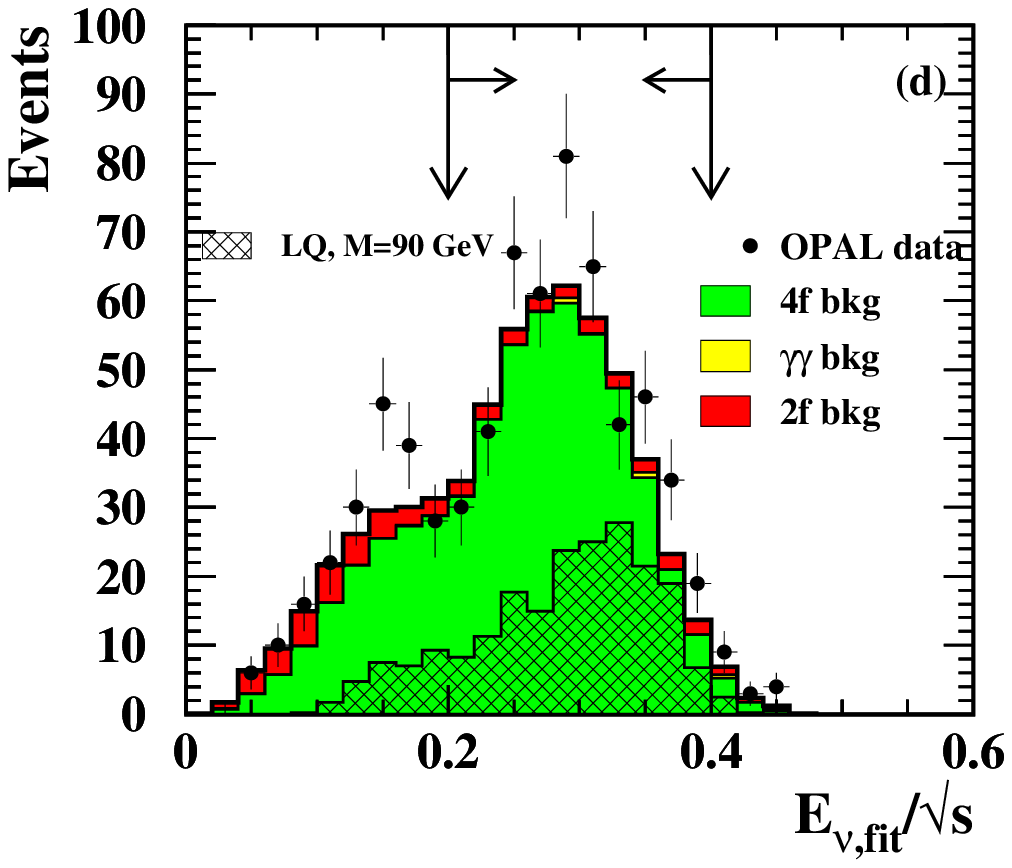,width=8cm,
                           bbllx=5pt,bblly=270pt,bburx=295pt,bbury=525pt}
         \end{center}
          \vspace*{-0.7cm}
        \end{minipage}
    \end{center}
    \begin{center}
       \begin{minipage}[t]{0.45\textwidth}
           \begin{center}
              \vspace*{-0.5cm}
                 \epsfig{file=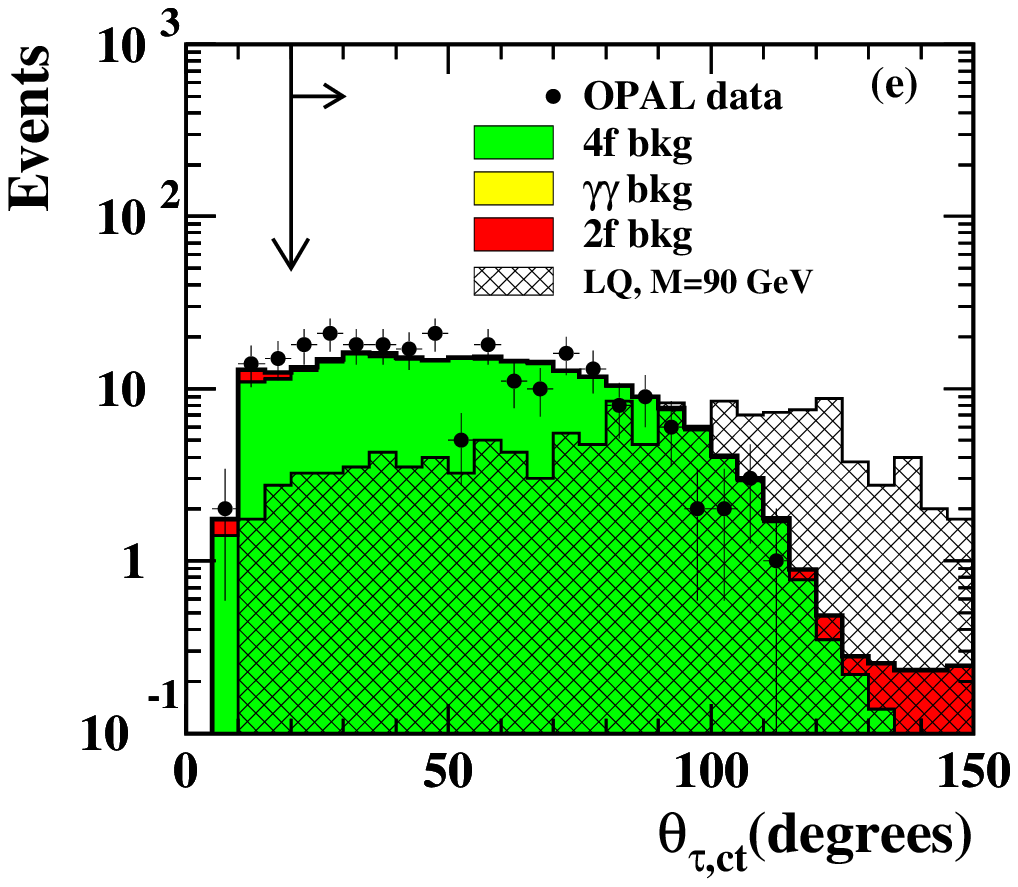,width=8cm,
                         bbllx=5pt,bblly=270pt,bburx=295pt,bbury=525pt}
             \end{center}
             \vspace*{-0.9cm}
       \end{minipage}
\hfill
       \begin{minipage}[t]{0.45\textwidth}
          \begin{center}
            \vspace*{-0.5cm}
              \epsfig{file=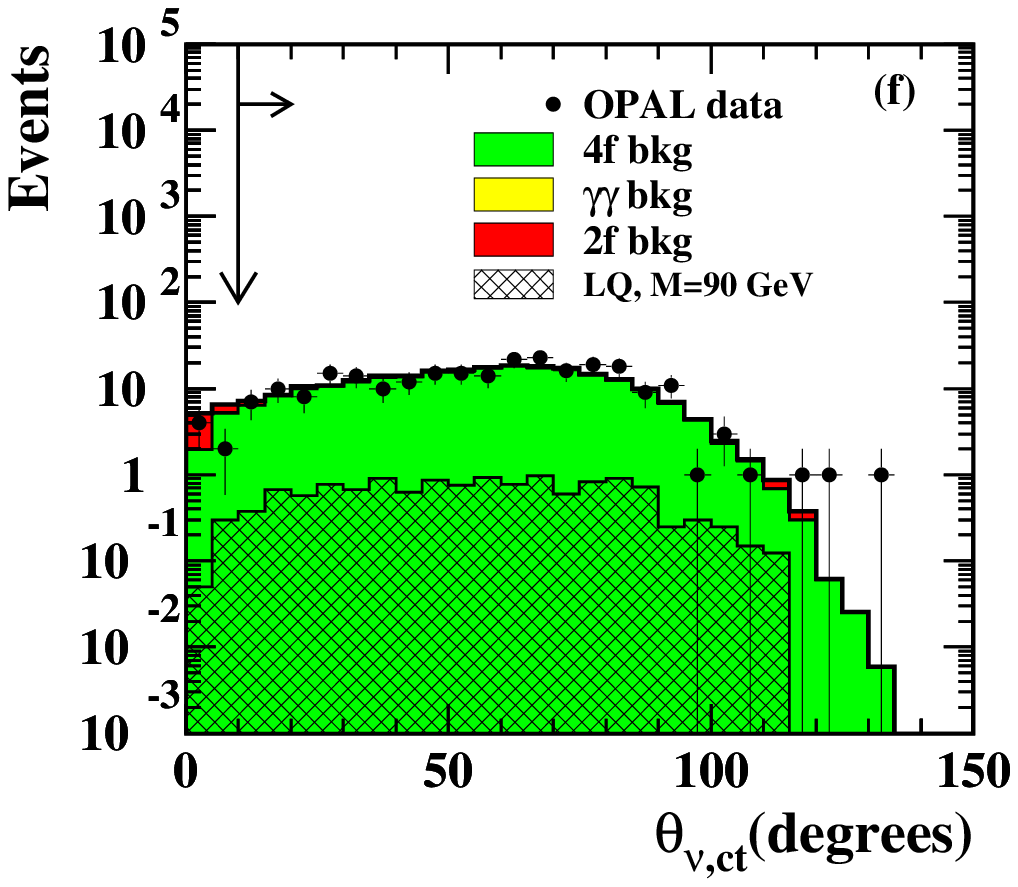,width=8cm,
                         bbllx=5pt,bblly=270pt,bburx=295pt,bbury=525pt}
         \end{center}
          \vspace*{-0.9cm}
       \end{minipage}
     \end{center}
\caption[1]{\sl Class {\bf B2}, the $\tau^{\pm} \nu {\rm q q}$ channel:
                 same as Figure~\ref{fig_vv}, but for class {\bf B2}. 
              \newline
                 {\bf (a)}~The output $\mathcal{O}_{\tau}$ 
                           from the neural network algorithm, 
                          for the tau 
                          with the highest output
                          in the event. 
                  {\bf (b)}~The invariant mass, ${\rm M_{jj,fit}}$ 
                            of the jet-jet system reconstructed
                            by the first kinematic fit   
                            described 
                            in cut {\bf (B2-5)}.
                             The first bin contains also the events failing 
                            the fit or with a  
                             probability smaller than 0.1. 
                 {\bf (c)--(d)}~The  scaled energies of the tau lepton
                                      and the neutrino, 
                                       $E_{\tau,{\rm fit}} / \sqrt{s}$ and
                                       $ E_{\nu,{\rm fit}} / \sqrt{s} $, 
                                      as calculated by the kinematic
                                       fit used to reconstruct the 
                                       leptoquark mass, after cut {\bf (B2-5)}.
                 {\bf (e)--(f)}~The angles between the leptons and 
                                the nearest charged track for the
                                 tau, $\theta_{\tau,{\rm ct}}$, and
                                 the neutrino, $\theta_{\nu,{\rm ct}}$,
                                 respectively, after cut {\bf (B2-6)}.  
             }
 \label{fig_tv}
 \end{figure}

Figure~\ref{fig_tv} shows some of the variables used in this selection.
The discrepancy between the
 observed data and the simulated SM background
in the distribution of the output from the neural net, $\mathcal{O}_{\tau}$,
Figure~\ref{fig_tv}(a), is
due to an excess of low energy tau candidates in the data.
These candidates are not  selected by cut {\bf (B2-6)}.
The excess in the data between 0.3$\sqrt{s}$ and
0.36$\sqrt{s}$ for the scaled energy of the tau (Figure~\ref{fig_tv}(c)),
predominantly stems from events taken at the lowest 
centre-of-mass energies.  
\vspace*{0.3cm}
\begin{table}[htb]
\begin{center}
{\footnotesize
\begin{tabular}{|c||c||c||c|c|c||c|}
\hline
\hline
 \multicolumn{7}{|c|}{\rule[-3mm]{0pt}{9mm}  
          {\normalsize {\boldmath  $ \tau^{\pm}\nu {\rm q q} $}}}  \\
\hline
\rule[-4mm]{0pt}{10mm} {\bf Cut} & {\bf Data}    &  {\bf Background}       
                      & 4f  &
                        $\gamma \gamma$  & 2f & 
                       {\boldmath $\varepsilon~(\%)$ }\\
\hline
\rule[-4mm]{0pt}{10mm} (B2-1) & 30206 & 29212.0  & 5884.0 & 362.2  &  22966.0  
                                                   & 92.5 \\
\rule[-4mm]{0pt}{7mm} (B2-2) & ~7049 & ~6580.0  & 4106.0 & ~57.2  &  ~2417.0  
                                                  & 84.8 \\
\rule[-4mm]{0pt}{7mm} (B2-3) & ~6731 & ~6304.0   & 4043.0 & ~52.5 &  ~2208.0  
                                                  & 83.9 \\
\rule[-4mm]{0pt}{7mm} (B2-4) & ~3048 & ~3038.0   & 2803.0 & ~16.1 &  ~~218.5  
                                                  & 64.0 \\
\rule[-4mm]{0pt}{7mm} (B2-5) & ~~699 & ~~620.4   & ~561.6 & ~~3.2 &  ~~~55.6  
                                                 & 40.4 \\
\rule[-4mm]{0pt}{7mm} (B2-6)&  ~~252 & ~~248.0   & ~236.4 & ~~0.6 &  ~~~11.0  
                                                 & 26.1 \\
\rule[-4mm]{0pt}{7mm} (B2-7)&  ~~216 & ~~211.1   & ~206.7 & ~~0.5 &  ~~~~3.9  
                                                 & 24.6 \\
\rule[-4mm]{0pt}{7mm} (B2-8)&  ~~~35 & ~~~36.0   & ~~34.4 & ~~0.1 &  ~~~~1.5  
                                                 & 18.0 \\
\hline \hline
\end{tabular}
 }
\caption[]{\sl Same as Table~\ref{tab:nevA}, but for 
                selection {\bf B2} ($\tau^{\pm}\nu {\rm qq}$ channel). 
            }
\label{tab:nevB2}
\end{center}
\end{table}
%
%
\subsection{The {\boldmath$l^{+} l^{-} {\rm q q}$} channel (class C)}
%
\indent
Signal events of this type are characterized 
by the presence of a pair of isolated high energy charged leptons of the same
generation and opposite charge. 
The missing energy of the events is small
for the first and second generation 
while, in the case of third generation,
a significant missing energy is expected  
because of neutrinos produced in the tau decays.
Different sets of cuts were applied to select events with electrons or muons
and to select events with taus. 
 \subsubsection{Electron and muon channels (class C1)}
\begin{itemize}
\item[{\bf ~~~(C1-1)}] 
                   The visible energy is required to satisfy  
                    $E_{\rm{vis}} > 0.75\sqrt{s}$. 
\item[{\bf ~~~(C1-2)}] The presence of at least one pair of 
                   identified electrons or muons with opposite 
                   charge is required.
                   The most energetic leptons of the same generation
                   and of opposite  charge are called the ``pair" 
                   in the following. 
\item[{\bf ~~~(C1-3)}]
                   The energy of the most energetic lepton 
                   of the pair, $E_{{\rm e1}}$ or $E_{{\rm \mu 1 }}$, 
                   has to exceed 0.15$\sqrt{s}$, while an energy
                   $E_{{\rm e2}}$ or $E_{{\rm \mu 2 }}$ 
                   of at least 0.1$\sqrt{s}$ is required for the other 
                   lepton.
\item[{\bf ~~~(C1-4)}] An isolation cut is applied by requiring that the angle 
                      between each lepton of the pair and the nearest 
                      charged track
                      is at least 10$^{\circ}$.
\item[{\bf ~~~(C1-5)}] After the exclusion of the tracks corresponding to the 
                     lepton pair, the event is 
                   forced into two jets. Events are rejected 
                   if  $\cos(\theta_{{\rm jj}})$ and $\cos(\theta_{{\rm ll}})$
                   are both 
                   smaller than $-0.8$, 
                  where $\theta_{{\rm jj}}$ and 
                   $\theta_{{\rm ll}}$ are the angles between the two jets and 
                   the two leptons respectively.
\item[{\bf ~~~(C1-6)}] Finally, a kinematic fit with five effective
                   constraints is applied to
                   reconstruct the leptoquark mass by requiring energy and
                   momentum conservation and constraining the two
                   lepton-jet pairs to have the same mass. Of the
                   two possible lepton-jet combinations the one
                   with the higher fit probability  is chosen.
                   Events are accepted if this probability is
                   larger than 10$^{-6}$, while the fitted mass
                   has to be at least 50 GeV. 
\end{itemize}
The numbers of events after each cut, together with the numbers of expected 
background events and the efficiencies for signal events corresponding
to ${\mathrm{M_{LQ}}}$~=~90~GeV, are shown in Table~\ref{tab:nevC1}.
Cuts {\bf (C1-2)} and  {\bf (C1-3)} greatly reduce all kinds of background.
Cut {\bf (C1-4)} totally suppresses
2f events.
The requirements {\bf (C1-5)} and  {\bf (C1-6)} are useful in further reducing 
four-fermion background.  
In the search for first generation leptoquarks, 20 
events are observed in the data while 
$12.8^{+2.5}_{-1.3}$ (stat.) are expected from Standard Model
background. For the second generation, 4 events are observed, 
the background expectation being $8.7^{+2.2}_{-0.7}$ (stat.). A contribution
of about $60\%$ to the total background is expected from 
${\rm Z}^{0}{\rm Z}^{0}$ events. 
The efficiency of the selection 
for signal events with leptoquarks
of mass M$_{{\rm LQ}}$ = 90 GeV is $(50.3 \pm 0.7({\rm stat.}))\%$ and 
$(62.8 \pm 0.7({\rm stat.}))\%$ for the first and second generation,
 respectively, at $\sqrt{s} = 206$ GeV.\\

\begin{figure}[p]
   \begin{center}
       \begin{minipage}[t]{0.45\textwidth}
          \begin{center}
             \vspace*{-2cm}
                \epsfig{file=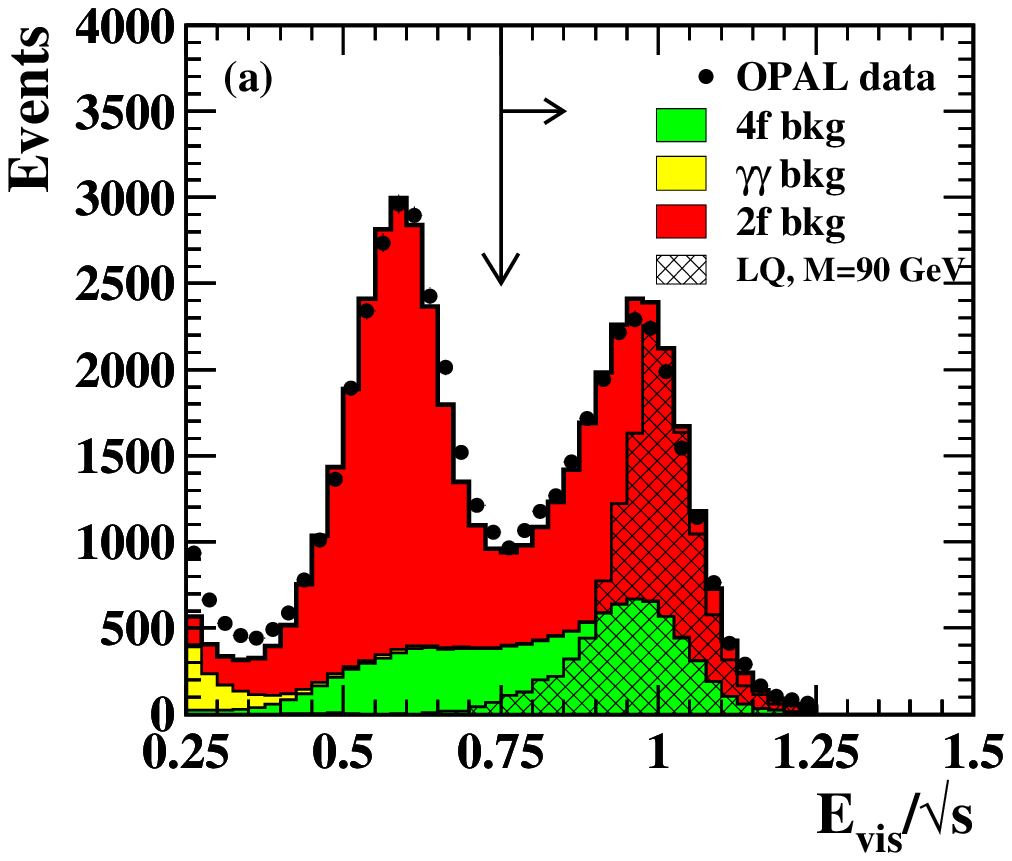,width=8cm,
                         bbllx=5pt,bblly=270pt,bburx=295pt,bbury=525pt}
             \end{center}
             \vspace*{-0.7cm}
       \end{minipage}
\hfill
       \begin{minipage}[t]{0.45\textwidth}
          \begin{center}
             \vspace*{-2cm}
                 \epsfig{file=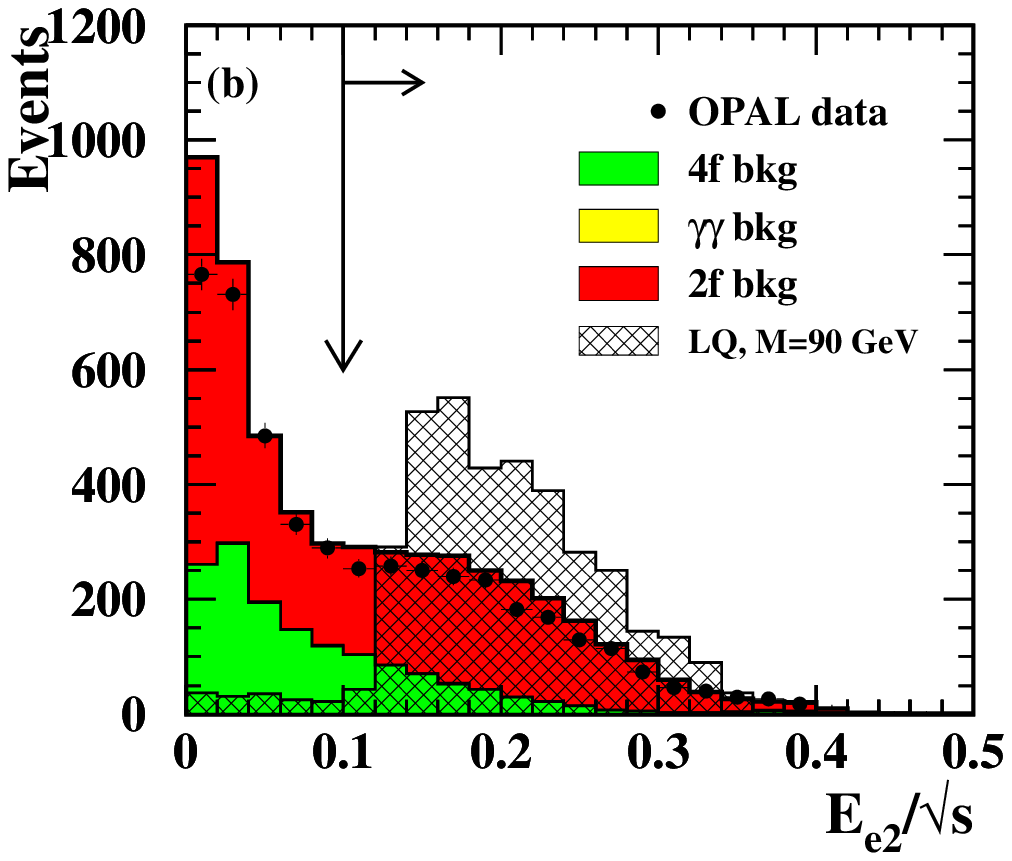,width=8cm,
                           bbllx=5pt,bblly=270pt,bburx=295pt,bbury=525pt}
          \end{center}
          \vspace*{-0.7cm}
        \end{minipage}
    \end{center}
    \begin{center}
       \begin{minipage}[t]{0.45\textwidth}
            \begin{center}
            \vspace*{-0.5cm}
                \epsfig{file=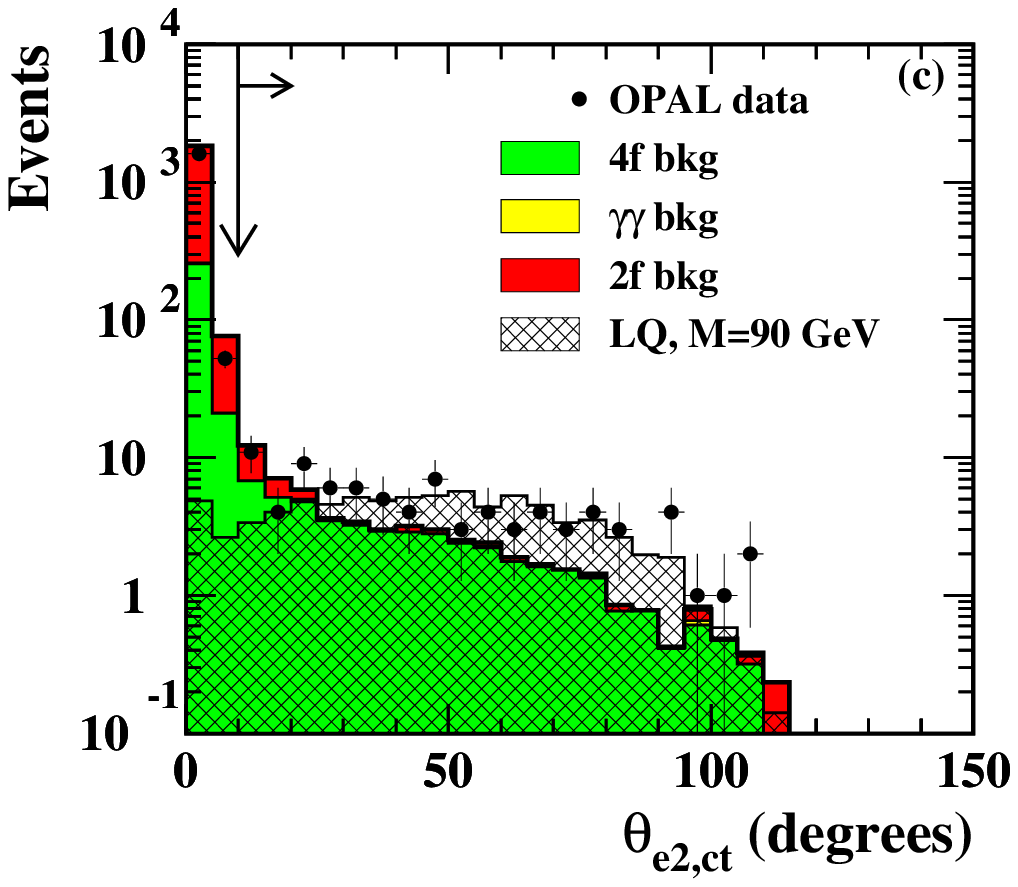,width=8cm,
                           bbllx=5pt,bblly=270pt,bburx=295pt,bbury=525pt}
             \end{center}
             \vspace*{-0.7cm}
         \end{minipage}
\hfill
       \begin{minipage}[t]{0.45\textwidth}
          \begin{center}
            \vspace*{-0.5cm}
              \epsfig{file=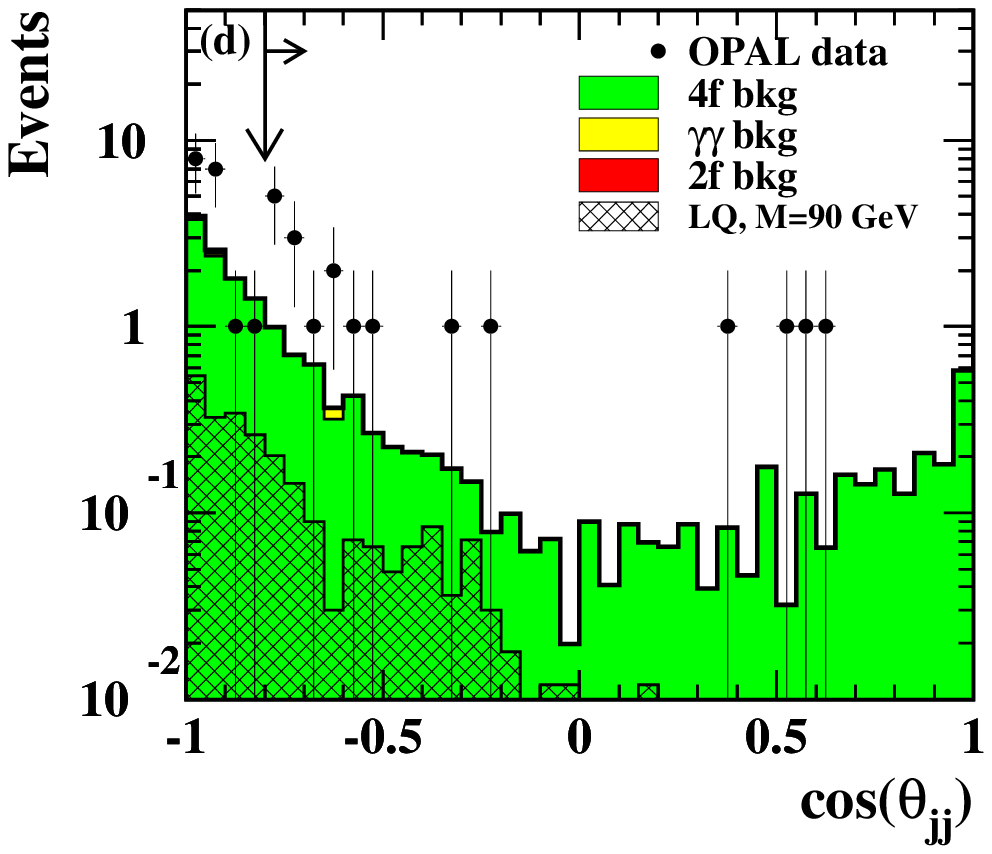,width=8cm,
                           bbllx=5pt,bblly=270pt,bburx=295pt,bbury=525pt}
         \end{center}
          \vspace*{-0.7cm}
       \end{minipage}
    \end{center}
    \begin{center}
       \begin{minipage}[t]{0.45\textwidth}
           \begin{center}
             \vspace*{-0.5cm}
               \epsfig{file=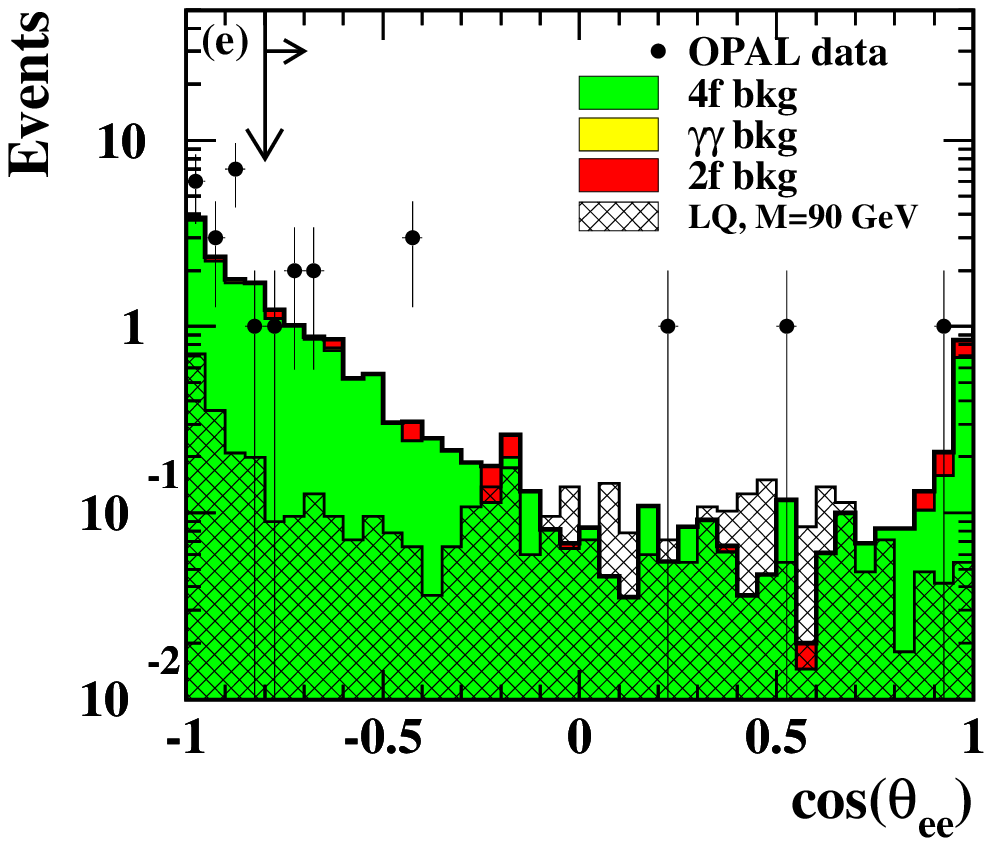,width=8cm,
                           bbllx=5pt,bblly=270pt,bburx=295pt,bbury=525pt}
           \end{center}
           \vspace*{-0.9cm}
       \end{minipage}
\hfill
       \begin{minipage}[t]{0.45\textwidth}
          \begin{center}
            \vspace*{-0.5cm}
                 \epsfig{file=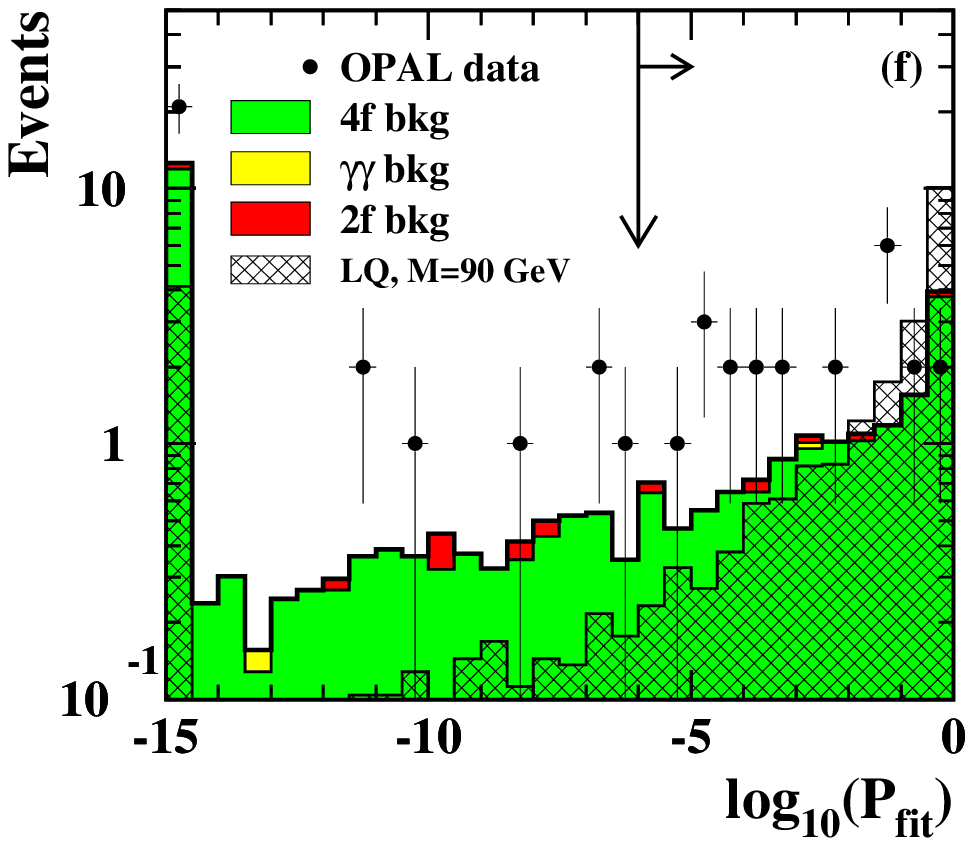,width=8cm,
                           bbllx=5pt,bblly=270pt,bburx=295pt,bbury=525pt}
           \end{center}
          \vspace*{-0.9cm}
       \end{minipage}
    \end{center}
\vspace*{-0.4cm} 
 \caption[1]{\sl Class {\bf C1}, the $l^{+} l^{-} {\rm q q}$ channel:
                  same as Figure~\ref{fig_vv}, but for class {\bf C1}.
                  All the distributions refer to the selection for first 
                  generation leptoquarks.
              \newline
                 {\bf (a)}~The scaled visible energy. 
                 {\bf (b)}~The scaled energy of the second most energetic 
                           electron in the event, after cut {\bf (C1-2)}.  
                 {\bf (c)}~The angle between the second most energetic
                           electron and the nearest charged track, 
                           after {\bf (C1-3)}. 
                 {\bf (d)}~The cosine of $\theta_{{\rm jj}}$, 
                           the angle between  
                           the two reconstructed jets, after cut {\bf (C1-4)}. 
                            The distribution does not contain
                            the events with $\cos(\theta_{{\rm ee}}) > -0.8$, 
                            which are
                            always selected by cut {\bf (C1-5)}, 
                            independently of the value of
                            $\cos(\theta_{{\rm jj}})$. 
                 {\bf (e)}~The cosine of $\theta_{{\rm ee}}$, 
                           the angle between  
                           the two most energetic electrons, 
                           after cut {\bf (C1-4)}.   
                           The distribution does not contain
                           the events with $\cos(\theta_{{\rm jj}}) > -0.8$, 
                            which
                           are  always selected by cut {\bf (C1-5)}, 
                           independently of the value of
                            $\cos(\theta_{{\rm ee}})$. 
                 {\bf (f)}~The logarithm of the probability of the fit
                            used to reconstruct the leptoquark mass. 
                              The first bin contains also the events failing 
                            the fit or with a probability 
                            smaller than  $10^{-15}$.
           }
 \label{fig_ll}
 \end{figure}

Figure~\ref{fig_ll} shows some of the variables used to select 
events belonging to
class {\bf C1}. 
\begin{table}[t]
\begin{center}
\hspace*{-0.75cm}
{\footnotesize
\begin{tabular}{|c||c||c||c|c|c||c|}
\hline
\hline
 \multicolumn{7}{|c|}{\rule[-3mm]{0pt}{9mm}  {\normalsize {\boldmath  
$ {\rm e}^{+}{\rm e}^{-} {\rm q q}$}}} \\ 
\hline
\rule[-3mm]{0pt}{9mm} {\bf Cut} & {\bf Data}    &  {\bf Background}         
                         & 4f  &
                           $\gamma \gamma$  & 2f & 
                           {\boldmath $\varepsilon~(\%)$ }\\
\hline
\rule[-3mm]{0pt}{9mm} (C1-1) & 22905 & 23093.0   & 6998.0  & 33.0  &  16062.0 
                             & 95.2  \\
\rule[-4mm]{0pt}{7mm} (C1-2) & ~4687 & ~5269.0   & 1475.0  & ~7.0  &  ~3787.0 
                             & 83.6  \\ 
\rule[-4mm]{0pt}{7mm} (C1-3) & ~1729 & ~1969.0   & ~322.9  & ~3.1  &  ~1643.0 
                             &  79.9  \\ 
\rule[-4mm]{0pt}{7mm} (C1-4) & ~~~67 & ~~~42.3   & ~~40.4  & ~0.2  &  ~~~~1.7 
                             &  70.4  \\
\rule[-4mm]{0pt}{7mm} (C1-5) & ~~~50 & ~~~32.5   & ~~31.0  & ~0.1  &  ~~~~1.4 
                             &  65.3  \\
\rule[-4mm]{0pt}{7mm} (C1-6) & ~~~20 & ~~~12.8  & ~~12.3  &$ < 0.1$ & ~~~~0.5 
                             &   50.3   \\
\hline \hline
  \multicolumn{7}{|c|}{\rule[-3mm]{0pt}{9mm}  
              {\normalsize {\boldmath  $ \mu^{+}\mu^{-} {\rm q q} $}}}  \\
\hline
\rule[-3mm]{0pt}{9mm} {\bf Cut} & {\bf Data}    &  {\bf Background}       
                       & 4f  &
                         $\gamma \gamma$  & 2f & 
                         {\boldmath $\varepsilon~(\%)$ }\\

\hline
\rule[-3mm]{0pt}{9mm} (C1-1) & 22905 & 23093.0   & 6998.0 & 33.0  & 16062.0   
                             & 91.7 \\
\rule[-4mm]{0pt}{7mm} (C1-2) & ~1821 & ~2047.0   & ~643.6 & ~1.0  & ~1402.0   
                             & 80.4 \\ 
\rule[-4mm]{0pt}{7mm} (C1-3) & ~~~68 & ~~~70.3   & ~~49.3 & ~0.0  & ~~~21.0   
                             &  77.4 \\ 
\rule[-4mm]{0pt}{7mm} (C1-4) & ~~~29 & ~~~28.0   & ~~28.0 & ~0.0  &~~~~0.0   
                             &  72.1 \\
\rule[-4mm]{0pt}{7mm} (C1-5) & ~~~21 & ~~~22.6   & ~~22.6 & ~0.0  & ~~~~0.0   
                             &  67.7 \\
\rule[-4mm]{0pt}{7mm} (C1-6) & ~~~~4 & ~~~~8.7   & ~~~8.7 & ~0.0  & ~~~~0.0   
                             &  62.8  \\
\hline \hline
\end{tabular}
 }
\caption[]{\sl Same as Table~\ref{tab:nevA}, but for  
                selection {\bf C1} ($ {\rm e}^{+}{\rm e}^{-} {\rm q q} $ 
                and $ \mu^{+}\mu^{-} {\rm q q }$ channels). 
           }
\label{tab:nevC1}
\end{center}
\end{table}

%
 \subsubsection{Tau channel (class C2)}
%
%
\begin{itemize}
\item[{\bf (C2-1)}] 
                   The visible energy must lie in the range
                   $ 0.45 <  E_{\rm{vis}} / \sqrt{s} < 0.95$.
\item[{\bf (C2-2)}] The presence of at least one pair of 
                   identified taus with opposite charge is required.
\item[{\bf (C2-3)}] 
                   For each electric charge
                   the tau candidate with the largest output from the neural
                   network algorithm is chosen, and the two outputs 
                   $\mathcal{O}_{\tau_{1}}$ and $\mathcal{O}_{\tau_{2}}$
                   are combined to form the two-tau probability :
                   \begin{center} {\large
                    $ \mathcal{P}_{\tau\tau} = 
                    \frac{\mathcal{O}_{\tau_{1}}\mathcal{O}_{\tau_{2}}}
                         {\mathcal{O}_{\tau_{1}}\mathcal{O}_{\tau_{2}}+
                         (1-\mathcal{O}_{\tau_{1}})(1-\mathcal{O}_{\tau_{2}})}
                    $ }
                   \end{center}
                    $ \mathcal{P}_{\tau\tau}$ is required to be at least 0.9.
\item[{\bf (C2-4)}] As in selection {\bf B2}, the energy and momentum of each
                   tau of the pair are calculated from the tracks
                   associated to the tau by the identification algorithm
                   and all the clusters in the calorimeters within a cone 
                   of half angle of 10$^{\circ}$ around the track with 
                   the largest momentum.
                   The taus are then removed and the event is forced into
                   a two-jet configuration. Then an equation similar
                   to the one described in cut {\bf (B2-5)}, but
                   containing the energies and momenta of the jets and taus, 
                   is solved. 
                  
                       The unchanged jet momenta and the rescaled tau momenta
                       are used as inputs to the  kinematic fit 
                       described in cut {\bf (C1-6)}     to reconstruct
                       the leptoquark mass.
                       As the energies of the taus are rescaled using
                       energy and momentum conservation, the effective
                       number of constraints in the fit is three. 
                       The events are selected if the fitted 
                       mass is larger than 50 GeV and the fit probability 
                       is larger than 10$^{-6}$.
 \end{itemize}
After cut {\bf (C2-4)}, the following selections similar 
to {\bf (C1-3)}-{\bf (C1-5)} are
applied using the energies and momenta of the taus and the jets 
obtained from the kinematic fit.
\begin{itemize}
\item[{\bf (C2-5)}] The energy of the most energetic tau  
                   of the pair has to exceed 0.15$\sqrt{s}$, while an energy 
                   of at least 0.1$\sqrt{s}$ is required for the other tau.
\item[{\bf (C2-6)}] The angle between a tau and the nearest charged
                   track not belonging to the tau itself, 
                   $\theta_{{\rm \tau 1, ct}}$ and
                    $\theta_{{\rm \tau 2, ct}}$, 
                   is required to be at least 20$^{\circ}$ for each
                   candidate.
\item[{\bf (C2-7)}] 
                   Events are rejected 
                   if  both $\cos(\theta_{{\rm jj}})$ 
                   and $\cos(\theta_{\tau\tau})$
                   are smaller than $-0.8$.
\end{itemize}
The numbers of events after each cut are shown in Table~\ref{tab:nevC2}, 
together with the numbers of expected 
background events and the efficiencies for signal events corresponding
to ${\mathrm{M_{LQ}}}$~=~90~GeV.
Cut {\bf (C2-3)} reduces in particular the background 
from 2f events. 
Cut {\bf (C2-4)} is efficient against each kind of background.
In the whole data sample 37 events survive the selection, in good agreement
with the number expected from Standard Model background, 
of $38.0^{+3.1}_{-2.0}$ (stat.), mostly due to ${\rm W}^{\pm}$ 
and ${\rm Z}^{0}$ pair production
processes (about $50\%$ and $20\%$).
At  $\sqrt{s} = 206$ GeV the efficiency for signal events for leptoquarks
of mass M$_{{\rm LQ}}$ = 90 GeV is $(33.3 \pm 0.7({\rm stat.}))\%$. \\  
 
\begin{figure}[p]
  \begin{center}
     \begin{minipage}[t]{0.45\textwidth}
        \begin{center}
           \vspace*{-2cm}
               \epsfig{file=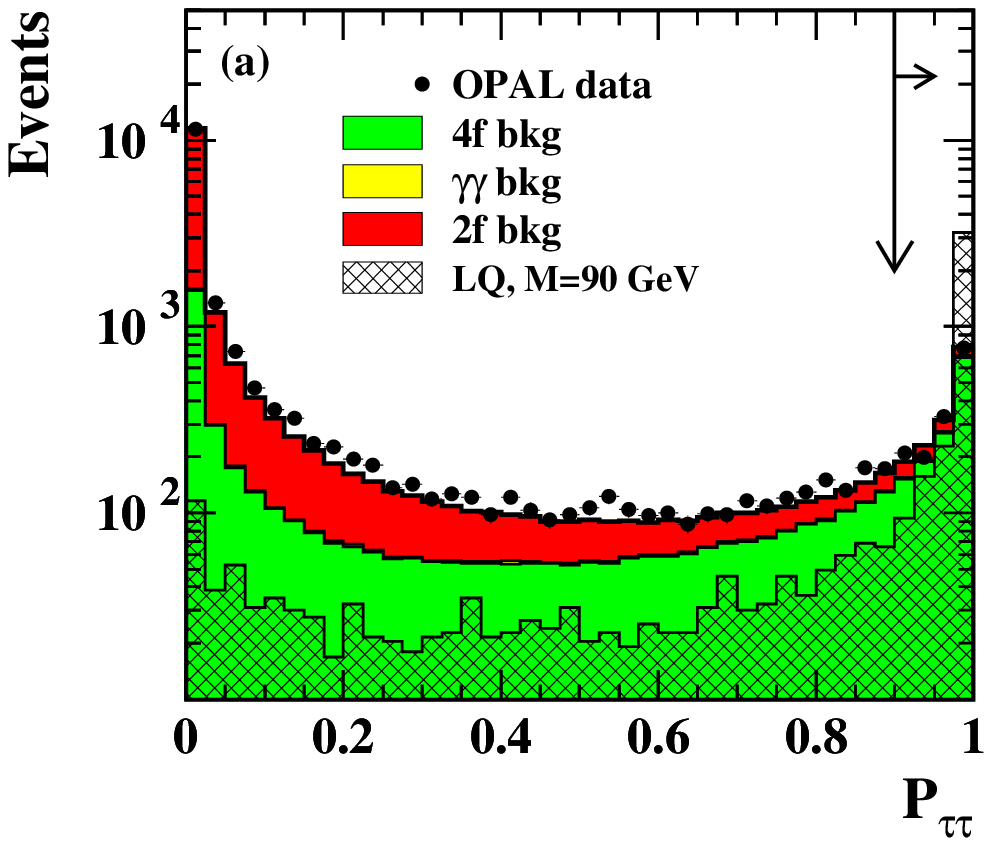,width=8cm,
                           bbllx=5pt,bblly=270pt,bburx=295pt,bbury=525pt}
         \end{center}
         \vspace*{-0.7cm}
     \end{minipage}
\hfill
     \begin{minipage}[t]{0.45\textwidth}
          \begin{center}
            \vspace*{-2cm}
                 \epsfig{file=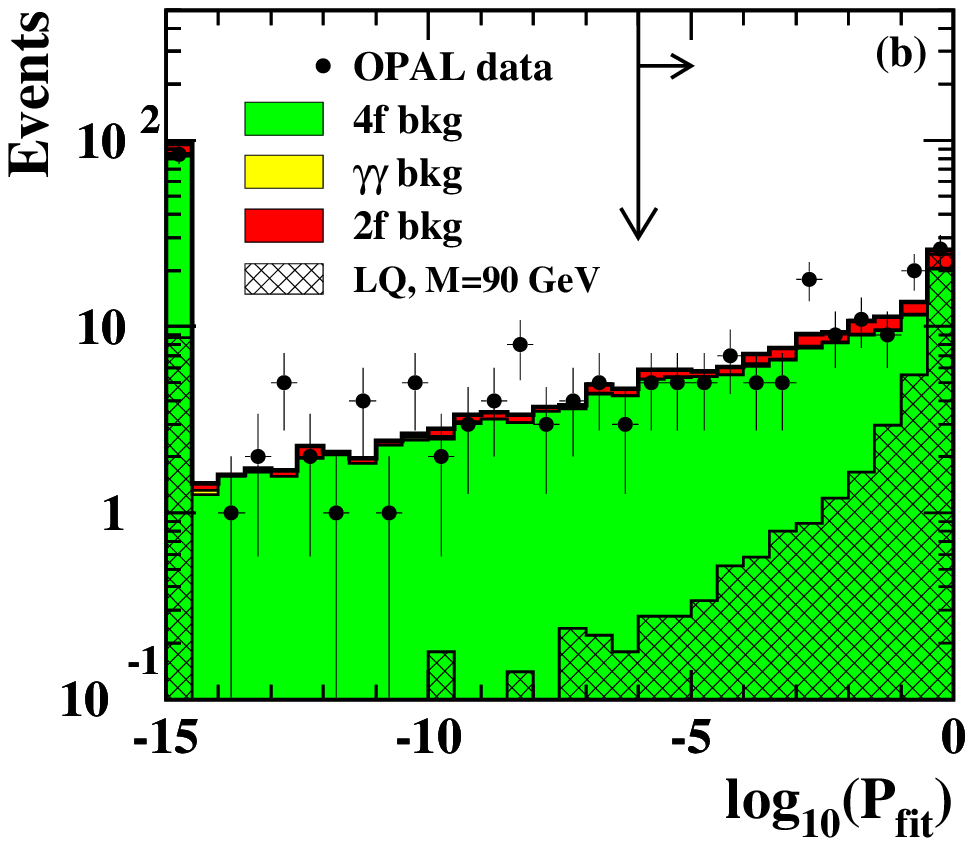,width=8cm,
                           bbllx=5pt,bblly=270pt,bburx=295pt,bbury=525pt}
           \end{center}
          \vspace*{-0.7cm}
     \end{minipage}
  \end{center}
  \begin{center}
     \begin{minipage}[t]{0.45\textwidth}
         \begin{center}
            \vspace*{-0.5cm}
                \epsfig{file=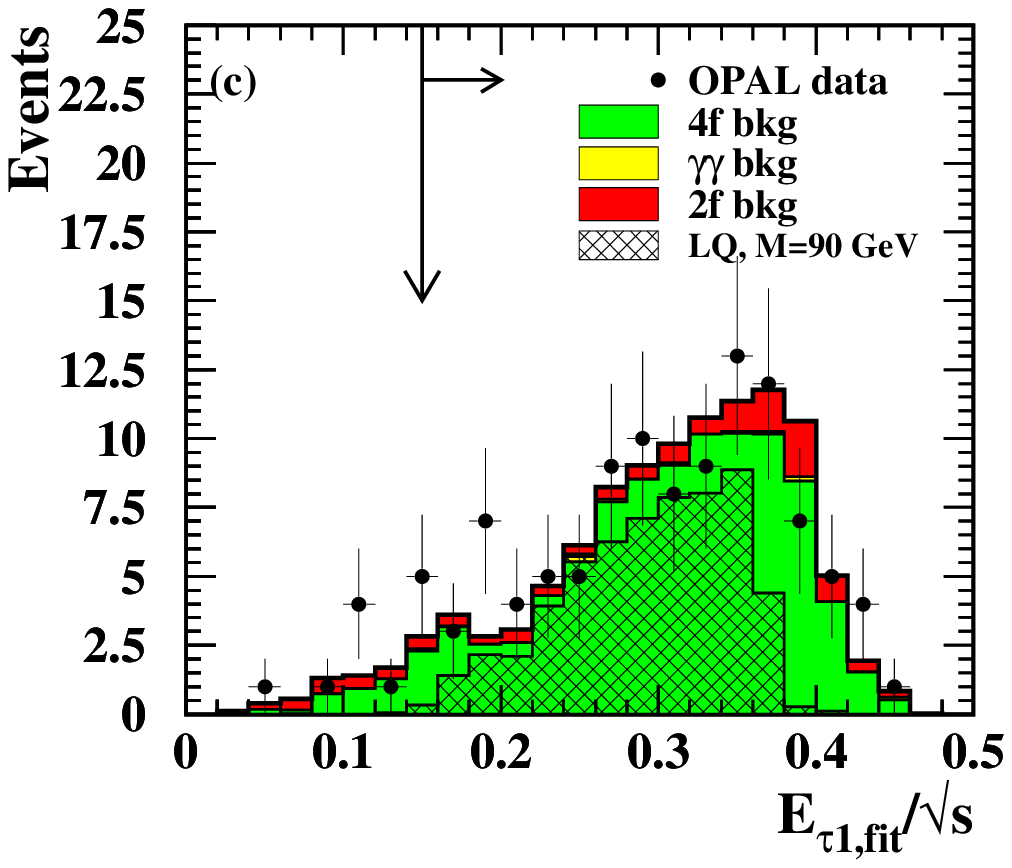,width=8cm,
                           bbllx=5pt,bblly=270pt,bburx=295pt,bbury=525pt}
          \end{center}
           \vspace*{-0.7cm}
     \end{minipage}
\hfill
     \begin{minipage}[t]{0.45\textwidth}
       \begin{center}
          \vspace*{-0.5cm}
              \epsfig{file=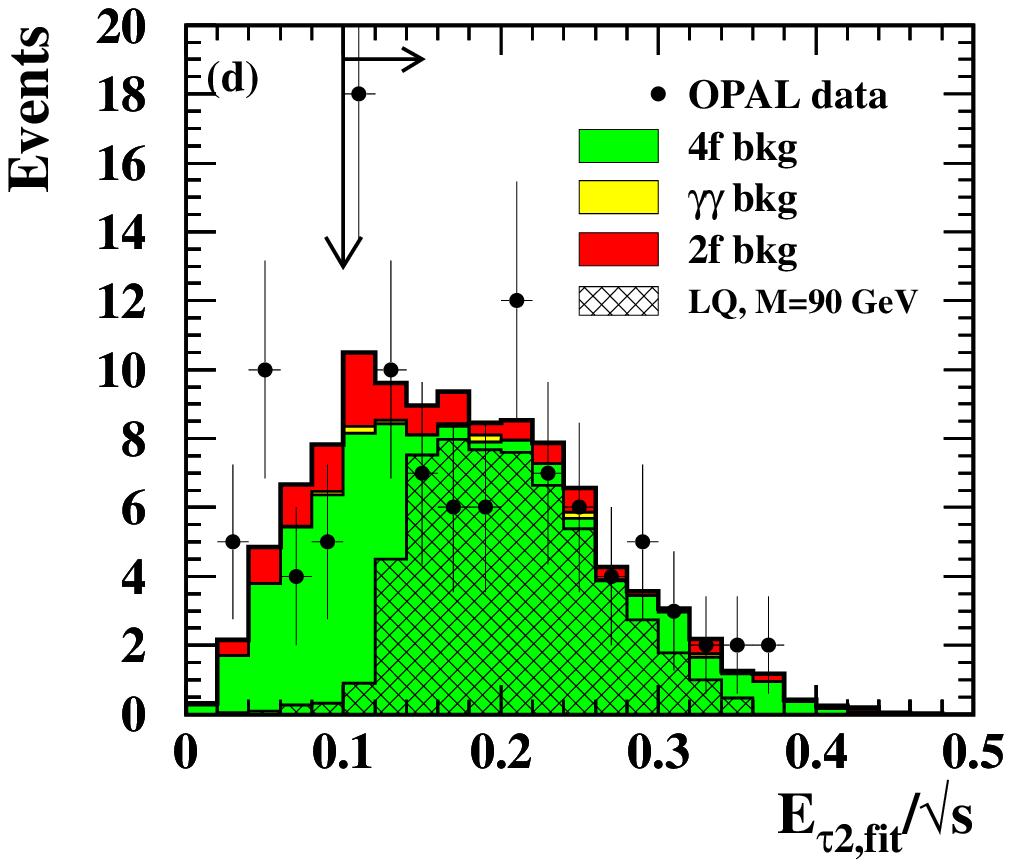,width=8cm,
                           bbllx=5pt,bblly=270pt,bburx=295pt,bbury=525pt}
       \end{center}
       \vspace*{-0.7cm}
      \end{minipage}
   \end{center}
   \begin{center}
     \begin{minipage}[t]{0.45\textwidth}
         \begin{center}
           \vspace*{-0.5cm}
              \epsfig{file=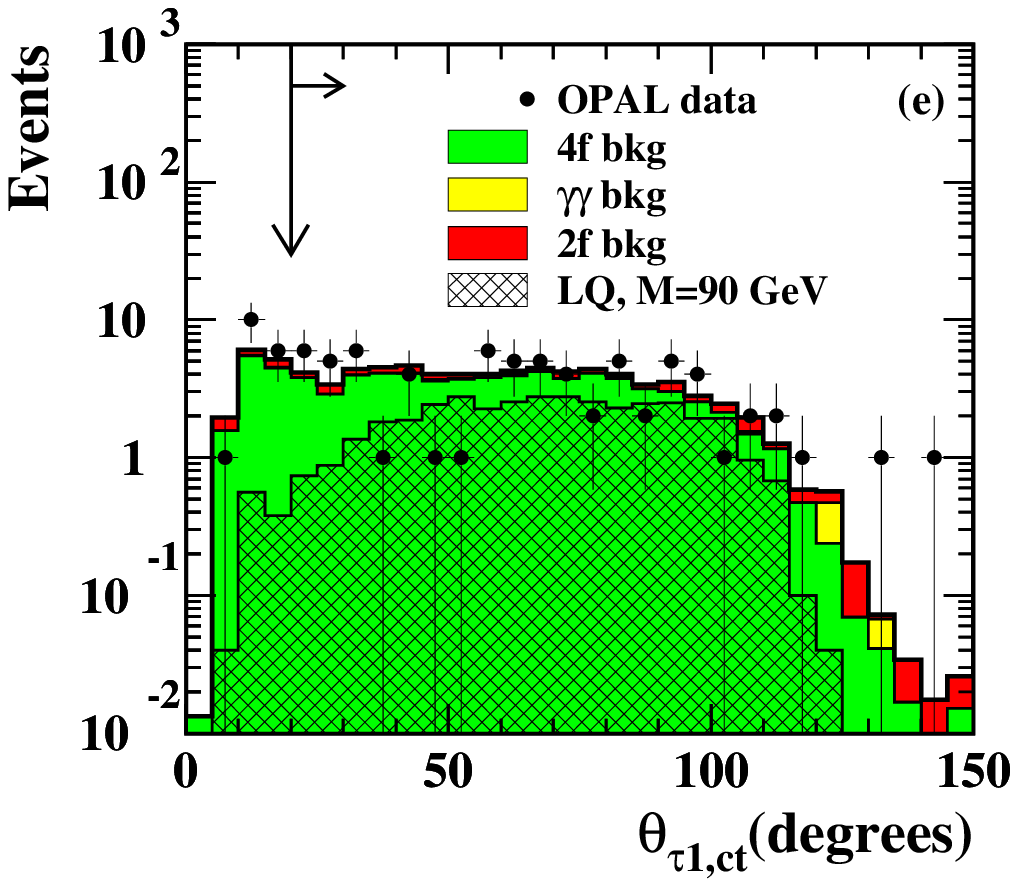,width=8cm,
                           bbllx=5pt,bblly=270pt,bburx=295pt,bbury=525pt}
             \end{center}
             \vspace*{-0.9cm}
      \end{minipage}
\hfill
       \begin{minipage}[t]{0.45\textwidth}
          \begin{center}
            \vspace*{-0.5cm}
                 \epsfig{file=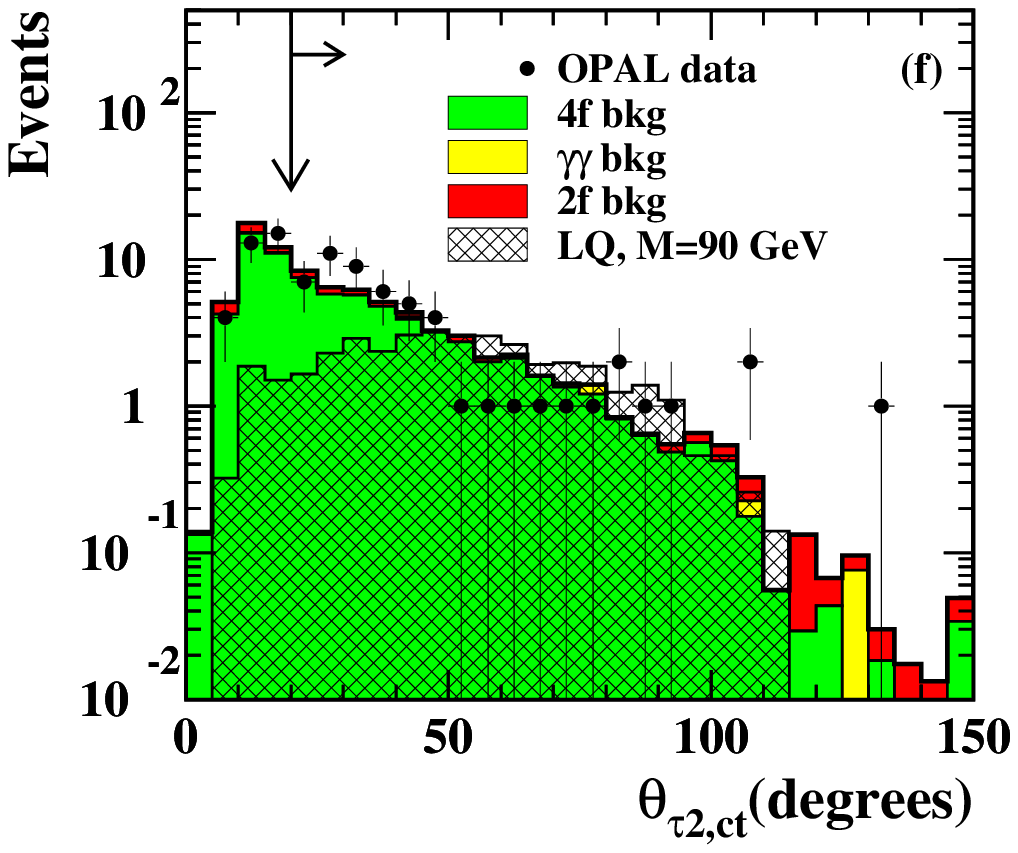,width=8cm,
                           bbllx=5pt,bblly=270pt,bburx=295pt,bbury=525pt}
         \end{center}
          \vspace*{-0.9cm}
      \end{minipage}
    \end{center}
 \caption[]{\sl Class {\bf C2}, the $\tau^{+} \tau^{-} {\rm q q}$ channel: 
                 same as Figure~\ref{fig_vv}, but for class {\bf C2}.
             \newline
                 {\bf (a)} The two-tau probability, 
                          $\cal{P}_{\tau\tau}$, defined in cut {\bf (C2-3)}.
                 {\bf (b)} The logarithm of the probability 
                            of the kinematic fit used to reconstruct the
                            leptoquark mass.
                            The first bin contains also 
                            the events failing 
                            the fit or with a probability lower 
                            than 10$^{-15}$.
               {\bf (c)-(d)} The scaled energies of the most energetic and
                             of the second most energetic tau in the event 
                             respectively, after cut {\bf (C2-4)}. 
                 {\bf (e)-(f)} The angles between the direction of 
                                the momenta of the most energetic and the 
                                second most energetic tau leptons and the 
                                nearest 
                                charged track, respectively,  
                                after cut {\bf (C2-5)}.
           }

 \label{fig_tt}
 \end{figure}

The distributions of some of the variables used in this selection are 
shown in Fig.~\ref{fig_tt}.

\vspace*{0.7cm}
\begin{table}[htbp]
\begin{center}
{\footnotesize
\begin{tabular}{|c||c||c||c|c|c||c|}
\hline
\hline
 \multicolumn{7}{|c|}{\rule[-3mm]{0pt}{9mm}  
          {\normalsize {\boldmath  $ \tau^{+}\tau^{-} {\rm q q} $}}}  \\
\hline
\rule[-3mm]{0pt}{9mm} {\bf Cut} & {\bf Data}    &  {\bf Background}        
                      & 4f  &
                        $\gamma \gamma$  & 2f & 
                        {\boldmath $\varepsilon~(\%)$ }\\
\hline
\rule[-3mm]{0pt}{9mm} (C2-1)&   35243  &  34573.0   & 7821.0 & 187.0 & 26565.0
                            & 90.3 \\
\rule[-3mm]{0pt}{6mm} (C2-2)&   20067  &  19393.0  & 5690.0 & 101.3
                                              & 13602.0
                               & 81.7 \\
\rule[-3mm]{0pt}{6mm} (C2-3)&   ~1503  &  ~~1506.0  & 1290.0 & ~24.7
                                              &  ~~191.4
                               & 61.5 \\
\rule[-3mm]{0pt}{6mm} (C2-4)&   ~~114   & ~~~108.1  & ~~~94.1 & ~~1.0 &~~~13.0
                                       &39.9 \\
\rule[-3mm]{0pt}{6mm} (C2-5)&    ~~~87  & ~~~~84.7  & ~~~75.5 & ~~0.9 & ~~~~8.3
                                        &39.2 \\
\rule[-3mm]{0pt}{6mm} (C2-6)&    ~~~41  & ~~~~41.1  & ~~~37.6 & ~~0.4&  ~~~~3.1
                                        & 34.7 \\
\rule[-3mm]{0pt}{6mm} (C2-7)&    ~~~37  & ~~~~38.0  & ~~~34.7 & ~~0.4&  ~~~~2.9
                                        & 33.3  \\
\hline \hline
\end{tabular}
 }
\caption[]{\sl Same as Table~\ref{tab:nevA}, but for 
                selection {\bf C2} ($\tau^{+}\tau^{-} {\rm q q}$ channel). 
          }
\label{tab:nevC2}
\end{center}
\end{table}
%
%
\section{Results}
%

No evidence for leptoquark pair production is observed 
in the data.
Table~\ref{tab:data_MC_compare} shows the numbers of events selected in
the  data together with the expectations from Monte Carlo simulations
of the background processes, including the errors, after all cuts  
for the different signal topologies. A separate comparison
is made for the data with $\sqrt{s} > 195$~GeV 
because the search for vector leptoquarks
includes only data collected at these energies. \\
\begin{table}[t]
{\footnotesize
 \begin{center}
\begin{tabular}{|cc||ccc||ccc|}
\hline \hline
 & $\sqrt{s}$  & \multicolumn{3}{c||}{\rule[-3mm]{0pt}{9mm}  189-209}  &  
              \multicolumn{3}{c|}{\rule[-3mm]{0pt}{9mm}  195-209}      \\
\rule[-3mm]{0pt}{9mm} Channel     &     &  Data &  &  Bkg  & Data &  & Bkg \\
\hline
{\rule[-3mm]{0pt}{9mm} $\nu\nu  {\rm qq}$}          &   & 28 & 
                                 & $22.8^{+2.7+3.5}_{-1.3-3.5} $   
                                 & 20 & & $15.2^{+1.8+2.3}_{-0.9-2.3}$  \\
{\rule[-3mm]{0pt}{7mm} ${\rm e}^{\pm} \nu {\rm  qq}$} & & 13 & 
                                & $13.7^{+2.4+5.8}_{-1.0-5.8}$     
                                & 10 & & $10.7^{+1.7+4.5}_{-0.7-4.5}$   \\
{\rule[-3mm]{0pt}{7mm} $\mu^{\pm} \nu  {\rm qq}$}     & & 26 & 
                                 &$24.5^{+2.5+4.8}_{-1.3-4.8}$     
                                 & 18 & & $19.6^{+1.8+3.9}_{-1.0-3.9}$  \\
{\rule[-3mm]{0pt}{7mm} $\tau^{\pm} \nu {\rm  qq}$}   &  & 35 & 
                                 & $36.0^{+2.7+8.1}_{-1.6-8.1}$     
                                 & 21 & & $27.7^{+1.9+6.3}_{-1.2-6.3}$   \\
{\rule[-3mm]{0pt}{7mm} ${\rm e^{+} e^{-}}   {\rm qq}$}& & 20 & 
                                & $12.8^{+2.5+4.6}_{-1.3-4.6}$    
                                & 15 & & $ 9.3^{+1.7+3.4}_{-0.9-3.4}$    \\
{\rule[-3mm]{0pt}{7mm} $\mu^{+} \mu^{-}  {\rm qq}$}&    &  4 & 
                                & $ 8.7^{+2.2+2.7}_{-0.7-2.7}$     
                                &  3 & &  $ 7.0^{+1.6+2.2}_{-0.5-2.2}$     \\
{\rule[-3mm]{0pt}{7mm} $\tau^{+} \tau^{-}  {\rm qq}$}&  & 37 & 
                                &$38.0^{+3.1+6.7}_{-2.0-6.7}$     
                                & 24 & & $24.6^{+2.1+4.4}_{-1.3-4.4}$     \\

\hline \hline
\end{tabular}
 \end{center}
}
 \caption{\sl The numbers of events observed in the  data compared to
              the numbers of expected background events from Monte Carlo, 
              in the different search channels
              considered.  
              The errors are statistical and systematic, respectively.} 
\label{tab:data_MC_compare}
\end{table}
\newline

A total error on the number
of expected background events of 19--46$\%$ is estimated assuming the 
following sources:
\begin{itemize} 
   \item The statistical uncertainty due to the limited number of simulated 
         Monte Carlo events (8--26$\%$).
   \item The uncertainty introduced by the Monte Carlo modelling of the 
         variables used in the selections (12--29$\%$).
         This is evaluated by displacing the cut value on a given 
         variable, $x$,
         from the original position $x_{0}$ to a new position  $x_{0}'$, 
         to reproduce on the
         simulated events the effect of the cut on the real data.  
         $x_{0}'$ is defined by
      \begin{center} {\large
         $ x_{0}' = \left( x_{0} - \langle x \rangle_{{\rm data}} \right) 
                             \frac{\sigma_{{\rm bkg}}}{\sigma_{{\rm data}}} +
                              \langle x \rangle_{{\rm bkg}} $ }
      \end{center}
      where $ \langle x \rangle_{{\rm data}}$,  
     $ \langle x \rangle_{{\rm bkg}}$,
      $\sigma_{{\rm data}}$ and $\sigma_{{\rm bkg}}$ are the mean values and
      the standard deviations of the distributions of the variable 
      $x$ for the data
      and the simulated background. These quantities are calculated by the
      distributions of $x$ given by the events surviving the cuts
      on all the other variables used in the selection.  It was checked
      that using the distributions of $x$ at other stages of the selection
      leads to negligible changes in the values of this uncertainty.       
       This procedure is repeated separately for each variable used in 
      the event
      selections and the change in the number of the expected background 
      events due to
      the displacement of the cut is taken as the systematic error
      from this source.  The different contributions are added in
      quadrature.
      The main contributions are due 
      to the fit probability and the reconstructed ${\rm W}$ boson mass in the
      selection of events of class {\bf B} first generation (21\%), 
      and to the scaled 
      muons' energies in the search for events of class {\bf C} 
      second generation
      (15\%).

      \item The error associated with the lepton identification method
      is evaluated by considering the difference between the number of
      expected events from Monte Carlo background and the number of
      events observed in the data when only the preselection  cuts 
      and the request
      for presence (or absence, for class {\bf A}) of leptons are 
      made in the different selections. 
      Depending on  the class of events,     
      this error is found to range from  3$\%$ ($\tau^{\pm} \nu {\rm q q}$ 
      channel) to 14$\%$ ($\mu^{\pm} \nu {\rm q q}$ channel). 
     \item Alternative Monte Carlo generators and fragmentation models are used
           to check the number of expected background events.
           The differences between the numbers obtained using these samples 
           and  the
           main Monte Carlo samples are taken as  systematic errors and 
           are found to
           contribute a 5--30$\%$, depending on the different selections. 
\end{itemize}

The error on the integrated luminosity of the data is less than 
0.5$\%$ at each energy and is neglected.  \\ 

\begin{figure}[p]
 \vspace*{-1.5cm}
  \begin{center}
  \end{center}
\vspace*{0.3cm}                  
  \begin{center}
        \begin{minipage}[t]{0.45\textwidth}
           \vspace*{-1cm}
            \begin{center}
                 {\large \hspace*{1cm} 
                   {\boldmath ${\rm e}^{\pm} \nu {\rm qq}$ }
                   \vspace*{-0.2cm} 
                 }
               \epsfig{file=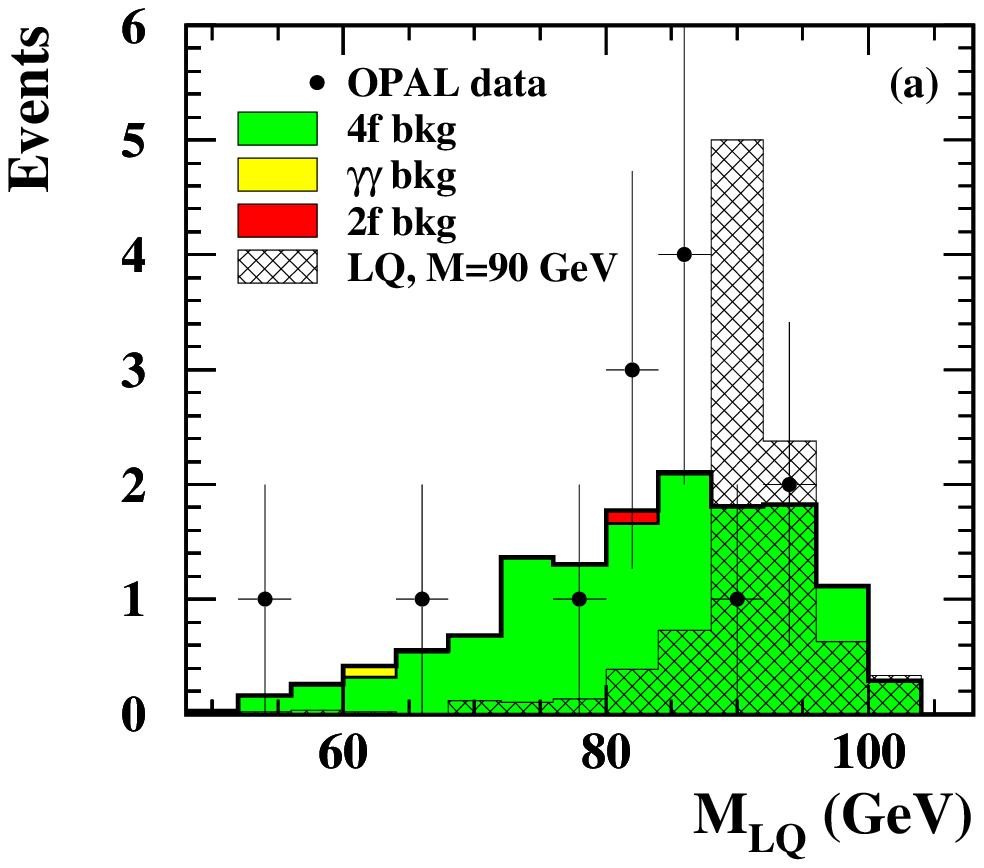,width=8cm,
                           bbllx=5pt,bblly=270pt,bburx=295pt,bbury=525pt}
             \end{center}
         \end{minipage}
\hfill
       \begin{minipage}[t]{0.45\textwidth}
          \vspace*{-1cm}
          \begin{center}
               {\large  \hspace*{1cm}   {\boldmath $\mu^{\pm} \nu {\rm qq}$ }
                 \vspace*{-0.2cm}}
               \epsfig{file=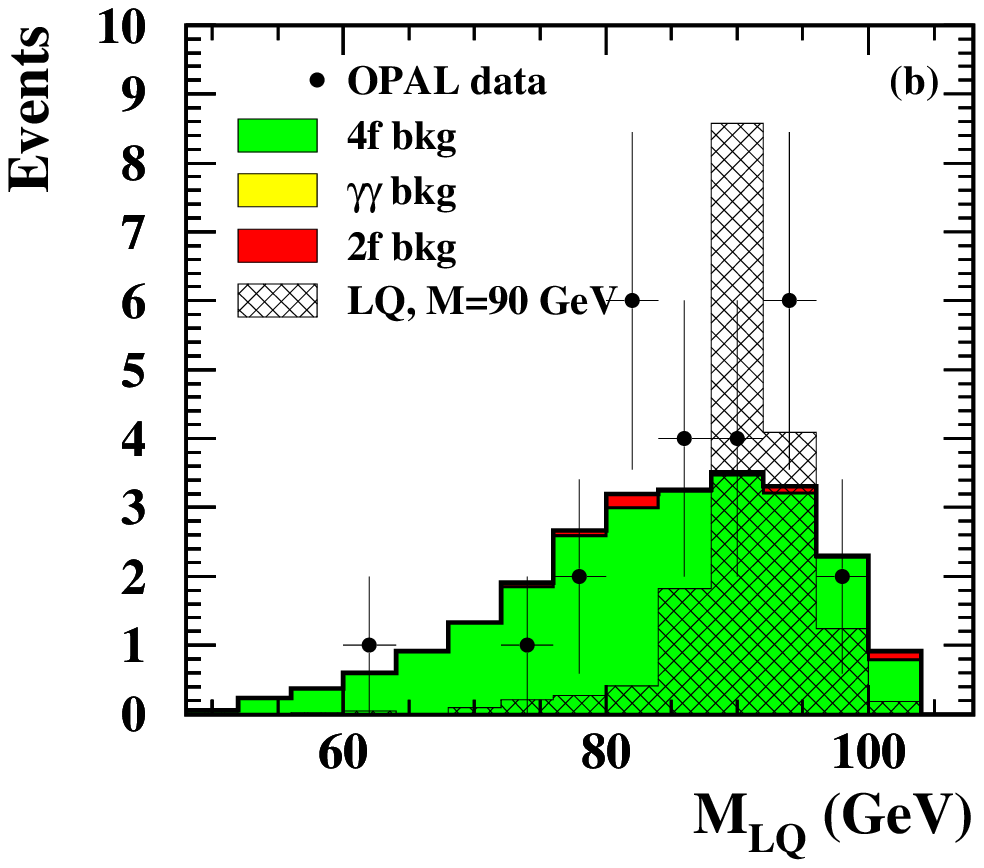,width=8cm,
                           bbllx=5pt,bblly=270pt,bburx=295pt,bbury=525pt}
           \end{center}
        \end{minipage}
    \end{center}
 \vspace*{0.1cm}
 \begin{center}
        \begin{minipage}[t]{0.45\textwidth}
             \vspace*{-0.5cm}
              \begin{center}
                 {\large \hspace*{1cm}    
                     {\boldmath $\tau^{\pm} \nu {\rm qq}$ }
                   \vspace*{-0.2cm}}
                \epsfig{file=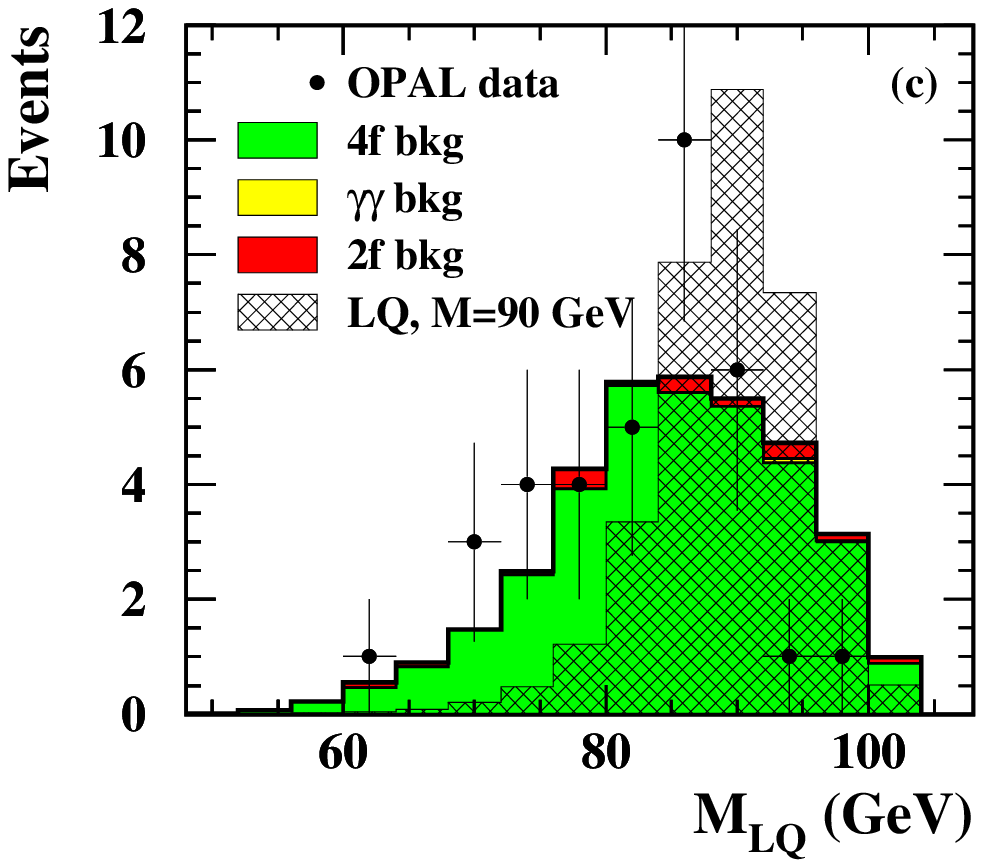,width=8cm,
                           bbllx=5pt,bblly=270pt,bburx=295pt,bbury=525pt}
             \end{center}
         \end{minipage}
\hfill
       \begin{minipage}[t]{0.45\textwidth}
           \vspace*{-0.5cm}
             \begin{center}
                  {\large  \hspace*{1cm}       
                       {\boldmath ${\rm e}^{+} {\rm e}^{-}  {\rm qq}$ }
                     \vspace*{-0.2cm}}
                 \epsfig{file=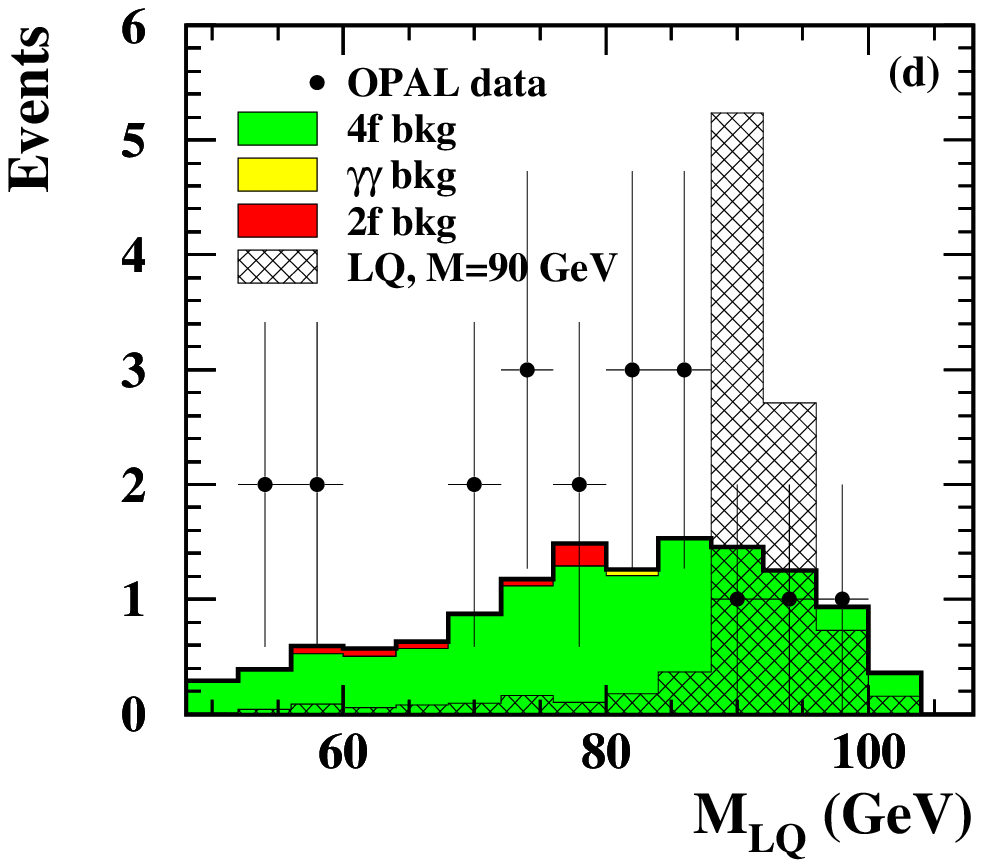,width=8cm,
                           bbllx=5pt,bblly=270pt,bburx=295pt,bbury=525pt}
         \end{center}
       \end{minipage}
 \end{center}
 \vspace*{0.1cm}
 \begin{center}
        \begin{minipage}[t]{0.45\textwidth}
             \vspace*{-0.5cm}
            \begin{center}
                 {\large  \hspace*{1cm}     
                     {\boldmath $ \mu^{+} \mu^{-}  {\rm qq}$ }
                  \vspace*{-0.2cm}}
                  \epsfig{file=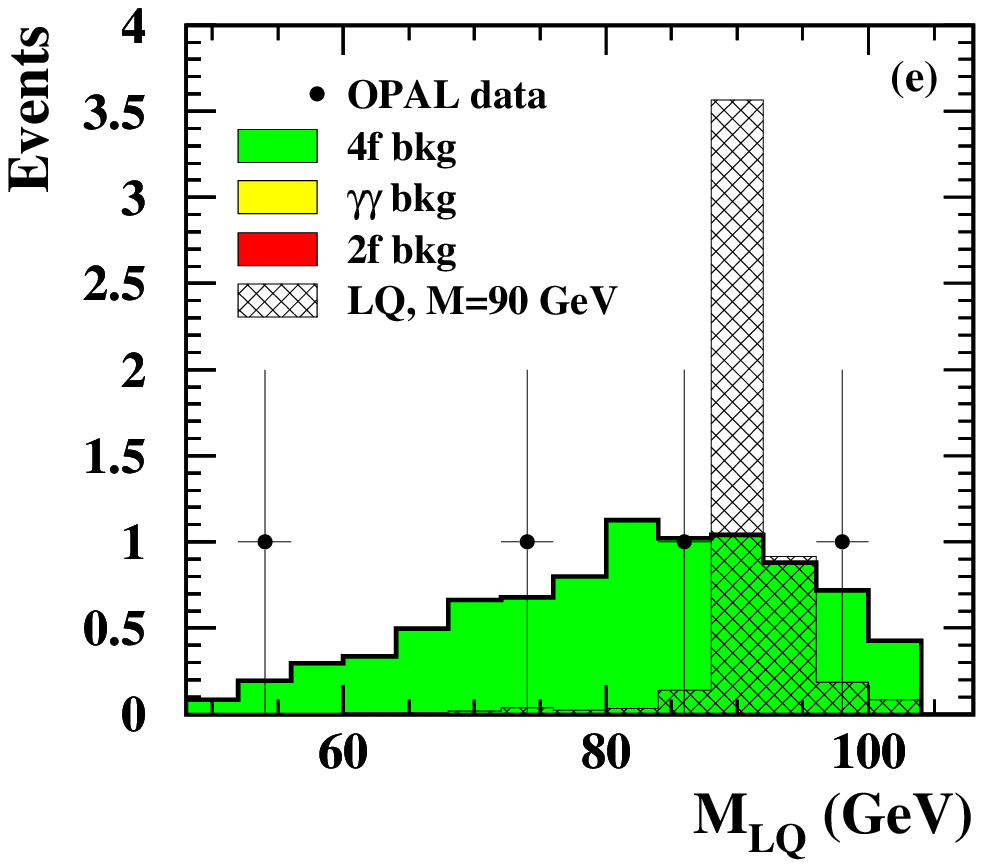,width=8cm,
                           bbllx=5pt,bblly=270pt,bburx=295pt,bbury=525pt}
             \end{center}
       \end{minipage}
\hfill
       \begin{minipage}[t]{0.45\textwidth}
           \vspace*{-0.5cm}
          \begin{center}
                 {\large  \hspace*{1cm}    
                     {\boldmath $ \tau^{+} \tau^{-}  {\rm qq}$ }
                   \vspace*{-0.2cm}} 
                 \epsfig{file=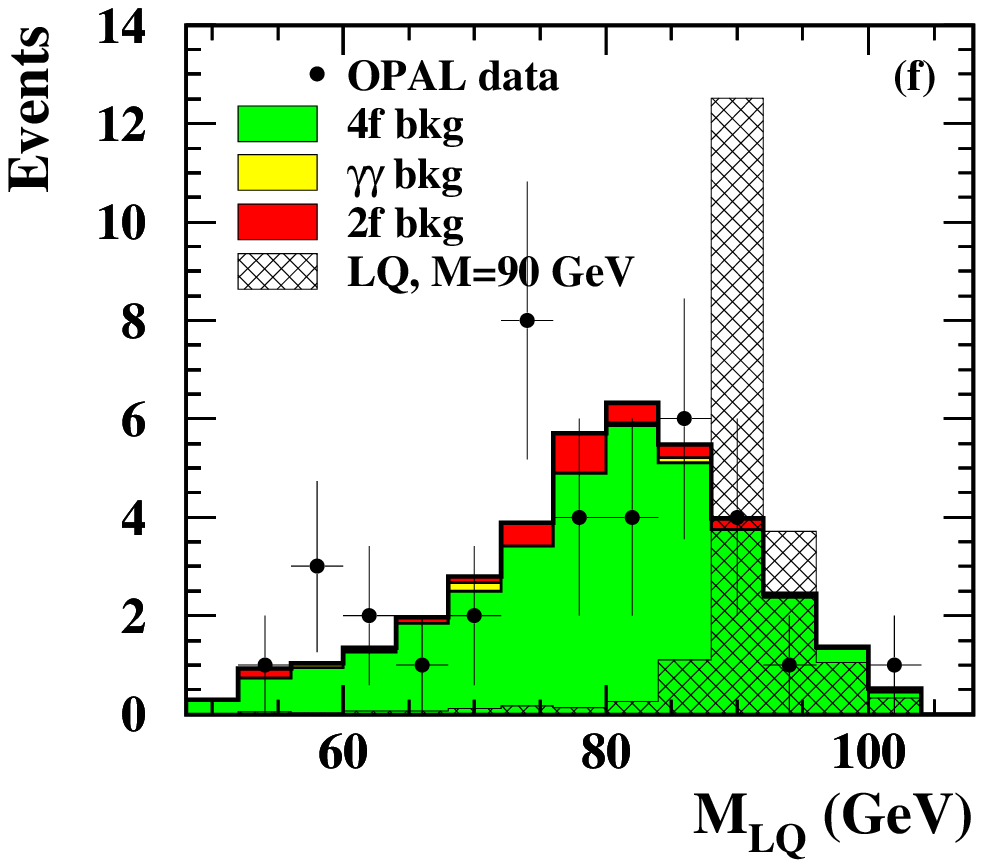,width=8cm,
                           bbllx=5pt,bblly=270pt,bburx=295pt,bbury=525pt}
         \end{center}
      \end{minipage}
 \end{center}
\caption{\sl The leptoquark masses reconstructed by the kinematic fits used in
             the selections for events of classes {\bf B} and {\bf C}. 
             The notation is the same
             as in Figure~\ref{fig_vv}.
        }
 \label{fig:mlq}
\end{figure}

In Figure~\ref{fig:mlq} the leptoquark mass reconstructed by the kinematic fits
is shown for all the events surviving the selections for classes {\bf B} 
and {\bf C}, for both the background and a simulated signal.
For a leptoquark mass ${\rm M_{LQ}} = 90$ GeV at the centre-of-mass energy of
206 GeV the mass resolution, obtained by a Gaussian fit to the peak
region, ranges from 1.3 GeV ($\mu^{+}\mu^{-} {\rm qq}$ channel) 
to 5.0 GeV ($\tau^{\pm}\nu {\rm qq}$ channel),
while the mean value of the reconstructed mass is between 89.8 
($\tau^{\pm}\nu {\rm qq}$) and 91.6 GeV ($e^{+}e^{-} {\rm qq}$). \\

The detection efficiencies for the different topologies of signal events, 
as functions of the leptoquark mass M$_{\mathrm{LQ}}$, are listed in 
Tables~\ref{effic_189_206_scalar} and~\ref{effic_189_206_vector} for 
scalar and vector
leptoquarks respectively, for the centre-of-mass energies 
where the signal was simulated. \\
\begin{table}[htb]
\begin{center}
{\footnotesize
\begin{tabular}{|l|c||r|r|r|r|r|r|r|c|c|c|}
\hline
  \multicolumn{2}{|c||}{ \rule[-3mm]{0pt}{9mm} M$_{\mathrm{LQ}}$ (GeV)} 
                           &\multicolumn{1}{c|}{50} & 
                            \multicolumn{1}{c|}{60} &
                           \multicolumn{1}{c|}{70}  &
                          \multicolumn{1}{c|}{75} & 
                            \multicolumn{1}{c|}{80} &
                           \multicolumn{1}{c|}{85}  &
                          \multicolumn{1}{c|}{90} & 
                            \multicolumn{1}{c|}{95} &
                           \multicolumn{1}{c|}{99}  &
                           \multicolumn{1}{c|}{102}    \\
\hline
\rule[-3mm]{0pt}{9mm}  Signal topology & Generation& \multicolumn{10}{c|}{} \\
\hline
\hline
%
   & &  \multicolumn{10}{c|}{\rule[-3mm]{0pt}{9mm}  $\sqrt{s}$~=~189 GeV}  \\
\hline
   Class A  & 1,2,3 & 11.4 & 17.7 & 23.9  & 26.9   & 30.2  & 34.4   & 37.4   
                    & --  & -- & --  \\
\hline
 Class B1 & 1     & 10.0 & 16.1 & 22.7  & 27.3   & 27.9  & 31.6   & 36.0   
                    & --  & -- & -- \\
 Class B1 & 2     & 13.5 & 22.8 & 29.6  & 33.2   & 35.5  & 39.9   & 41.9   
                    & --  & -- & -- \\
 Class B2 & 3     & 1.9 & 5.3 & 9.9  & 12.3   & 15.7  & 19.0   & 22.1   
                    & --  & -- & --   \\
\hline
 Class C1 & 1     & 30.2 & 37.4 & 44.5  & 48.2   & 48.3  & 51.6    & 53.7  
                   & -- & --  & --   \\
 Class C1 & 2     & 35.1 & 42.9 & 52.0  & 57.1   & 61.1  & 65.5    & 66.2  
                   & -- &  -- & --   \\
 Class C2 & 3     & 20.0 & 26.2 & 31.0  & 30.7   & 34.0  & 33.7    & 35.3  
                   & -- &  --  & --   \\
\hline
\hline
  & &  \multicolumn{10}{c|}{\rule[-3mm]{0pt}{9mm}    $\sqrt{s}$~=~196 GeV}  \\
\hline
 Class A  & 1,2,3 & 10.3 & 16.3 & 22.4   & 26.4  & 28.1 & 30.5 & 33.8  & 37.7 
                   & -- & --  \\
\hline
 Class B1 & 1     & 8.9 & 14.7 & 20.9   & 21.9  & 25.6 & 28.6 & 32.3   & 35.3 
                    & -- & --  \\
 Class B1 & 2     & 11.7 & 21.2 & 28.6   & 31.3  & 34.2 & 36.7 & 41.3   & 43.5
                    & -- & --  \\
 Class B2 & 3     & 1.3 & 4.4 & 8.5    &  11.1  & 13.1 & 17.4 & 19.9   & 23.1
                    & -- & --   \\
\hline
 Class C1 & 1  & 28.9 & 36.1 & 41.8   & 45.1  & 49.7  & 51.2  & 53.3   & 54.3 
                    & --   &  --  \\
 Class C1 & 2   & 33.7 & 41.1 & 49.4   & 52.9  & 61.1  & 62.5  &  65.4  & 67.8
                    & --  &  --  \\
 Class C2 & 3   & 18.7 & 25.7 & 27.8   & 29.1  & 32.2  & 32.8  &  33.1  & 34.4
                    & -- &  --  \\
\hline
\hline
  & &  \multicolumn{10}{c|}{ \rule[-3mm]{0pt}{9mm}  $\sqrt{s}$~=~200 GeV}   \\
\hline
  Class A  & 1,2,3 & 9.7 & 15.1 &  21.0  & 23.7  & 26.6  & 29.5 & 32.1 & 34.9
                   & 38.2 & --  \\
\hline
 Class B1 & 1     & 8.7 & 14.2 & 18.9  & 23.8  & 25.0  & 29.3   & 30.2 & 34.0
                   & 35.4 & --  \\
 Class B1 & 2     & 11.4 & 20.7 & 27.7  & 31.2  & 32.8  & 37.0   & 37.8 & 42.5
                   & 43.1 & --  \\
 Class B2 & 3     & 1.2 & 3.9 & 7.1   &  11.3  & 13.0  & 17.7   & 18.3 & 22.9
                   & 25.4 & --   \\
\hline
 Class C1 & 1     & 27.8 & 35.1 & 39.8  & 44.7  & 46.8   & 48.8  & 51.7 & 54.4
                   & 55.6 & --   \\
 Class C1 & 2     & 32.8 & 40.6 &  48.7  & 53.5  & 57.3   & 62.1 & 65.8 & 67.5
                   & 66.7 & --   \\
 Class C2 & 3     & 18.0 & 25.0 & 26.9  & 29.4  & 31.7   & 32.0  & 33.3 & 33.6
                   & 33.6 & --    \\
\hline
\hline
  & &  \multicolumn{10}{c|}{\rule[-3mm]{0pt}{9mm}   $\sqrt{s}$~=~206 GeV}   \\
\hline
 Class A  & 1,2,3 & 9.1 & 13.5 & 18.9  & 21.1  & 24.8  & 27.4   & 30.2 & 33.4
                  & 34.2 & 38.1  \\
\hline
 Class B1 & 1     & 8.6 & 13.8 & 18.2  & 21.2  & 23.7  & 25.8   & 29.1 & 31.9
                  & 32.9 & 33.1  \\
 Class B1 & 2     & 10.8 & 20.8 & 25.1  & 28.7  & 32.4  & 34.5   & 36.9 & 38.5
                  & 39.6 & 41.7  \\
 Class B2 & 3     &  1.4 & 2.9 & 6.8   &  8.5  & 10.9  & 13.7   & 18.6 & 20.8
                  & 22.2 & 22.9   \\
\hline
 Class C1 & 1     & 26.0 & 33.8 & 39.5  & 42.6  & 46.8   & 48.0  & 50.0& 52.9
                  & 54.1 & 54.4   \\
 Class C1 & 2     & 31.3 & 40.2 & 45.4  & 51.8  & 57.6   & 59.3 & 63.0 & 66.3
                  & 65.7 & 66.6   \\
 Class C2 & 3     & 16.9 & 25.0 & 25.9  & 29.4  & 31.3   & 32.5 & 32.7 & 35.2
                  & 34.0 & 34.7    \\
\hline
\hline

\end{tabular}
}
\caption[ ]{\sl The percentage detection efficiencies for 
                scalar leptoquarks for the various selections 
                as functions of the leptoquark mass, 
                M$_{\mathrm{LQ}}$, and  the centre-of-mass energy.
           }
\label{effic_189_206_scalar}
\end{center}
\end{table}
\begin{table}[htb]
\begin{center}
{\footnotesize
\begin{tabular}{|l|c||r|r|r|r|r|c|c|c|}
\hline
\multicolumn{2}{|c||}{ \rule[-3mm]{0pt}{9mm}M$_{\mathrm{LQ}}$ (GeV)} &
                           \multicolumn{1}{c|}{70}  &
                          \multicolumn{1}{c|}{75} & 
                            \multicolumn{1}{c|}{80} &
                           \multicolumn{1}{c|}{85}  &
                          \multicolumn{1}{c|}{90} & 
                            \multicolumn{1}{c|}{95} &
                           \multicolumn{1}{c|}{99}  &
                           \multicolumn{1}{c|}{102}    \\
\hline
\rule[-3mm]{0pt}{9mm}  Signal topology & Generation&\multicolumn{8}{c|}{}  \\
\hline 
\hline
   & &  \multicolumn{8}{c|}{\rule[-3mm]{0pt}{9mm}  $\sqrt{s}$~=~196 GeV}   \\
\hline
 Class A  & 1,2,3 & 22.3  & 25.5   & 28.9  & 31.9   & 35.9   & 40.8  
                  & -- & --  \\
\hline
 Class B1 & 1     & 17.9  & 21.3   & 27.0  & 29.5   & 31.2   & 33.4  
                  & -- & -- \\
 Class B1 & 2     & 25.3  & 28.3   & 32.5  & 35.5   & 38.1   & 41.4  
                  & -- & -- \\
 Class B2 & 3     &  7.9  & 10.3   & 13.8  & 18.3   & 21.6   & 23.7  
                  & -- & --   \\
\hline
 Class C1 & 1     & 40.5  & 45.5   & 49.2  & 50.1    & 51.4  & 54.5 
                  & --  & --   \\
 Class C1 & 2     & 49.4  & 55.1   & 59.5  & 62.7    & 64.1  & 66.8 
                  &  -- & --   \\
 Class C2 & 3     & 28.5  & 30.8   & 32.4  & 33.8    & 34.9  & 33.9 
                  &  --  & --   \\
\hline
\hline
   & &  \multicolumn{8}{c|}{\rule[-3mm]{0pt}{9mm}   $\sqrt{s}$~=~200 GeV}  \\
\hline
 Class A  & 1,2,3 & 21.4  & 24.4   & 28.0  & 30.8   & 34.0   & 39.1  & 42.7 
                  & --  \\
\hline
 Class B1 & 1     & 17.4  & 19.2   & 25.8  & 29.2   & 30.0   & 33.3  & 33.5 
                  & -- \\
 Class B1 & 2     & 24.5  & 26.9   & 31.4  & 34.8   & 36.7   & 40.6  & 42.3 
                  & -- \\
 Class B2 & 3     &  7.3  & 9.4   & 12.1  & 17.1   & 20.4   & 23.4  & 23.9 
                  & --   \\
\hline
 Class C1 & 1     & 39.1  & 44.1   & 48.8  & 49.9    & 50.5  & 52.9 & 56.3  
                  & --   \\
 Class C1 & 2     & 47.7  & 53.8   & 58.1  & 62.4    & 63.3  & 65.6 &  68.0 
                  & --   \\
 Class C2 & 3     & 27.8  & 30.2   & 32.0  & 33.2    & 35.0  & 34.7 &  33.0  
                  & --   \\
\hline
\hline
  & &  \multicolumn{8}{c|}{\rule[-3mm]{0pt}{9mm}    $\sqrt{s}$~=~206 GeV}  \\
\hline
Class A  & 1,2,3 & 19.8   & 22.0  & 25.3 & 28.5 & 32.4   & 34.4  & 37.6 & 41.1
                \\
\hline
 Class B1 & 1     & 16.1   & 17.9  & 22.3 & 24.7 & 26.9  & 30.9  & 32.4 & 31.9
                \\
 Class B1 & 2     & 21.5   & 26.5  & 29.6 & 31.3 & 34.7  & 38.9  & 40.4 & 40.7
                 \\
 Class B2 & 3     &  6.3    & 8.9  & 11.7 & 14.2 & 17.4  & 21.1  & 22.4 & 24.4
                 \\
\hline
 Class C1 & 1     & 40.5   & 43.4  & 47.1  & 48.6  & 50.7   & 51.3 & 54.1   
                  & 54.6  \\
 Class C1 & 2     & 46.2   & 51.0  & 55.6  & 60.0  &  62.6  & 66.6 & 66.4   
                  &  67.2  \\
 Class C2 & 3     & 27.0   & 28.1  & 31.0  & 34.0  &  33.9  & 34.0 & 35.1   
                  &  34.2  \\
\hline
\hline
%
\end{tabular}
}
\caption[ ]{\sl Same as Table~\ref{effic_189_206_scalar}, but for
                vector leptoquarks.} 
\label{effic_189_206_vector}
\end{center}
\end{table}

The systematic uncertainty on the signal efficiency  
is evaluated to be 8--31$\%$ depending
on the signal topology and the leptoquark mass. 
This is estimated by taking into account the following sources 
(the quoted errors are relative):
\begin{itemize}
\item The statistical uncertainty due to the limited number of 
      simulated signal 
      events lies in the range of 1--28$\%$.
\item In the region between two simulated leptoquark masses, the
      value of the efficiency is calculated by a linear
      interpolation. The error associated with this procedure
      is estimated to be 2--8$\%$. 
\item The uncertainty introduced by the Monte Carlo modelling 
      of the variables
      used in the selections 
        contributes a systematic error between 3 and 29\%.
       The largest relative effects are due 
      to the scaled energies of the muons, $ E_{\mu 1} / \sqrt{s}$ and  
      $ E_{\mu 2} / \sqrt{s} $,  
      for events of class {\bf C}, second generation, (up to 28\%),
      to the scaled lepton energies, 
      $E_{\tau,{\rm fit}} / \sqrt{s}$ and $E_{\nu,{\rm fit}} /
      \sqrt{s}$, for events of class {\bf B}, third generation, (up to
      25\%) and to the fit probability and the reconstructed leptoquark mass
      (up to 19\%), for events of the same class. 
      Most of the other selection variables contribute uncertainties of
      less than 10\%.
\item  The error associated with the lepton identification method (3--14$\%$). 
\item The uncertainty due to the flavour of the final state quarks
      contributes an error of  2--8\%. 
      This is evaluated by comparing the efficiencies
      corresponding to all the final states 
      with different quark flavours simulated for a given decay channel,    
      characterized by the lepton flavour 
      (for example
      ${\rm e^{+}e^{-} u\overline{u}}$, ${\rm e^{+}e^{-} d\overline{d}}$,
      ${\rm e^{+}e^{-} c\overline{c}}$, ${\rm e^{+}e^{-} s\overline{s}}$ and
      ${\rm e^{+}e^{-} b\overline{b}}$ for class {\bf C}, first generation).
      The value of the efficiency 
      is taken to be the mean value, and the largest
      difference between the mean and the single contributions is taken as a
      systematic error. 
\item In the range of values of the $\lambda$ couplings covered 
      by this analysis
      the produced leptoquarks may hadronize before decaying.
      This process is not simulated by the standard signal Monte Carlo.
      The systematic error on the detection efficiencies
      associated with the fragmentation model 
      is estimated to be 2--4$\%$, evaluated by using MC samples 
      with pair produced
      scalar quarks ({\em squarks}) with R-parity violating decays.
      These events have features similar to events of class {\bf C} but 
      in these samples the hadronization step is simulated 
      before the squark decay.  
      The efficiencies obtained by applying the selection for class {\bf C} to 
      these events are compared to those obtained using the 
      corresponding standard leptoquark samples
      and the differences are taken as the systematic errors.
      Moreover, for the classes of events where the leptoquark
      mass is reconstructed, the mass distributions obtained by using
      the different samples are also compared and the mean of the 
      absolute value of
      the difference between
      the contents of corresponding bins in the two distributions is taken as
      a systematic error. This contribution is estimated to be 3--8\%.

\item The data sample is divided into 10 energy bins, as shown in 
      Table~\ref{tab:lumin}. However the signal is not simulated 
      at each energy. 
      At centre-of-mass energies, $\sqrt{s}$, where no simulation exists, 
      the efficiency 
      for a given leptoquark mass is inferred from the sample at the nearest
      simulated energy, $\sqrt{s'}$.  The efficiency is assumed to be
      the same as the efficiency for the mass point at $\sqrt{s'}$
       with the same Lorentz boost, that is
       $\varepsilon(\sqrt{s},~\mathrm{M_{{\rm LQ}}})$=
      $\varepsilon(\sqrt{s'},~\sqrt{s'/s}~\mathrm{M_{{\rm LQ}}})$.
      The error associated with this assumption is calculated
      by comparing the efficiencies obtained for corresponding masses  
      at the energies at which the signal is simulated.
      The difference is taken as the error and it is found to be 2--7\%.
\end{itemize}
The polarization of tau leptons from leptoquark decay is not considered
in the simulation of tau decay in the signal events belonging to
 $\tau^{+}\tau^{-} {\rm qq}$ and  $\tau^{\pm} \nu {\rm q q}$ channels. 
However it has been checked
that the effect on the detection efficiencies for vector leptoquarks 
is negligible.  
All the above  errors are considered to be independent 
and added in quadrature. \\

For the purpose of setting limits, the events  
are divided into different search channels
by considering their centre-of-mass
energy, the decay channel and, for events of classes {\bf B} and {\bf C}, 
the reconstructed leptoquark mass divided into 1 GeV bins.    
The confidence level for the existence of a signal is calculated following the
method described in~\cite{likelrat}.
A test statistic is defined which expresses how signal-like the data are. 
The confidence  levels are computed from the value of the test statistic  of 
the observed data and its expected
distributions in a large number of simulated experiments under two
hypotheses: the background-only ($b$) hypothesis and the signal+background 
($s + b$) hypothesis. \\

The test statistic chosen is the likelihood ratio, $Q$,  
the ratio of the probability of observing
the data given the $s + b$ hypothesis to the probability of observing the data
given the $b$ hypothesis. 
As all the search channels are considered to be statistically independent
and to obey Poisson statistics, the likelihood ratio can be computed as
\begin{center} 
$\displaystyle  
        Q = {\rm e}^{ -s_{{\rm tot}}} \prod_{i}(1+ s_{i}/b_{i})^{n_{i}}$ 
\end{center} 
where
$n_{i}$, $b_{i}$ and $s_{i}$ are the number of observed candidates, 
the expected background
and the expected signal in channel $i$ respectively and 
$ s_{{\rm tot}}=\sum_{i}s_{i}$.\\
 
The confidence level for the $b$ hypothesis is $1-{\rm CL}_{b}$, 
representing the fraction of  background-only experiments which would produce
a value of $Q$ more signal-like than the observed data:
\begin{center}
$ 1-{\rm CL}_{b} = P(Q > Q_{obs} | b)$.
\end{center}
If the data agreed perfectly with the expectation from the background-only
hypothesis, a value of  $1-{\rm CL}_{b}=0.5$ would be obtained. A lower 
value indicates an excess of events in the data; 
a higher value indicates a deficit.
Similarly, the agreement of the data with the $s + b$ hypothesis is
tested by the confidence level CL$_{s+b}$, defined as
\begin{center} 
CL$_{s+b} = P(Q \leq Q_{obs} | s+b)$ 
\end{center}
which can be used to exclude the $s + b$ hypothesis when it has a small value.
However, in the case of a large downward fluctuation of the
observed background, this procedure may exclude a signal for which there is
no sensitivity. \\

To reduce this possibility the ratio
\begin{center}

   ${\rm CL}_{s} =  {\rm CL}_{s+b}/ {\rm CL}_{b} $ 

\end{center} 
is used to set limits instead.
A signal is therefore considered {\em excluded} at the  95$\%$ confidence 
level if  ${\rm CL}_{s} < 0.05$. \\

The expected signal $s_{i}$ depends
on the electroweak quantum numbers of each leptoquark and on the unknown 
leptoquark mass.
The assumption is made that for each scenario only one state 
contributes to the cross-section. Therefore, for each state in the model, 
 ${\rm CL}_{s}$ and  $1-{\rm CL}_{b}$ must be 
calculated as a function of M$_{\mathrm{LQ}}$.
In the cases of $S_{0}(-1/3)$, $S_{1/2}(-2/3)$, $V_{0}(-2/3)$ 
and  $V_{1/2}(-1/3)$
the value of the branching ratio into a charged lepton and a quark, 
$\beta$, is not predicted in the model either and
exclusion limits are therefore functions of both  M$_{\mathrm{LQ}}$ 
and $\beta$.\\   

The statistical and systematic uncertainties on the expected number of 
events both for signal and  
background  are incorporated in the calculations of the confidence levels  
as suggested in~\cite{tomjunk}. The probability of observing $n_{i}$ events
in channel $i$, and the corresponding value of the test-statistic $Q$, are 
integrated over possible values of $s_{i}$ and $b_{i}$ given
by their uncertainties, assuming Gaussian distributions, with a lower tail
cut-off at zero, so that negative $s_{i}$ or $b_{i}$ are not allowed. In this
approach the errors on $s_{i}$ and $b_{i}$ within a channel 
and between channels are considered to be uncorrelated.
 \newline
  
\begin{figure}[t]
     \vspace*{-1.2cm}
     \begin{center}
     \end{center}
      \vspace*{-0.8cm}
      \begin{center}
        \hspace*{-1.2cm}
           \epsfig{file=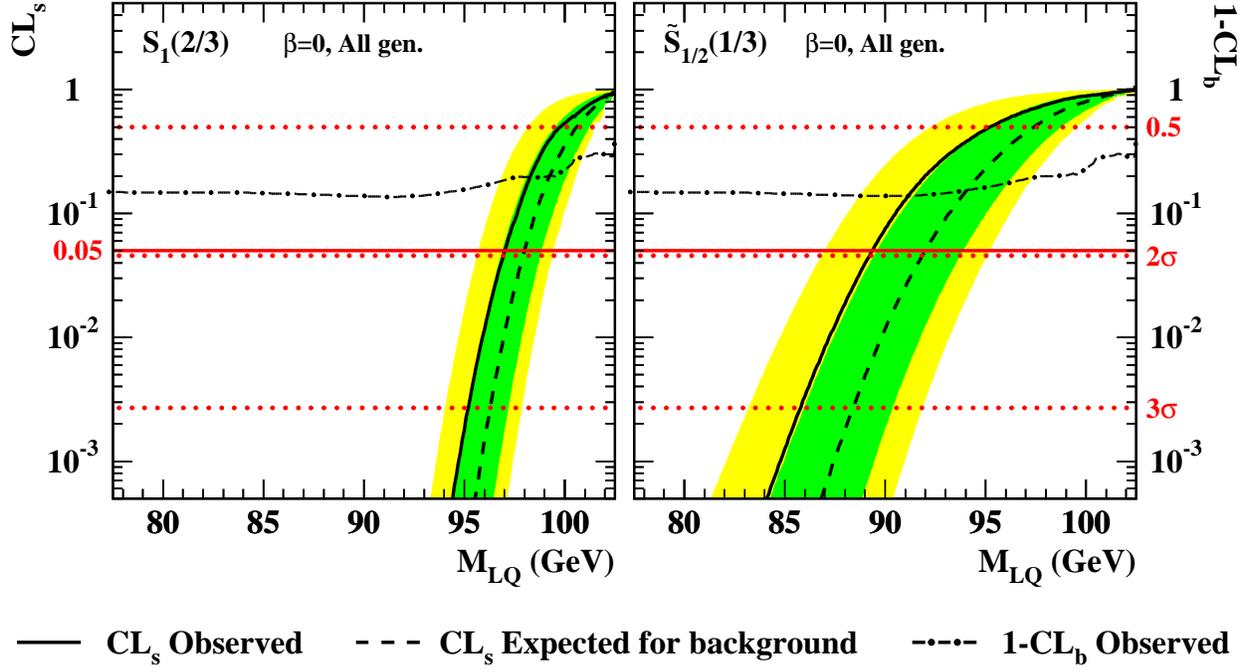,width=13cm,
                       bbllx=40pt,bblly=245pt,bburx=475pt,bbury=520pt}
      \end{center}
\vspace*{0.4cm}
 \caption[]{\sl  The value of CL$_{s}$ 
                 as a function of the mass, for  scalar  
                  leptoquarks $S_{1}(2/3)$ and $\tilde{S}_{1/2}(1/3)$ 
                  with $\beta =0$.
                 The observations for the data are shown with solid
                 lines. The shaded bands indicate the 68\% and 95\%
                 probability intervals with respect to the median
                 expectation in the absence of a signal (dashed lines). 
                 The mass values corresponding to the intersection of the 
                 observed CL$_{s}$  with the horizontal
                 solid line at CL$_{s}=0.05$
                 represent the exclusion limits at 95$\%$ CL.
                 The dash-dotted line shows the observed values for 
                 the Confidence Level
                 $1-{\rm CL}_{b}$; its median expectation in the  
                 background hypothesis (0.5) 
                 and
                 the levels for $2\sigma$ and $3\sigma$ deviations 
                 from this value
                 correspond to the horizontal dotted lines.
           }
 \label{fig_limits_beta0_scalar}
 \end{figure}

\begin{figure}[b]
    \vspace*{0.4cm}
      \begin{center}
     \end{center}
      \vspace*{-0.8cm}
       \begin{center}
            \hspace*{-1.2cm}
            \epsfig{file=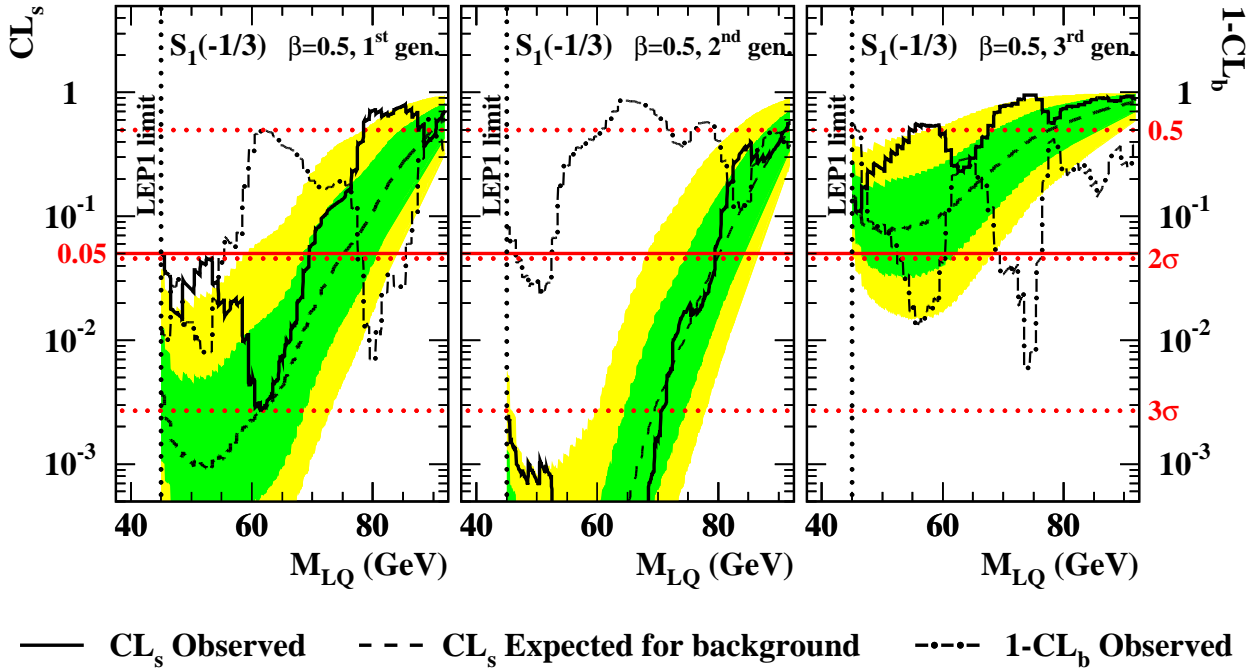,width=13cm,
                       bbllx=40pt,bblly=245pt,bburx=475pt,bbury=520pt}
       \end{center}
\vspace*{0.4cm}
 \caption[]{\sl Same as Figure~\ref{fig_limits_beta0_scalar}, but for
                   $S_{1}(-1/3)$
                   with $\beta =0.5$.
           }  
 \label{fig_limits_s3_1}
\end{figure}

\begin{figure}[t]
  \vspace*{-1cm}
    \begin{center}
      \hspace*{-1.2cm}
            \epsfig{file=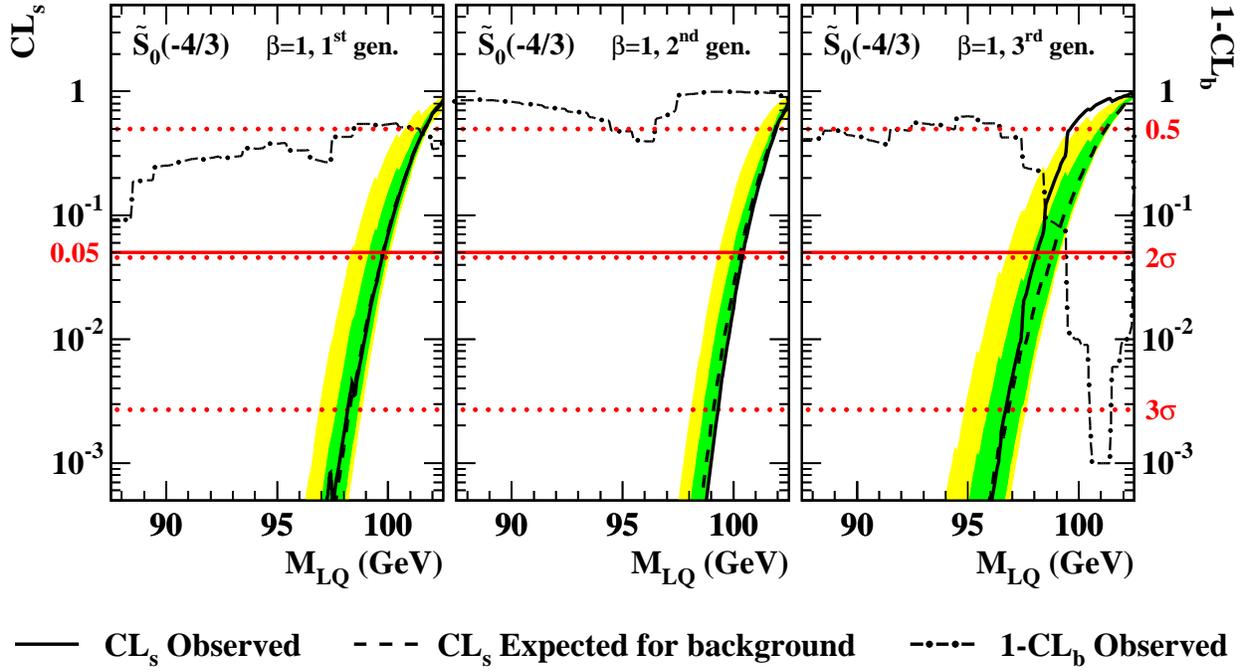,width=13cm,
                       bbllx=40pt,bblly=245pt,bburx=475pt,bbury=520pt}
  \end{center}
\vspace*{0.5cm}
 \caption[]{\sl Same as Figure~\ref{fig_limits_beta0_scalar}, but for   
                $\tilde{S}_{0}(-4/3)$  with $\beta =1$.}
 \label{fig_limits_s1t}
\end{figure}

\begin{figure}[b]
  \vspace*{-1cm}
  \begin{center}
      \hspace*{-1.2cm}
                \epsfig{file=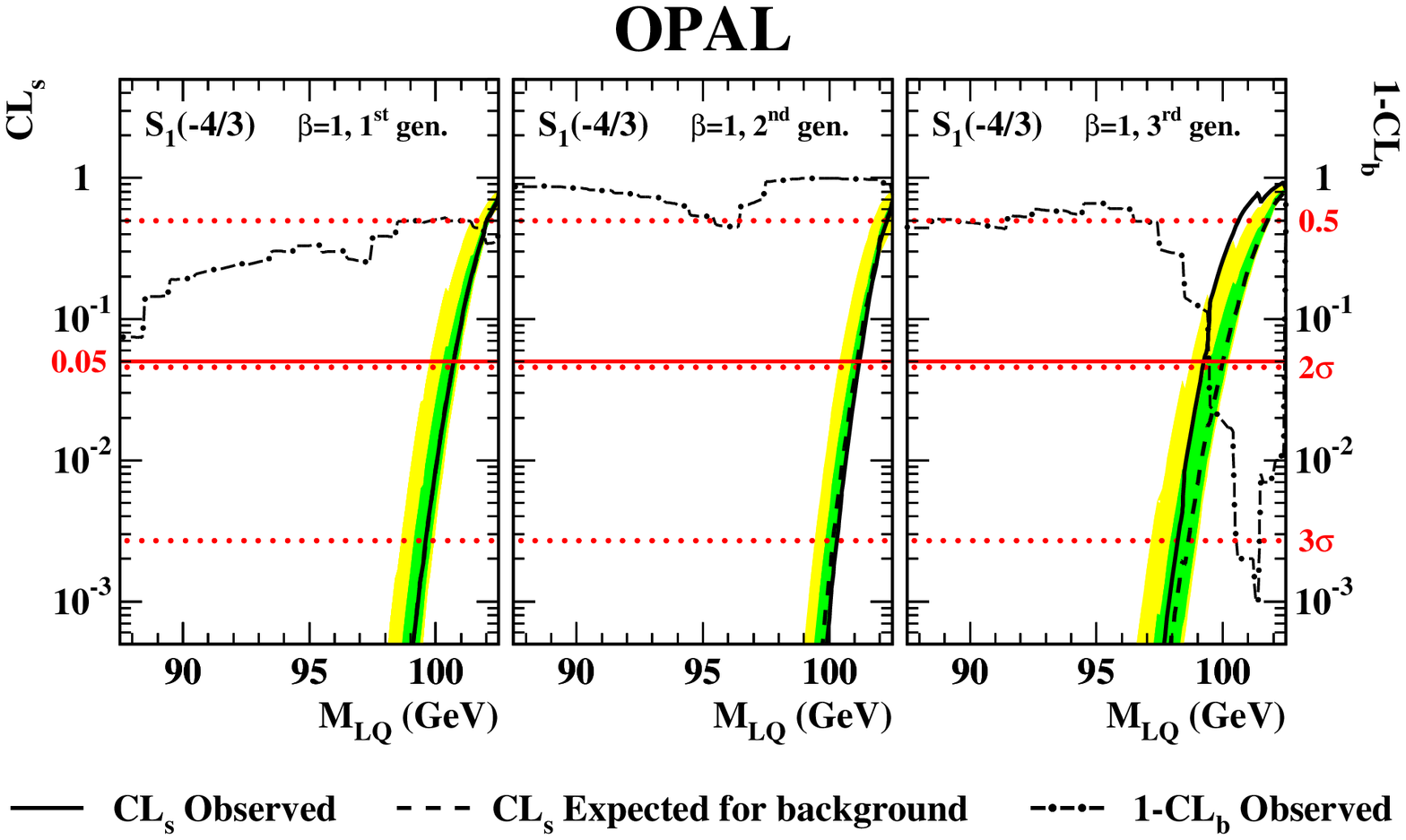,width=13cm,
                       bbllx=40pt,bblly=245pt,bburx=475pt,bbury=520pt}
   \end{center}
 \vspace*{0.5cm}
 \caption[]{\sl Same as Figure~\ref{fig_limits_beta0_scalar}, but for   
                 $S_{1}(-4/3)$ with $\beta =1$.}
 \label{fig_limits_s3_4}
 \end{figure}

\begin{figure}[t]
  \vspace*{-1cm}
  \begin{center}
      \hspace*{-1.2cm}
              \epsfig{file=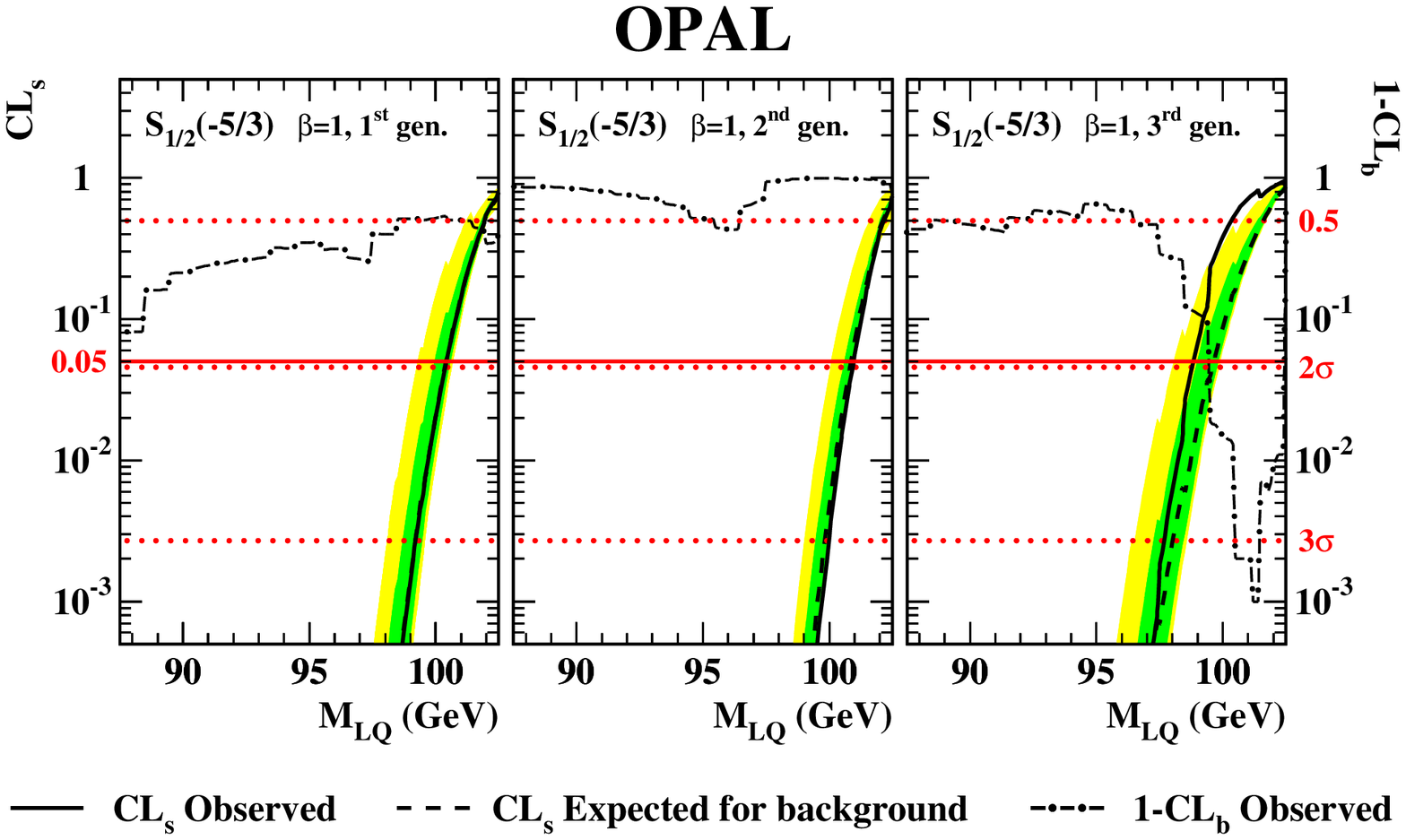,width=13cm,
                       bbllx=40pt,bblly=245pt,bburx=475pt,bbury=520pt}
  \end{center}
 \vspace*{0.5cm}
 \caption[]{\sl Same as Figure~\ref{fig_limits_beta0_scalar}, but for   
                 $S_{1/2}(-5/3)$ with $\beta =1$.
           }
  \label{fig_limits_r2_5}
\end{figure}

\begin{figure}[b]
  \vspace*{-1cm}
   \begin{center}
      \hspace*{-1.2cm}
           \epsfig{file=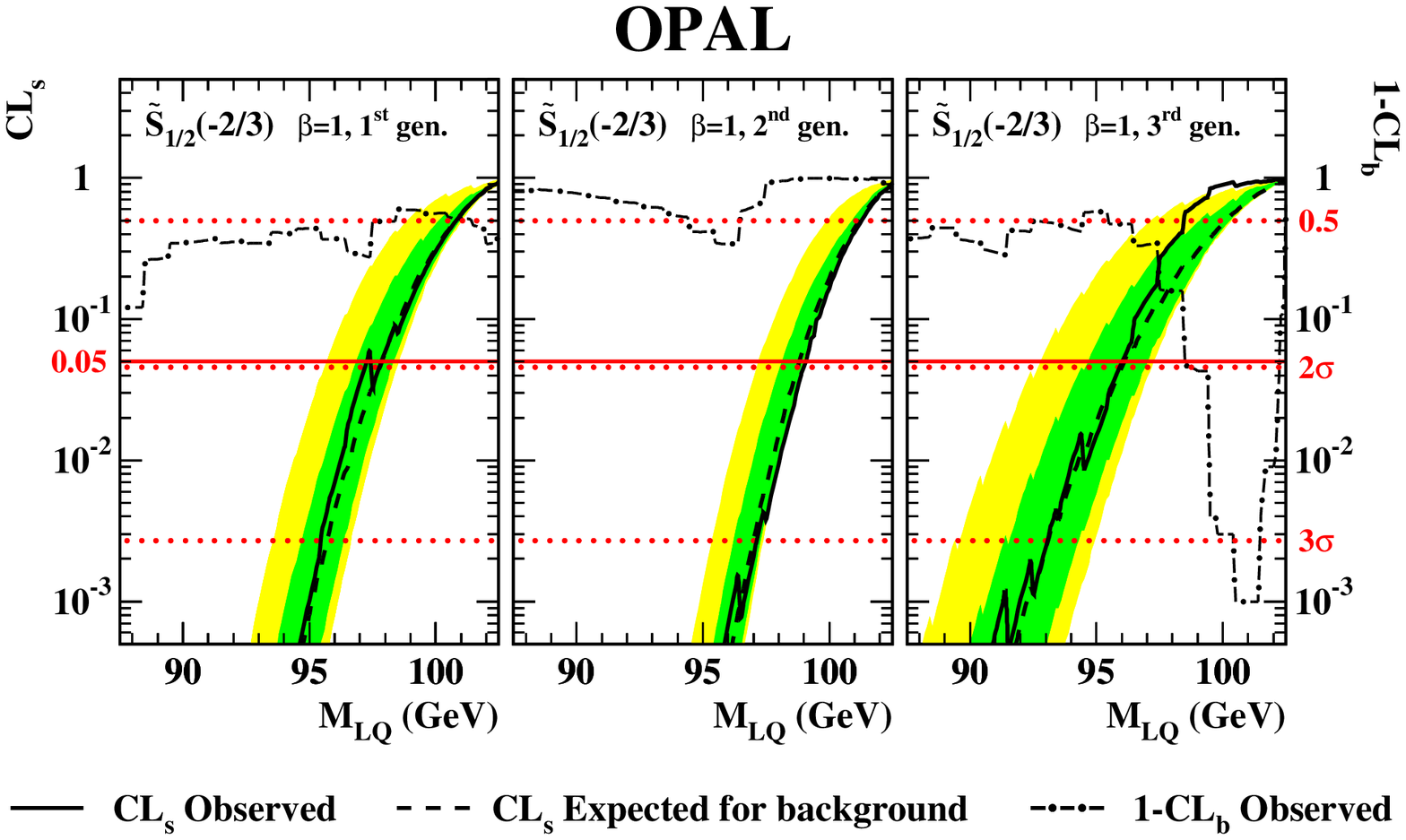,width=13cm,
                    bbllx=40pt,bblly=245pt,bburx=475pt,bbury=520pt}
  \end{center}
 \vspace*{0.5cm}
 \caption[]{\sl Same as Figure~\ref{fig_limits_beta0_scalar}, but for   
                 $\tilde{S}_{1/2}(-2/3)$ with $\beta =1$.
           }
 \label{fig_limits_r2t_2}
 \end{figure}

\begin{figure}[p]
   \vspace*{-1cm}
    \begin{center}
     \end{center}
     \vspace*{-0.8cm}
     \begin{center}
        \hspace*{-1.2cm}
          \epsfig{file=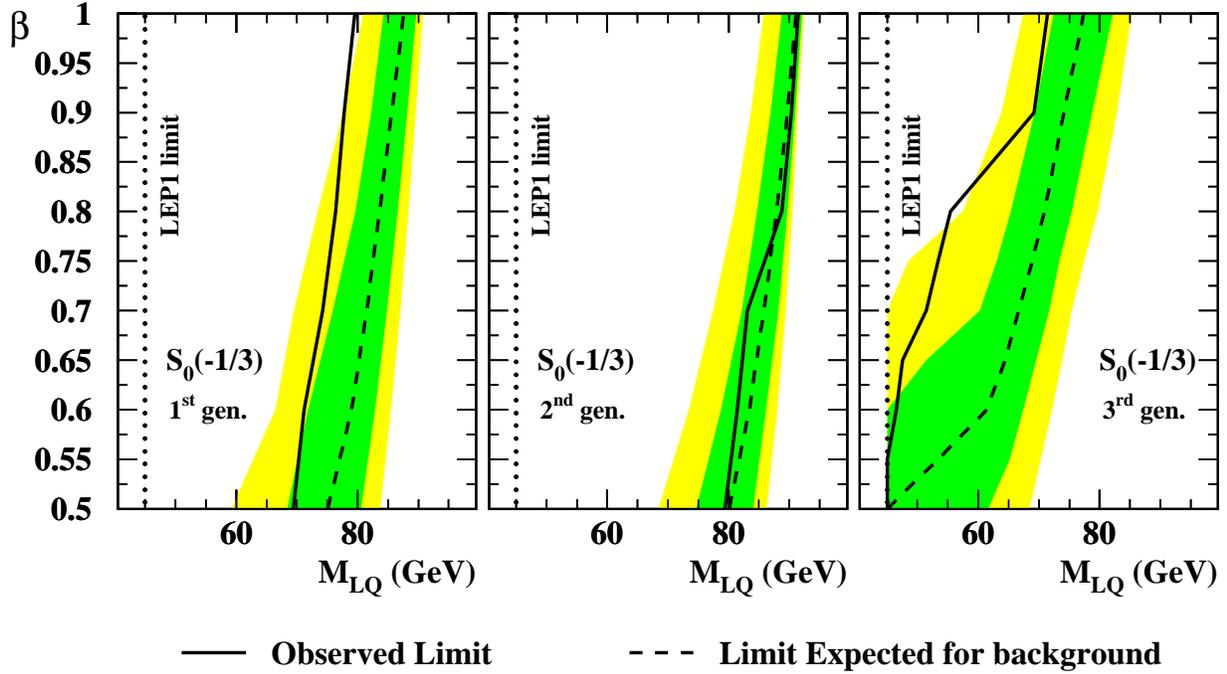,width=13cm,
                        bbllx=40pt,bblly=245pt,bburx=475pt,bbury=520pt}
      \end{center}
     \vspace*{0.5cm}
 \caption[]{\sl The 95$\%$ CL exclusion curves in the 
                  plane $\beta$~{\it vs}~M$_{\mathrm{LQ}}$
                  for $S_{0}(-1/3)$ with possible values of $\beta$ in 
                 the range $\left[0.5 , 1\right]$.
                 The observations for the data are shown with solid
                 lines. The shaded bands indicate the 68\% and 95\%
                 probability intervals with respect to the median
                 expectation in the absence of a signal (dashed lines).  
                 The excluded region is to the left
                  of the solid curve.
            }
 \label{fig_betam_limits_s1}
\end{figure}

\begin{figure}[b]
    \begin{center}
     \end{center}
    \vspace*{-1cm}
    \begin{center}
       \hspace*{-1.2cm}
          \epsfig{file=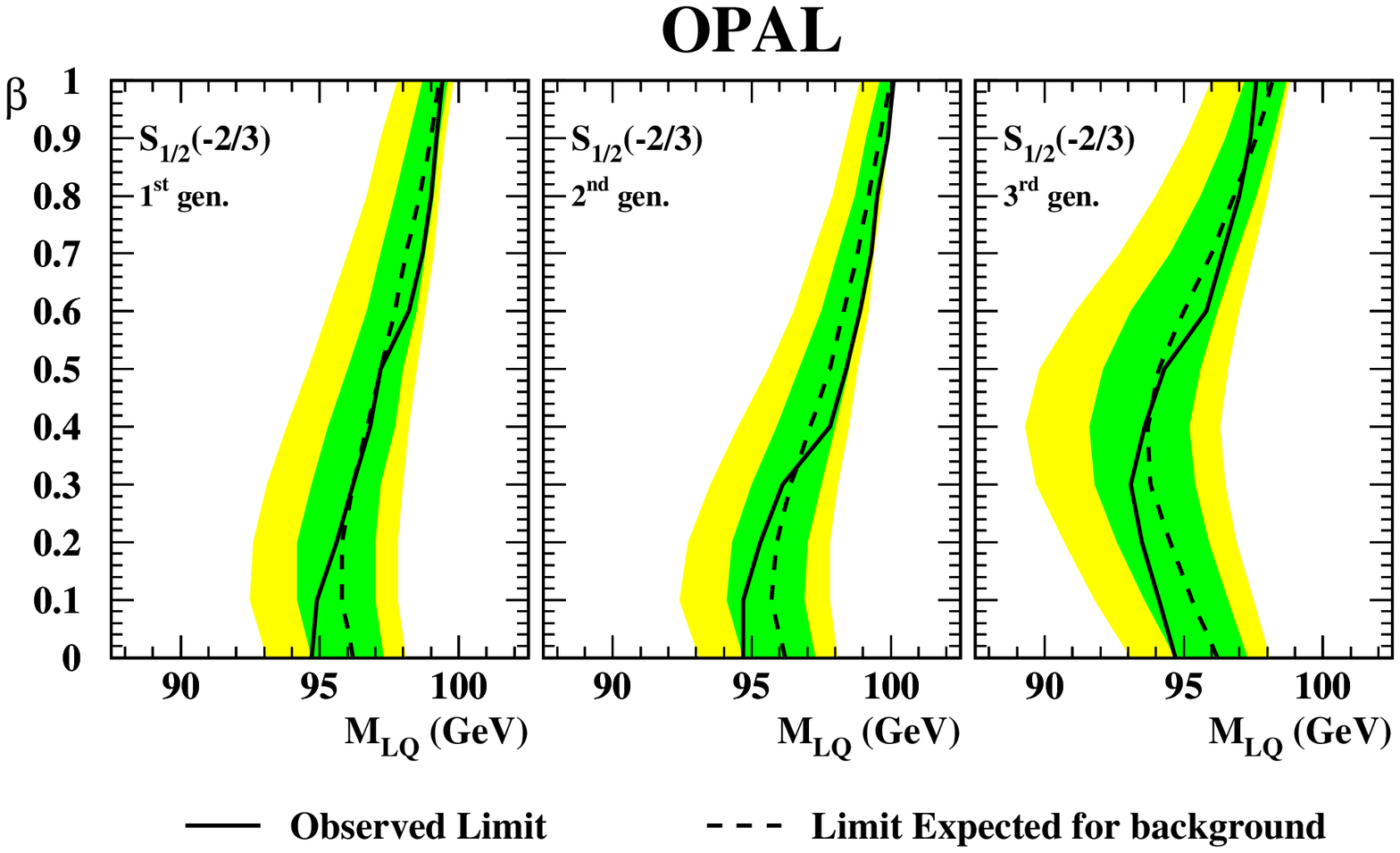,width=13cm,
                       bbllx=40pt,bblly=245pt,bburx=475pt,bbury=520pt}
    \end{center}
    \vspace*{0.5cm}
 \caption[]{\sl Same as Figure~\ref{fig_betam_limits_s1}, 
                  but for ${S}_{1/2}(-2/3)$
                 with possible values of $\beta$ in the range 
                 $\left[0 , 1\right]$.
           }
 \label{fig_betam_limits_r2_2}
 \end{figure}

\begin{figure}[p]
     \begin{center}
     \end{center}
     \vspace*{-1cm}
    \begin{center}
       \hspace*{-1.2cm}
          \epsfig{file=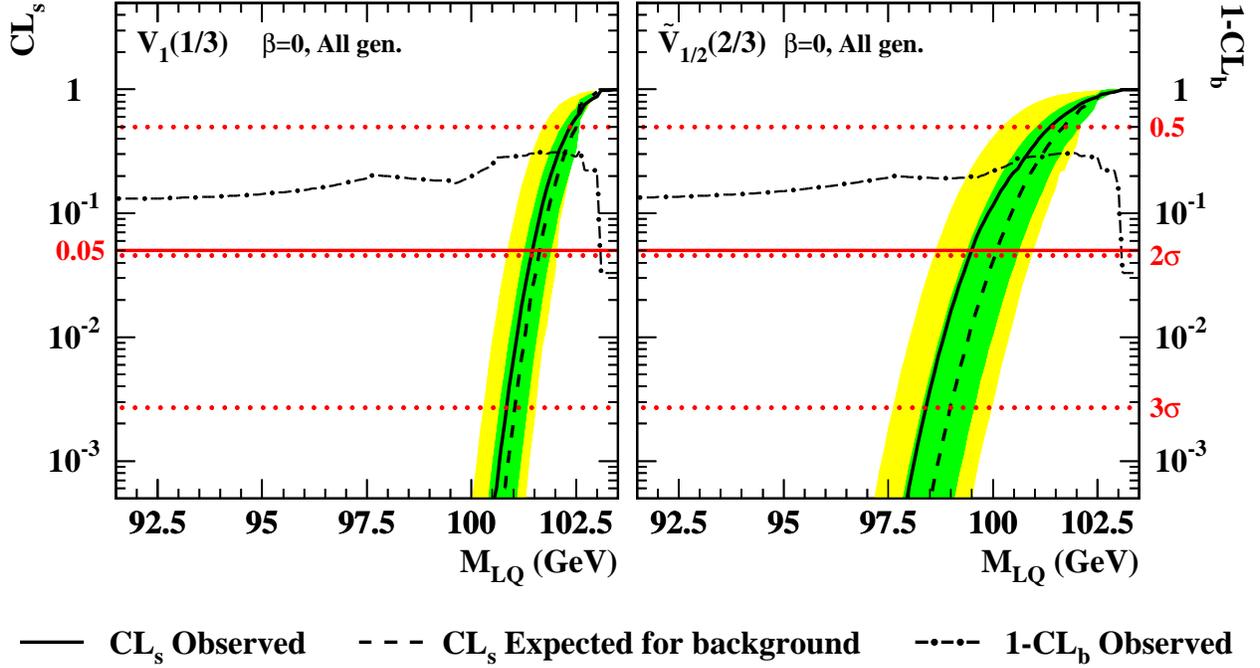,width=13cm,
                       bbllx=40pt,bblly=245pt,bburx=475pt,bbury=520pt}
    \end{center}
    \vspace*{0.5cm}
 \caption[]{\sl  Same as Figure~\ref{fig_limits_beta0_scalar}, but for
                  vector leptoquarks $V_{1}(1/3)$ and $\tilde{V}_{1/2}(2/3)$
                  with $\beta =0$.
           }
 \label{fig_limits_beta0_vector}
\end{figure}

\begin{figure}[b]
      \begin{center}
     \end{center}
     \vspace*{-1cm}
     \begin{center}
       \hspace*{-1.2cm}
           \epsfig{file=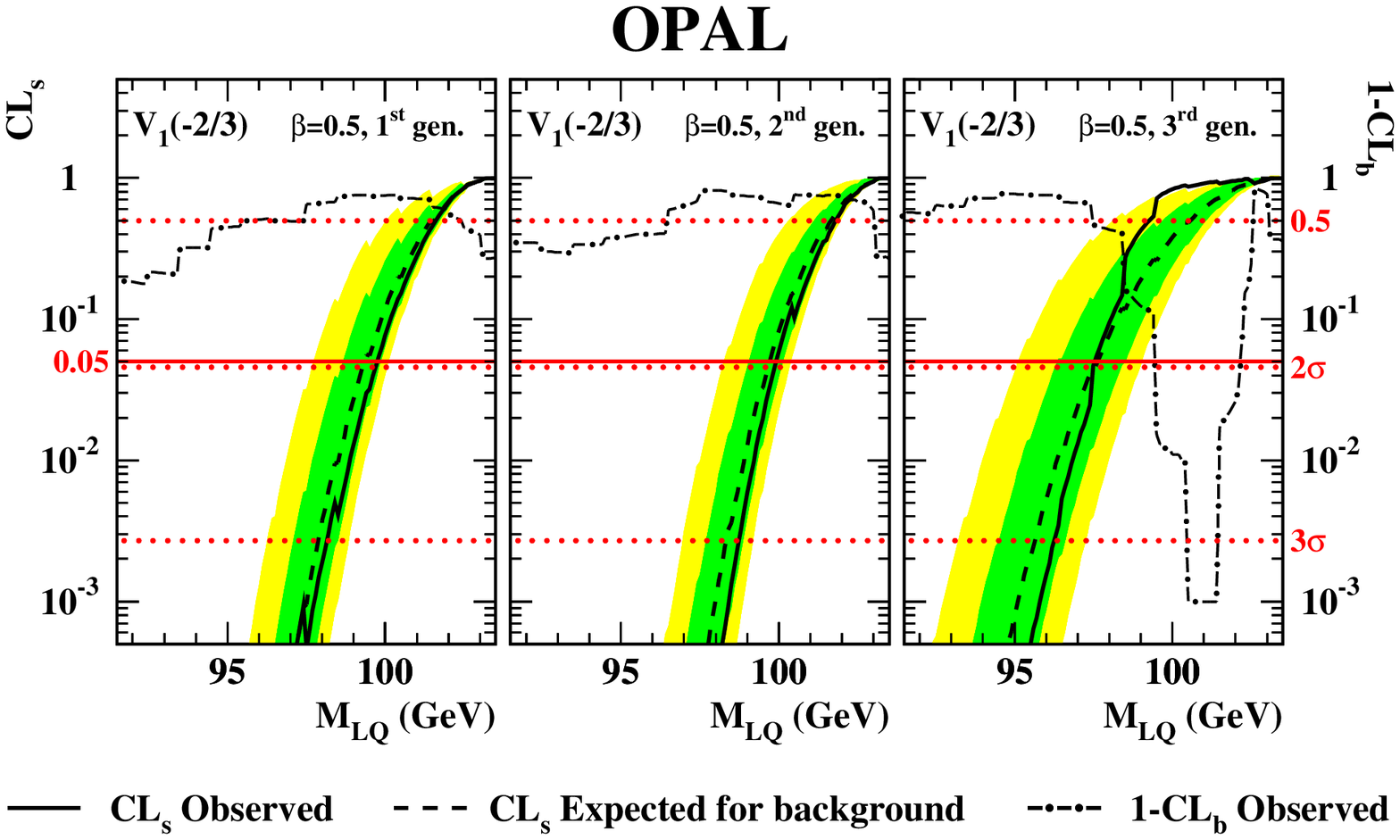,width=13cm,
                   bbllx=40pt,bblly=245pt,bburx=475pt,bbury=520pt}

     \end{center}
     \vspace*{0.5cm}
  \caption[]{\sl Same as Figure~\ref{fig_limits_beta0_vector}, but for
                   $V_{1}(-2/3)$ with $\beta =0.5$.
              }  
 \label{fig_limits_u3_2}
\end{figure}

\begin{figure}[p]
  \vspace*{-1cm}
     \begin{center}
         \hspace*{-1.2cm}
              \epsfig{file=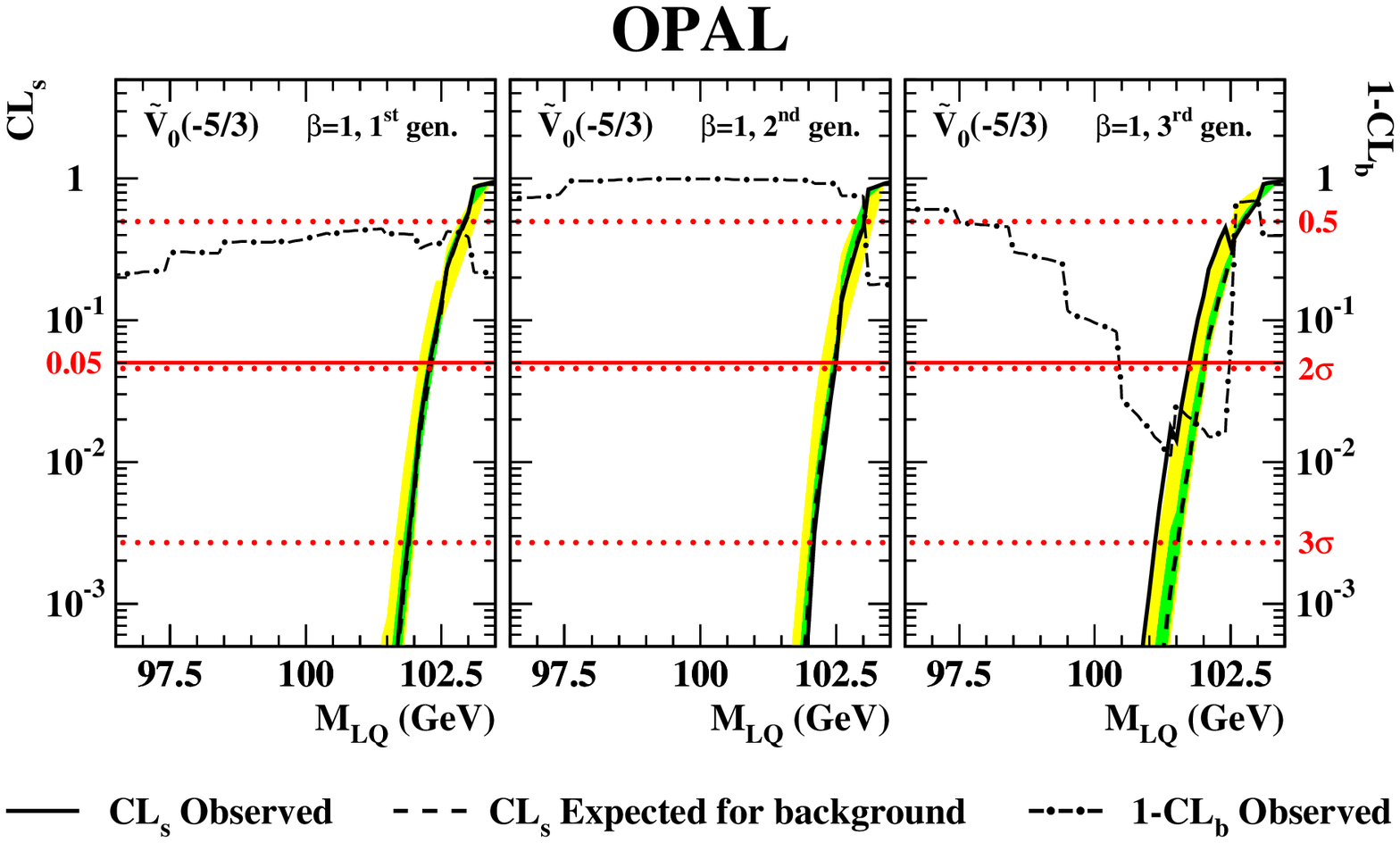,width=13cm,
                       bbllx=40pt,bblly=245pt,bburx=475pt,bbury=520pt}
     \end{center}
     \vspace*{0.5cm}
 \caption[]{\sl Same as Figure~\ref{fig_limits_beta0_vector}, but for   
                $\tilde{V}_{0}(-5/3)$  with $\beta =1$.
           }
\label{fig_limits_u1t}
\end{figure}

\begin{figure}[b]
   \vspace*{-1cm}
      \begin{center}
         \hspace*{-1.2cm}
           \epsfig{file=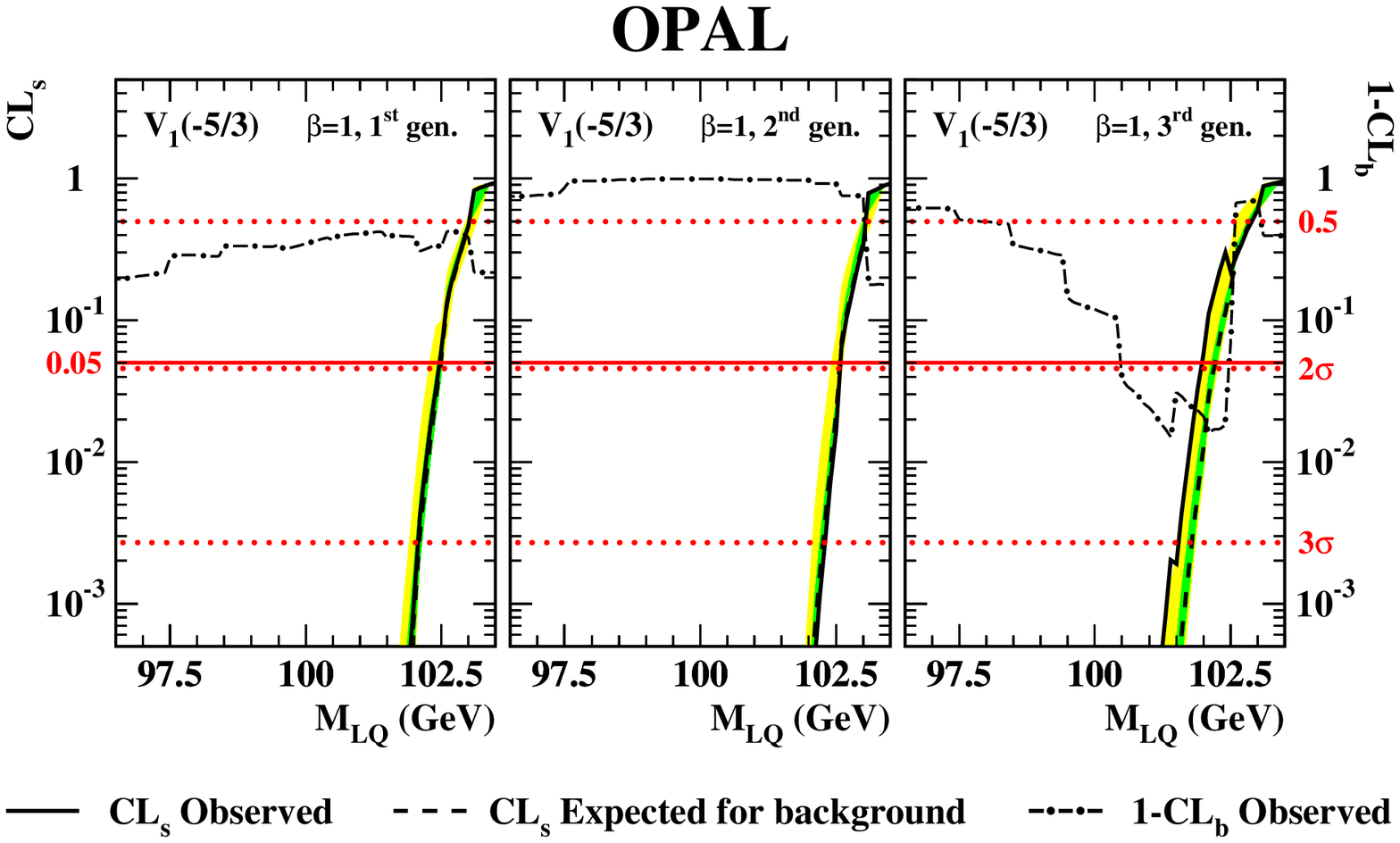,width=13cm,
                       bbllx=40pt,bblly=245pt,bburx=475pt,bbury=520pt}
      \end{center}
       \vspace*{0.5cm}
 \caption[]{\sl Same as Figure~\ref{fig_limits_beta0_vector}, but for   
                $V_{1}(-5/3)$  with $\beta =1$.
           }       
\label{fig_limits_u3_5}
\end{figure}

\begin{figure}[p]
  \vspace*{-1cm}
   \begin{center}
         \hspace*{-1.2cm}
              \epsfig{file=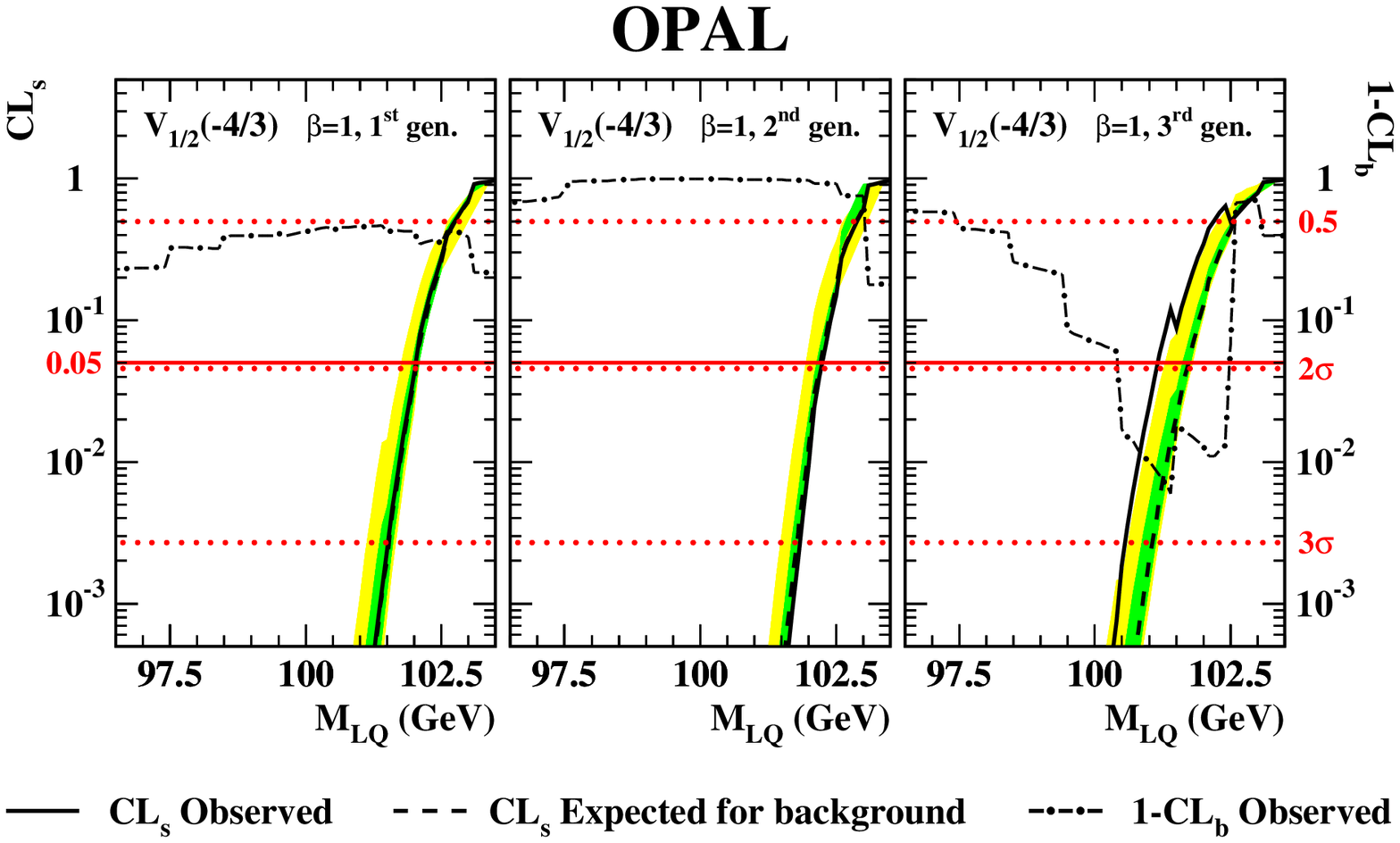,width=13cm,
                       bbllx=40pt,bblly=245pt,bburx=475pt,bbury=520pt}
   \end{center}
\vspace*{0.5cm}
 \caption[]{\sl Same as Figure~\ref{fig_limits_beta0_vector}, but for   
                $V_{1/2}(-4/3)$  with $\beta =1$.
           }
\label{fig_limits_v2_4}
\end{figure}

\begin{figure}[b]
  \vspace*{-1cm}
    \begin{center}
      \hspace*{-1.2cm}
             \epsfig{file=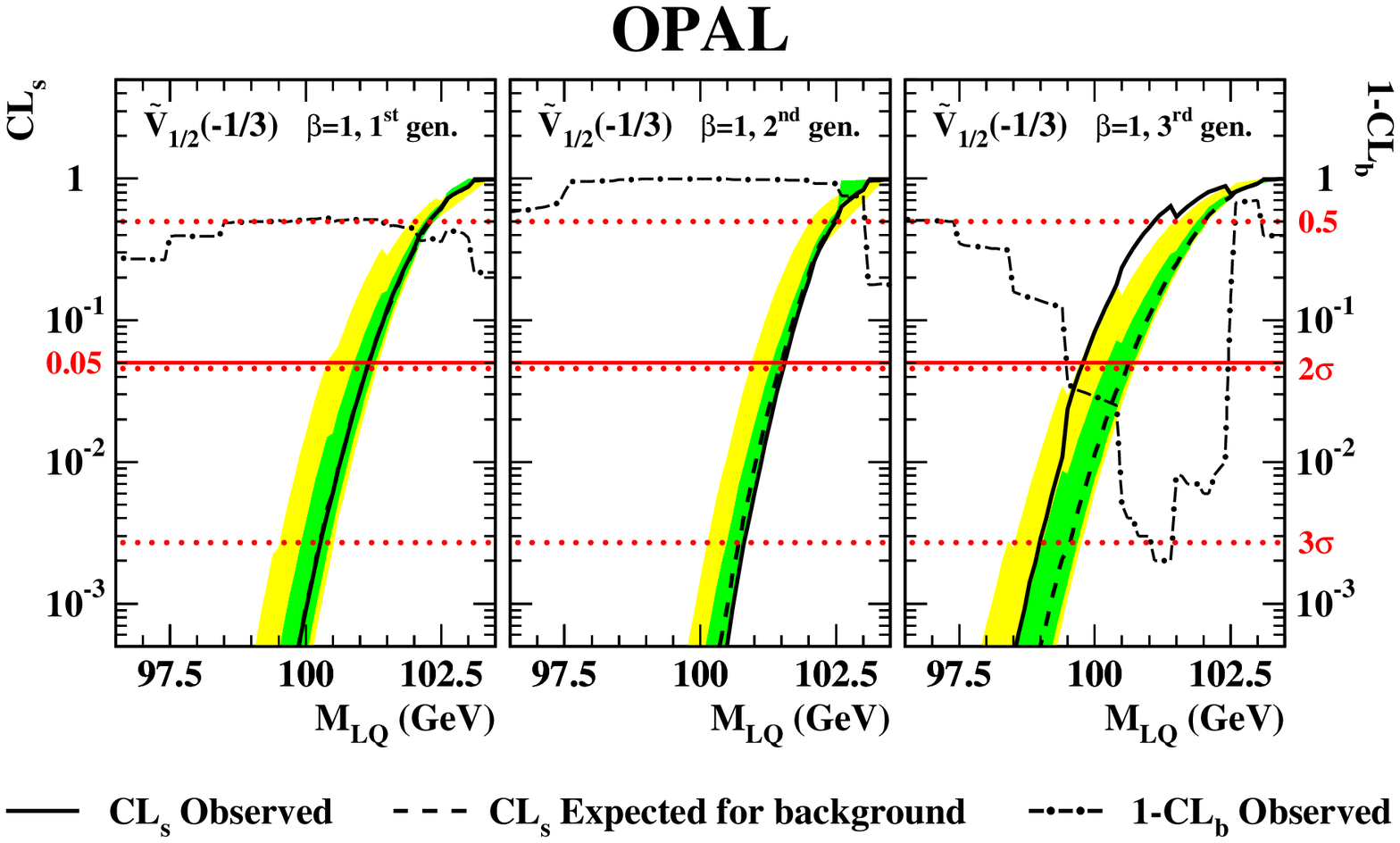,width=13cm,
                       bbllx=40pt,bblly=245pt,bburx=475pt,bbury=520pt}
    \end{center}
 \vspace*{0.5cm}
 \caption[]{\sl Same as Figure~\ref{fig_limits_beta0_vector}, but for   
                 $\tilde{V}_{1/2}(-1/3)$ with $\beta =1$.
           }
 \label{fig_limits_v2t_1}
\end{figure}

\begin{figure}[p]
  \vspace*{-1cm}
   \begin{center}
         \hspace*{-1.2cm}
              \epsfig{file=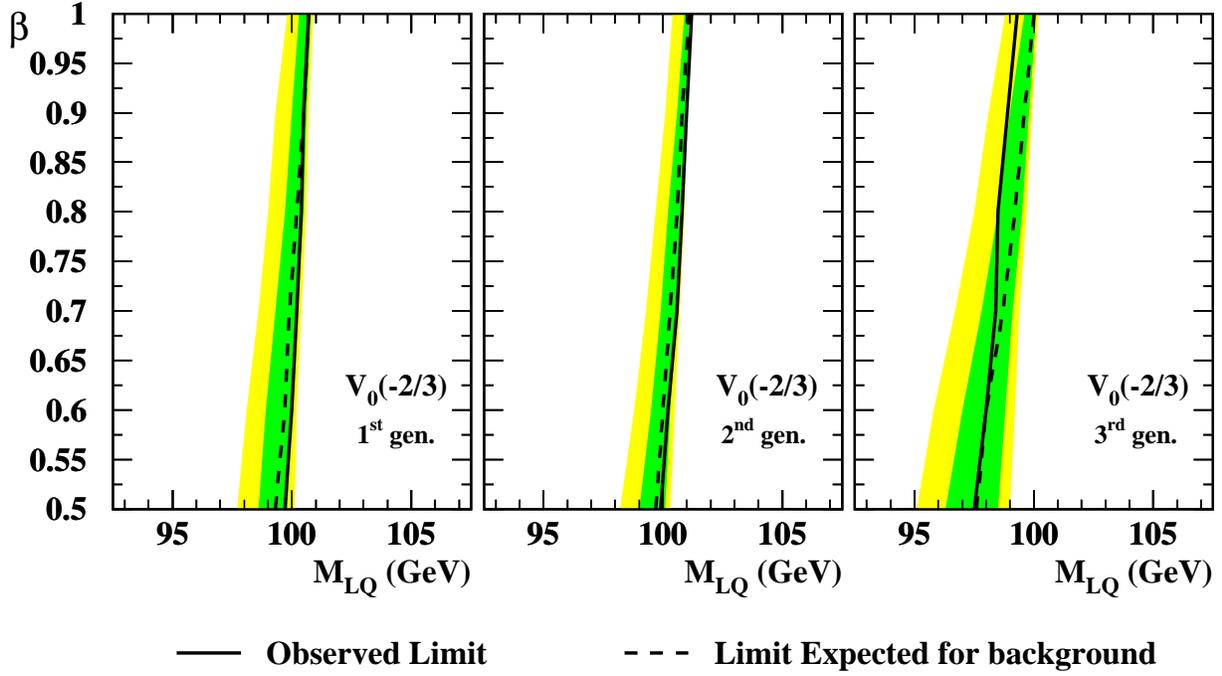,width=13cm,
                       bbllx=40pt,bblly=245pt,bburx=475pt,bbury=520pt}
   \end{center}
\vspace*{0.5cm}
 \caption[]{\sl Same as Figure~\ref{fig_betam_limits_s1},
                 but for vector leptoquark $V_{0}(-2/3)$ 
                 with possible values of $\beta$ in 
                 the range $\left[0.5 , 1\right]$.
           }
 \label{fig_betam_limits_u1}
\end{figure}

\begin{figure}[b]
   \vspace*{-1cm}
     \begin{center}
       \hspace*{-1.2cm}
             \epsfig{file=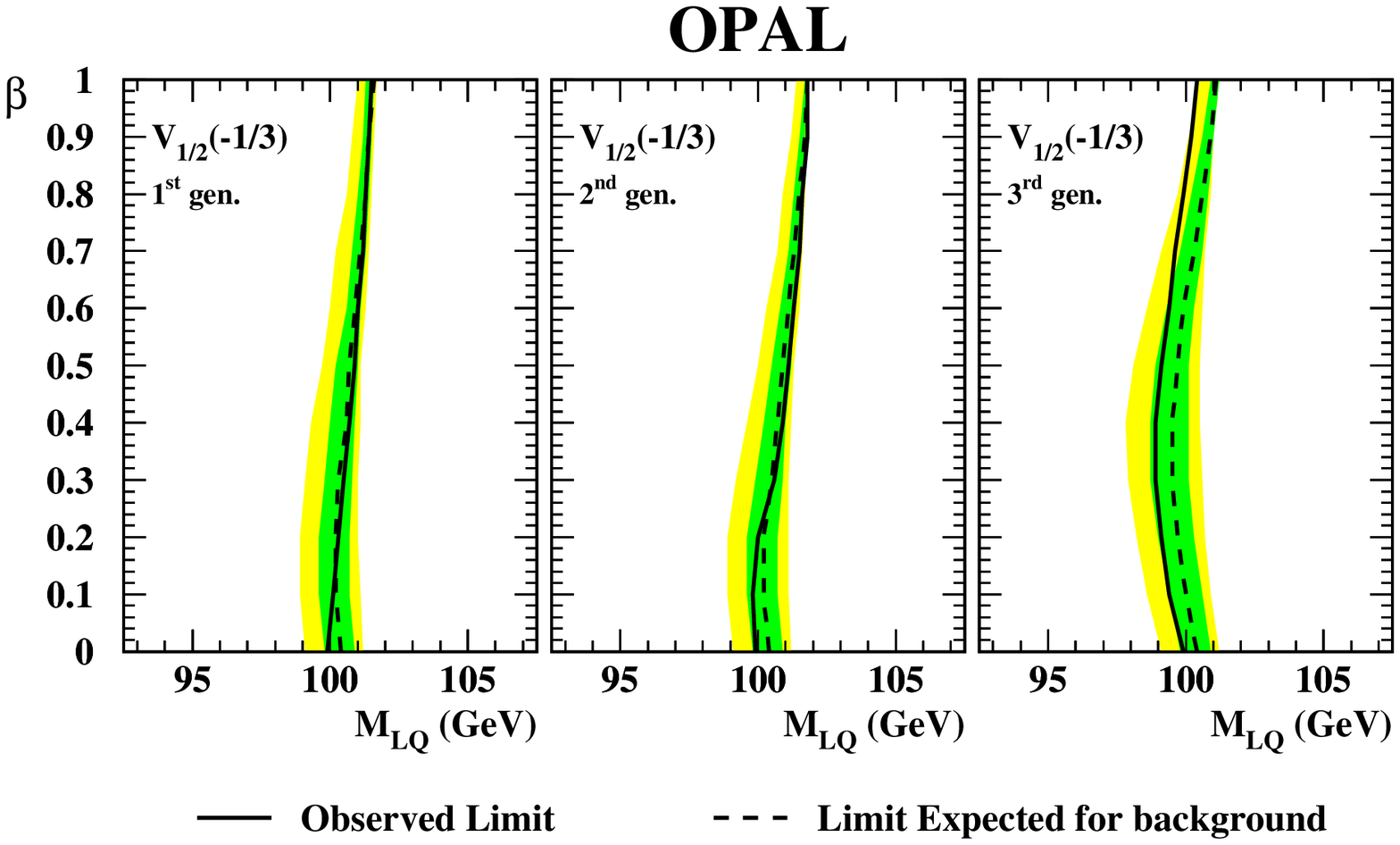,width=13cm,
                       bbllx=40pt,bblly=245pt,bburx=475pt,bbury=520pt}
   \end{center}
    \vspace*{0.5cm}
 \caption[]{\sl Same as Figure~\ref{fig_betam_limits_u1}, 
                  but for ${V}_{1/2}(-1/3)$
                 with possible values of $\beta$ in the range 
                 $\left[0 , 1\right]$.
              }
 \label{fig_betam_limits_v2_1}
 \end{figure}

 Figures~\ref{fig_limits_beta0_scalar}--\ref{fig_limits_r2t_2}
  and \ref{fig_limits_beta0_vector}--\ref{fig_limits_v2t_1} 
 show the
 values of CL$_{s}$ as a function of the leptoquark mass, M$_{\mathrm{LQ}}$,
 for  scalar and vector leptoquarks with the branching ratio
 $\beta$ predicted in the model. 
The lower limit at the  95$\%$ CL on M$_{\mathrm{LQ}}$ corresponds to the 
intersection with the line at  CL$_{s} = 0.05$.
In the same figures the curves representing the values of  $1-{\rm CL}_{b}$
are also shown.
In a Gaussian approximation
a value  $1-{\rm CL}_{b} = 4.55 \times 10^{-2}$ 
would indicate a $2\sigma$ excess beyond the background 
median expectation and $1-{\rm CL}_{b} =2.7 \times 10^{-3}$ would
indicate a $3\sigma$ excess. The vertical scales on the right-hand side of 
Figures~\ref{fig_limits_beta0_scalar}--\ref{fig_limits_r2t_2}
and \ref{fig_limits_beta0_vector}--\ref{fig_limits_v2t_1} 
correspond to this approximation. 
 The regions excluded in the  $\beta-{\rm M}_{\mathrm{LQ}}$ plane of the
 states $S_{0}$, $S_{1/2}(-2/3)$,  $V_{0}$ and $V_{1/2}(-1/3)$, 
 whose $\beta$ depend on  the relative weights of the unknown 
 left and right $\lambda$ couplings, are 
 shown in 
 Figures~\ref{fig_betam_limits_s1},~\ref{fig_betam_limits_r2_2},
 \ref{fig_betam_limits_u1} and~\ref{fig_betam_limits_v2_1}. 
 The mass limits obtained 
 are summarized in 
 Table~\ref{tab:limits}.
 Because of the very small cross-section and the lower efficiency
of the selection for the $\tau^{\pm}\nu {\rm q q}$ channel, 
this search can only improve 
 previous lower limits on the mass  of 
 the third generation state 
 $S_0(-1/3)$
 over a part of the $\beta$ range, while for the third generation 
state $S_1(-1/3)$ with $\beta = 0.5$
 no improvement is possible.   
\begin{table}[htb]
\begin{center}
{\footnotesize
           \begin{tabular}{|ccrcc|ccccc|}
\hline \hline
 \rule[-2.5mm]{0pt}{7mm}$\mathrm{LQ}$ & & $Q_{{\rm em}}$ & &$\beta$ 
                         & \multicolumn{5}{c|}{Generation} \\
 \rule[-2.5mm]{0pt}{5mm}             & &             & &  
                         & 1   &  &   2  & &  3  \\
  \hline \hline
\rule[-2.5mm]{0pt}{7.5mm} $S_{0}$
                      &    & -1/3 & &[0.5,1] &  69 
                                                  & & 
                                                79
                                                  & & 
                                                45$(\ast)$
                                            \\
  \hline
\rule[-2.5mm]{0pt}{7.5mm}$\tilde{S}_{0}$
                      &     & -4/3 &  &  1    & 99
                                              &  & 
                                                100 
                                              & & 98  \\
  \hline
\rule[-2.5mm]{0pt}{7.5mm}
                      &      & 2/3  & &   0    
    & \multicolumn{2}{r}{$\longleftarrow ---$} &  97   
                       & \multicolumn{2}{l|}{$ --- \longrightarrow$} \\
\rule[-2.5mm]{0pt}{7.5mm}$S_{1}$
              &      & -1/3 & &  0.5 & 69   &  & 79  & & 45$(\ast)$     \\
\rule[-2.5mm]{0pt}{7.5mm}
                      &     & -4/3 &  &  1    & 100 &  & 101  & & 99  \\
  \hline
\rule[-2.5mm]{0pt}{7.5mm}
                      &      & -2/3 & &  [0,1]  & 
                                                    94
                                                        & & 
                                                    94
                                                        & & 
                                                    93 
                                                \\
                     $S_{1/2}$   & &    &    &         &    &  &     &  &   \\
\rule[-2.5mm]{0pt}{7.5mm}
                      &      & -5/3 & &   1    & 100  & & 100  & & 98  \\
 \hline
\rule[-2.5mm]{0pt}{7.5mm}
                      &      & 1/3 & &    0 
    & \multicolumn{2}{r}{$\longleftarrow ---$} &  89 
                    & \multicolumn{2}{l|}{$ --- \longrightarrow$}  \\
              $\tilde{S}_{1/2}$&   & &  &         &      &     &  &   &    \\
\rule[-2.5mm]{0pt}{7.5mm}
                      &        & -2/3 & &   1 & 97  & &   99  & & 96   \\ 
%
%
    \hline \hline
\rule[-2.5mm]{0pt}{7.5mm} $V_{0}$
                       &      & -2/3 & & [0.5,1] &    99  
                                              & &     99  
                                              & &     97  
                                                 \\
    \hline
\rule[-2.5mm]{0pt}{7.5mm}$\tilde{V}_{0}$
                       &       & -5/3 & &   1    &102  & &102  & & 101  \\ 
   \hline
\rule[-2.5mm]{0pt}{7.5mm}
                       &      & 1/3  & &   0    
       & \multicolumn{2}{r}{$\longleftarrow ---$} & 101   
                 &  \multicolumn{2}{l|}{$ --- \longrightarrow$} \\
\rule[-2.5mm]{0pt}{7.5mm}$V_{1}$
                       &      & -2/3 & &   0.5   & 99 & & 99  & & 97  \\ 
\rule[-2.5mm]{0pt}{7.5mm}
                       &      & -5/3 & &  1  &102  & & 102  & & 101 \\
  \hline
\rule[-2.5mm]{0pt}{7.5mm}
                       &      & -1/3 & & [0,1]  &  99  
                                              & &  99 
                                              & &  98    
                                               \\
                     $V_{1/2}$
                             &     & &  &         &     & &    &  &      \\
\rule[-2.5mm]{0pt}{7.5mm}
                      &       & -4/3 & &   1    & 102  & & 102 & & 101  \\
  \hline
\rule[-2.5mm]{0pt}{7.5mm}
                      &       & 2/3  & &   0    
    &  \multicolumn{2}{r}{$\longleftarrow ---$}      & 99  
                    & \multicolumn{2}{l|}{$ --- \longrightarrow$}  \\
                     $\tilde{V}_{1/2}$& & &    &     &    & &    &   &     \\
\rule[-2.5mm]{0pt}{7.5mm}
                      &        & -1/3 & &    1 &101  & & 101 & & 99  \\ 
  \hline \hline
\end{tabular}
  }
\caption{\sl The $95\%$ CL lower limits on scalar and vector 
             leptoquarks masses, 
             in GeV,  as 
             obtained from the present analysis. 
             $\beta$ is the branching ratio into a charged lepton and a quark.
             Limits obtained by OPAL using LEP1 data
             are marked with ($\ast$).
        }
\label{tab:limits}
\end{center}
\end{table}
%
%
\section{Conclusions}
%
The data collected with the OPAL detector at $\sqrt{s}$ between
189 and 209 GeV,
corresponding to a total integrated luminosity of  596~pb$^{-1}$, 
are analysed to search for events with pair produced  leptoquarks of
all three generations.
The present analysis covers the region of small values of the couplings
$\lambda$ to fermions (from 10$^{-6}$ to 10$^{-2}$).
No significant signal-like excess 
with respect to Standard Model predictions is found 
in the data. 
Lower mass limits are set for scalar and vector leptoquarks under 
the assumption that,
for each scenario, 
only one leptoquark state contributes to the cross-section.
The present results improve most of the previous LEP lower limits
on  leptoquark masses derived from 
searches for events due to the pair production process~\cite{LEP1,mypaper}  
by 10--25 GeV, depending on the leptoquark quantum numbers.
%
%

%
%
%
%
\vspace*{-3cm}
%
%
%
%
%
%
%
\vspace*{-0.3cm} 
%
%
%
%
\vspace*{-2cm}
%
%
%
%
%
%
%
%
%
%
%
%
%
\clearpage
\newpage
%
%
%
%
\clearpage
\newpage
%
%
%
%
\clearpage
\newpage
%
%
%
%
%
\clearpage
\newpage
%
%
%
\clearpage
\newpage
%
%
%
%
%
\end{document}